\documentclass[5p,twocolumn,number,sort&compress,times]{elsarticle}

\journal{Nuclear Instruments and Methods in Physics Research A}

\usepackage{datetime}

\usepackage{graphics}
\usepackage{color}
\usepackage{graphicx}
\usepackage{amssymb}
\usepackage{amsmath}
\usepackage{epstopdf}
\usepackage{xspace}

\usepackage{mathbbol}
\usepackage{wasysym} 
\usepackage{bbm}     
 
\usepackage{rotating}
\usepackage{dcolumn}
\usepackage{bm}
\usepackage{longtable}
\usepackage{multirow}
\usepackage{enumitem}

\usepackage{ifpdf}
\ifpdf
\usepackage[pdftex]{hyperref}
\else
\usepackage[hypertex]{hyperref}
\fi

\hypersetup{
 pdftitle={},%
 pdfauthor={},%
 pdfsubject={},%
 pdfkeywords={},%
 pdfstartview={},%
 bookmarksopen=true, breaklinks=true, debug=true,
 colorlinks=true, linkcolor=blue, citecolor=red, urlcolor=red,
 hyperfigures=true
}

\def\figwid{3.50in}

\def\FThing#1#2{#1~\ref{#2}}
\def\FThings#1#2{\FThing{#1s}{#2}}

\def\Thing#1#2{#1.~\ref{#2}}
\def\Things#1#2{\Thing{#1s}{#2}}

\def\Sec#1{\Thing{Sect}{#1}}

\def\Fig#1{\Thing{Fig}{#1}}
\def\Figs#1{\Things{Fig}{#1}}

\def\Figure#1{\FThing{Figure}{#1}}
\def\Figures#1{\FThings{Figure}{#1}}

\def\Tab#1{\FThing{Table}{#1}}
\def\Tabs#1{\FThings{Table}{#1}}

\def\Eq#1{Eq.~(\ref{#1})}
\def\Eqs#1{Eqs.~(\ref{#1})}

\def\beq{\begin{equation}}
\def\eeq{\end{equation}}
\def\beqa{\begin{eqnarray}}
\def\eeqa{\end{eqnarray}}
\def\non{\nonumber}

\newcommand{\mi}  {~\ensuremath{{\mu\mathrm{m}}     }}
\newcommand{\nm}  {~\ensuremath{{\mathrm{nm}}     }}
\newcommand{\ns}  {~\ensuremath{{\mathrm{ns}}     }}

\newcommand{\mm}  {~\ensuremath{{\mathrm{mm}}     }}

\newcommand{\ie}   {{\it i.e.,}~}
\newcommand{\eg}   {{\it e.g.,}~}
\newcommand{\etal} {{\it {et al.}}}

\newcommand{\kev}  {~\ensuremath{{\mathrm{keV}}     }}
\newcommand{\gev}  {~\ensuremath{{\mathrm{GeV}}     }}

\newcommand{\enr}{\epsilon}
\newcommand{\ecrit}{\ensuremath{\enr_c}}
\newcommand{\gtsq}{{\gamma^2\,\theta^2}}
\newcommand{\amp}{\ensuremath{A_b}}
\newcommand{\siz}{\ensuremath{{\sigma_b}}}
\newcommand{\sizta}{\ensuremath{{\left<{\sigma_b}\right>}}}
\newcommand{\sizts}{\ensuremath{{\Sigma_b}}}
\newcommand{\sizcts}{\ensuremath{{\Sigma_{bc}}}}

\newcommand{\dtr}{\ensuremath{{d}}}
\newcommand{\off}{\ensuremath{{y_b}}}
\newcommand{\Eb}{\ensuremath{{E_b}}}

\newcommand{\ap}{\ensuremath{{a^{\,\prime}}}}
\newcommand{\yp}{\ensuremath{{y^{\,\prime}}}}

\newcommand{\rp}{\ensuremath{{r^{\,\prime}}}}
\newcommand{\lp}{\ensuremath{{L^{\,\prime}}}}
\newcommand{\thetap}{\ensuremath{{\theta^{\,\prime}}}}
\def\vee#1{\ensuremath{\mathcal{V}(\,#1\,)}}
\def\prfv#1#2{\ensuremath{\mathcal{G}(\,#1,\,#2\,)}}
\def\prf#1{\ensuremath{\mathcal{F}(\,#1\,)}}
\def\fcn#1#2#3#4#5{\ensuremath{\mathcal{H}(#1;#2,#3,#4,#5\,)}}
\def\hfcn#1#2#3{\ensuremath{\mathcal{\hat{H}}(#1;#2,#3\,)}}
\newcommand{\fitf}{\fcn{\,\yp}{\amp}{\siz}{\off}{\dtr}}
\newcommand{\pfcn}{\amp\,\hfcn{\,\yp-M\off - \dtr}{\siz}{\off}}
\newcommand{\basicf}{\hfcn{\,\yp}{\siz}{\off}}
\newcommand{\basicjk}{\ensuremath{\mathcal{\hat{H}}_{jk}(\yp)}}
\def\ds#1{{\displaystyle #1}}

\def\pj#1{\ensuremath{\mathcal{P}_j(#1)}}
\newcommand{\srd}{\ensuremath{\mathcal{S}_d(\enr)}}
\newcommand{\srf}{\ensuremath{\mathcal{S}_f(\enr)}}
\newcommand{\pjs}{\pj{\siz}}
\newcommand{\pjd}{\pj{\siz+\delta}}
\newcommand{\qs}{\ensuremath{\mathcal{Q}(\siz)}}
\newcommand{\qz}{\ensuremath{\mathcal{Q}_0}}
\newcommand{\pz}{\ensuremath{\mathcal{P}_0}}
\newcommand{\dsd}{\ensuremath{\mathcal{D}(\siz,\delta)}}

\newcommand{\Eimg}{\ensuremath{{{\bf E}_{\rm img}}}}
\newcommand{\Esyn}{\ensuremath{{{\bf E}_{\rm syn}}}}
\newcommand{\Eprp}{\ensuremath{{{E}_\perp}}}
\newcommand{\Epll}{\ensuremath{{{E}_\parallel}}}
\newcommand{\uprp}{{\ensuremath{\hat{\mathrm{\bf u}}_\perp}}}
\newcommand{\upll}{{\ensuremath{\hat{\mathrm{\bf u}}_\parallel}}}

\newcommand{\sth}{{\ensuremath{\sigma_\theta}}}

\def\NF{{\tt NF}}
\def\Di{{\tt Di}}
\def\Mo{{\tt Mo}}
\def\Al{{\tt Al}}

\def\MATLAB{{\tt MATLAB}\textregistered}
\def\ROOT{{\tt ROOT}}
\def\FFTW{{\tt FFTW}}
\def\CPP{{\tt C++}}
\def\EXCEL{{\tt EXCEL}\textregistered}
\def\MINUIT{{\tt MINUIT}}
\def\prftt{{\tt prf}}
\def\PRFtt{{\tt PRF}}

\begin{document}

\title{
Vertical beam size measurement in the CESR-TA $e^+e^-$ storage ring\\
using x-rays from synchrotron radiation}

\renewcommand{\thefootnote}{\fnsymbol{footnote}}

\author{J.~P.~Alexander}
\author{A.~Chatterjee}
\author{C.~Conolly}
\author{E.~Edwards}
\author{M.~P.~Ehrlichman}
\author{E.~Fontes}
\author{B.~K.~Heltsley\corref{cor1}}\ead{bkh2@cornell.edu}
\author{W.~Hopkins\fnref{fnhop}}
\author{A.~~Lyndaker}
\author{D.~P.~Peterson}
\author{N.~T.~Rider}
\author{D.~L.~Rubin}
\author{J.~Savino\fnref{fnsav}}
\author{R.~Seeley, and}
\author{J.~Shanks}
\address{Cornell University, Ithaca, NY 14853, USA}
\author{J.~W.~Flanagan}
\address{High Energy Accelerator Research Organization (KEK), Tsukuba, Japan}
\cortext[cor1]{Corresponding author}
\fntext[fnhop]{Current address: Department of Physics, University of
  Oregon, Eugene, OR 97403, USA}
\fntext[fnsav]{Current address: National Superconducting Cyclotron Laboratory,
               Michigan State University,
               East Lansing, MI 48824, USA}

\begin{abstract}
{\small
We describe the construction and operation of an x-ray beam size 
monitor (xBSM), a device measuring $e^+$ and $e^-$ beam sizes in the 
CESR-TA storage ring using synchrotron radiation. The device can measure
vertical beam sizes of $10-100~\mu$m on a turn-by-turn, bunch-by-bunch
basis at $e^\pm$ beam energies of $\sim2~$GeV. At such beam energies 
the xBSM images x-rays of $\epsilon\approx$1-10$~$keV 
($\lambda\approx 0.1-1~\nm$) that emerge from a hard-bend magnet 
through a single- or multiple-slit (coded aperture) optical element
onto an array of 32 InGaAs photodiodes with 50$~\mu$m pitch.
Beamlines and detectors are entirely in-vacuum, enabling single-shot 
beam size measurement down to below 0.1$~$mA ($2.5\times10^9$ particles) 
per bunch and inter-bunch spacing of as little as 4$~$ns.
At $E_{\rm b}=2.1~$GeV, systematic precision of $\sim 1~\mu$m is achieved 
for a beam size of $\sim12~\mu$m; this is expected to scale as 
$\propto 1/\sigma_{\rm b}$ and $\propto 1/E_{\rm b}$.
Achieving this precision requires comprehensive alignment 
and calibration of the detector, optical elements, and x-ray beam.
Data from the xBSM have been used to extract characteristics
of beam oscillations on long and short timescales, and
to make detailed studies of low-emittance tuning, intra-beam scattering, 
electron cloud effects, and multi-bunch instabilities.
}
\end{abstract}
\begin{keyword}
{\small
electron beam size \sep x-ray diffraction \sep pinhole
\sep coded aperture \sep synchrotron radiation \sep electron storage ring
}
\end{keyword}

\date{\today}
\maketitle

\renewcommand{\thefootnote}{\arabic{footnote}}

{\small \tableofcontents}
\newpage

\section[Introduction]{Introduction}

Precision measurement of vertical bunch size plays an increasingly important
role~\cite{ref:flan} in the design and operation of the current and future 
generation of electron storage rings. Such facilities include an $e^+e^-$ 
collider (Super-KEKB~\cite{ref:KEKB}), an $e^-$-ion collider (EIC~\cite{ref:eRHIC}), 
a multitude of present and future light 
sources~\cite{ref:borland,ref:lightsources} 
(\eg PEPLS~\cite{ref:PEPLS}), and a positron damping ring for an $e^+e^-$ 
linear collider (ILC~\cite{ref:ILCTDR} or CLIC~\cite{ref:CLIC}).
For each of these machines, performance (\eg luminosity or brightness) 
is directly related to the vertical emittance (and thus the vertical beam 
size). By providing the operator with real-time vertical beam size 
information, the accelerator can be tuned in a predictable, stable, and 
robust manner. Optical transition radiation~\cite{ref:otr,ref:otrtwo},
optical diffraction radiation~\cite{ref:odr,ref:odrtwo,ref:odrthree}, and
synchrotron radiation imaging have been successfully used
for such measurements using a variety of techniques. Synchrotron
radiation can be used in the visible with a streak camera~\cite{ref:streak},
an interferometer~\cite{ref:srint,ref:srinttwo,ref:srintthree}, 
or a vertical polarization monitor~\cite{ref:vertpol};
x-rays can be used with a pinhole 
camera~\cite{ref:elle,ref:ewald,ref:diamond},
refractive lenses~\cite{ref:refr}, Fresnel zone 
plates~\cite{ref:fzp,ref:fzptwo,ref:fzpthree},
or a Bragg polarimeter~\cite{ref:bragg}.
Challenges persist in obtaining precision at low beam size and current, 
as well as ability to distinguish among closely spaced 
bunches in extended trains, for which electron cloud effects can be large.

\begin{table}[b]
    \caption{Parameters of the Cornell Electron
    Storage Ring Test Accelerator (CESR-TA).}\label{tab:cesr_params}
\setlength{\tabcolsep}{0.65pc}
\begin{center}
\small\selectfont
\begin{tabular}{ll}
\hline\hline
\rule[10pt]{-1mm}{0mm}
Parameter & Value\\
\hline
        Circumference & 768.4~m \\
        Circulation Time  & 2.563~$\mu$s\\
        Circulation Frequency  & 390.1~kHz\\
        Beam Energy            & 2.085 (1.5-5.3)\gev\\
        Species           & $e^+$ \underline{or} $e^-$ \\
        RF Frequency      & 500~MHz \\
        Harmonic Number   & 1281 \\
        Bunch Spacing     & $\ge$4 ns \\
        Bunch Population  & 0.1-10$\times 10^{10}$  \\
        Number of Bunches/Turn & $\le45$ \\
        Horizontal emittance & 2.6~nm \\
        Vertical emittance & $\ge$10~pm \\
        Longitudinal bunch size (rms) & 10-15\mm \\
        Horizontal bunch size (rms) & 170-300\mi \\
        Vertical bunch size (rms) & 10-100\mi \\
\hline\hline
\end{tabular}
\end{center}
\end{table}

In 2008 the Cornell Electron Storage Ring (CESR) was reconfigured as 
a Test Accelerator (CESR-TA~\cite{ref:CTA}) with the intent of studying 
electron and positron beam dynamics. Parameters for the CESR
storage ring are shown in \Tab{tab:cesr_params}. The CESR storage 
ring is uniquely equipped for examining single-beam phenomena limiting 
the performance of next-generation colliders and storage rings. As a 
former collider which operated in both the $c\bar c$ and $b\bar b$ threshold
regions, CESR is able to circulate either electrons or positrons, and to
do so with beam energies from 1.5-5.3~GeV, enabling controlled measurements of 
charge- and energy-dependence.

One component of the CESR-TA program is the xBSM (x-ray Beam Size Monitor).
The desired specifications on vertical beam size instrumentation pose 
significant technical challenges: attaining sensitivity to beam size at the
$\sim$10-100\mi\ level with precision at the few\mi\ level; 
bunch-by-bunch, turn-by-turn measurements on a train of bunches with 
inter-bunch spacing of 4\ns; 
single-shot operation for 0.1-10~mA bunch current;
and fast feedback on beam size measurement to facilitate real-time tuning.
This article describes the xBSM conceptual basics and
data analysis methods in \Sec{sec:meth},
apparatus in \Sec{sec:app}, operational procedures and experience
in \Sec{sec:ops},
measurement capabilities in \Sec{sec:results}, and important 
conclusions in \Sec{sec:conc}.

\section[Methods \& tools]{Methods \& tools}
\label{sec:meth}

Each xBSM installation images x-rays from a hard-bend magnet through 
a single-  or multiple-slit optical element\footnote{We refer to a 
single slit optical element as a {\it pinhole} (PH) 
and multiple-slit element as {\it coded aperture} (CA).}
onto a detector, a 32-pixel array of photodiodes read out with fast 
electronics. Before describing the
apparatus in detail in \Sec{sec:app}, this section will review the conceptual
basis of the device, with particular attention to design principles
and tools as well as analysis techniques.

\subsection{X-ray spectrum}
\label{sec:spectrum}

The design and performance of optical and detector elements depend
strongly upon the x-ray spectrum. Here we review the formalism relevant
to synchrotron radiation.
The radiated power spectrum of a charged particle in a uniform 
magnetic field is given by the well-known formulae developed by  
Larmor~\cite{ref:Larmor}, Li\'{e}nard~\cite{ref:Lienard}, 
Wiechert~\cite{ref:Wiechert}, and, most prominently,
Schott~\cite{ref:Schott}, which were later summarized in modern form by
Schwinger~\cite{ref:Schwinger} and Jackson~\cite{ref:Jackson}.
The synchrotron radiation spectrum is a direct function of beam energy $E_b$, 
angle $\theta$ out of the orbital plane, instantaneous magnetic bend radius
$\rho$, and photon energy 
$\enr\equiv\hbar\omega=hc/\lambda$, where $h$ ($\hbar$) is the (reduced) 
Planck constant, $\omega$ is the radial frequency, $\lambda$ 
is the wavelength, and $c$ is the speed of light.

We define the Lorentz factor $\gamma$, the wave number $k$, 
the peak of the incident power spectrum \ecrit\  
(also known as the critical energy\footnote{The 
definition of the critical energy 
$\ecrit$ used here is half that of Jackson~\cite{ref:Jackson}.}),
and a useful dimensionless quantity $\xi$ as
\beqa
\label{eqn:ecrit}
\bigskip
\gamma\quad \equiv && \dfrac{E_b}{m_ec^2}\; ,\non\\
k\quad \equiv &&\dfrac{2\pi}{\lambda} = \dfrac{\enr}{\hbar c}\; ,\non\\
\ecrit\quad \equiv && \gamma^3\;\dfrac{3\hbar c}{2\rho}\;{\rm\; ~,~and}\non\\
\xi\quad \equiv && \dfrac{\enr}{2\ecrit}\left(1+\gtsq\right)^{3\over 2}\; ,
\eeqa
where $m_e$ is the electron mass.
Then the complex amplitude can be written as
\beqa
\label{eqn:SR}
\Esyn(r,\,\enr ,\theta ; \gamma,\rho)\; =
\dfrac{e^{ikr}}{\sqrt{r}}\left(\Epll \;\;\upll + 
\Eprp \;\;\uprp\right)
\eeqa
and the differential radiated power as
\beqa
\label{eqn:power}
\dfrac{d^{\,2}W}{d\theta\, d\enr}(\enr,\theta) = \left|\Epll \;\;\upll + 
\Eprp \;\;\uprp\right|^2\; ,
\eeqa
where $r$ is the two-dimensional distance from the emission point, and
\upll\ and \uprp\ are orthogonal polarization unit vectors into and out 
of the plane of motion, respectively. The coefficients of the
two orthogonal components of the amplitude are given by
\beqa
\label{eqn:SRpol}
\Epll &\equiv& E_0\; K_{2/3}(\xi)\; {\rm ~and}\non\\
\Eprp &\equiv& E_0\; K_{1/3}(\xi)\;
\dfrac{\gamma\,\theta}{\sqrt{1 + \gtsq}}\; ,{\rm ~where}\non\\
E_0&\equiv& \dfrac{\enr}{\ecrit}\; \gamma \left( 1 + \gtsq \right)
\,\sqrt{\dfrac{3\alpha}{4\pi^2}}\; ,
\eeqa
The $K_\nu$ are modified Bessel functions of the second kind, and
$\alpha$ is the fine structure constant.
The second term, $\Eprp$, vanishes at $\theta=0$, and remains small 
for $\gamma\,\theta\ll 1$. The angle in the horizontal plane has already 
been integrated out in \Eq{eqn:SR}, which can be justified by noting
that for the dimensions of the apparatus used here and 
the relevant wavelengths, the horizontal and vertical 
contributions to pathlength are decoupled.
Hence the radial dependence of the amplitude 
is $1/\sqrt{r}$, appropriate for radiation in
two dimensions, rather than the usual $1/r$ dependence
for a spherical wave.

\begin{figure}[tb]
   \includegraphics*[width=\figwid]{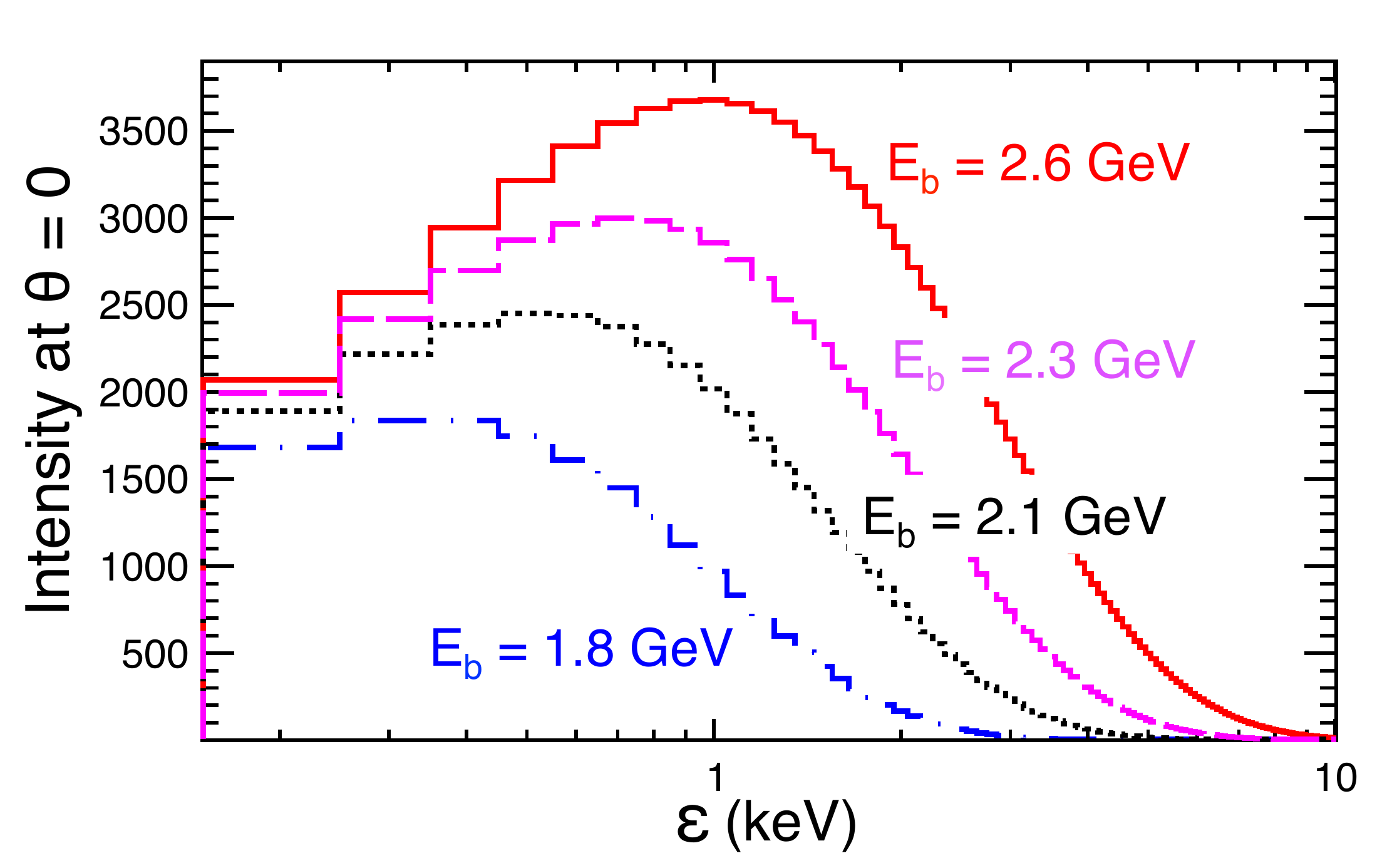}
   \caption{X-ray intensity spectrum (in arbitrary units) from synchrotron 
            radiation incident on the 
            CESR-TA xBSM optical element at $\theta=0$ 
            for several $e^\pm$ beam energies used at CESR-TA,
            using \Eq{eqn:power}.
}
\label{fig:SR}
\end{figure}

\begin{figure}[tb]
   \includegraphics*[width=\figwid]{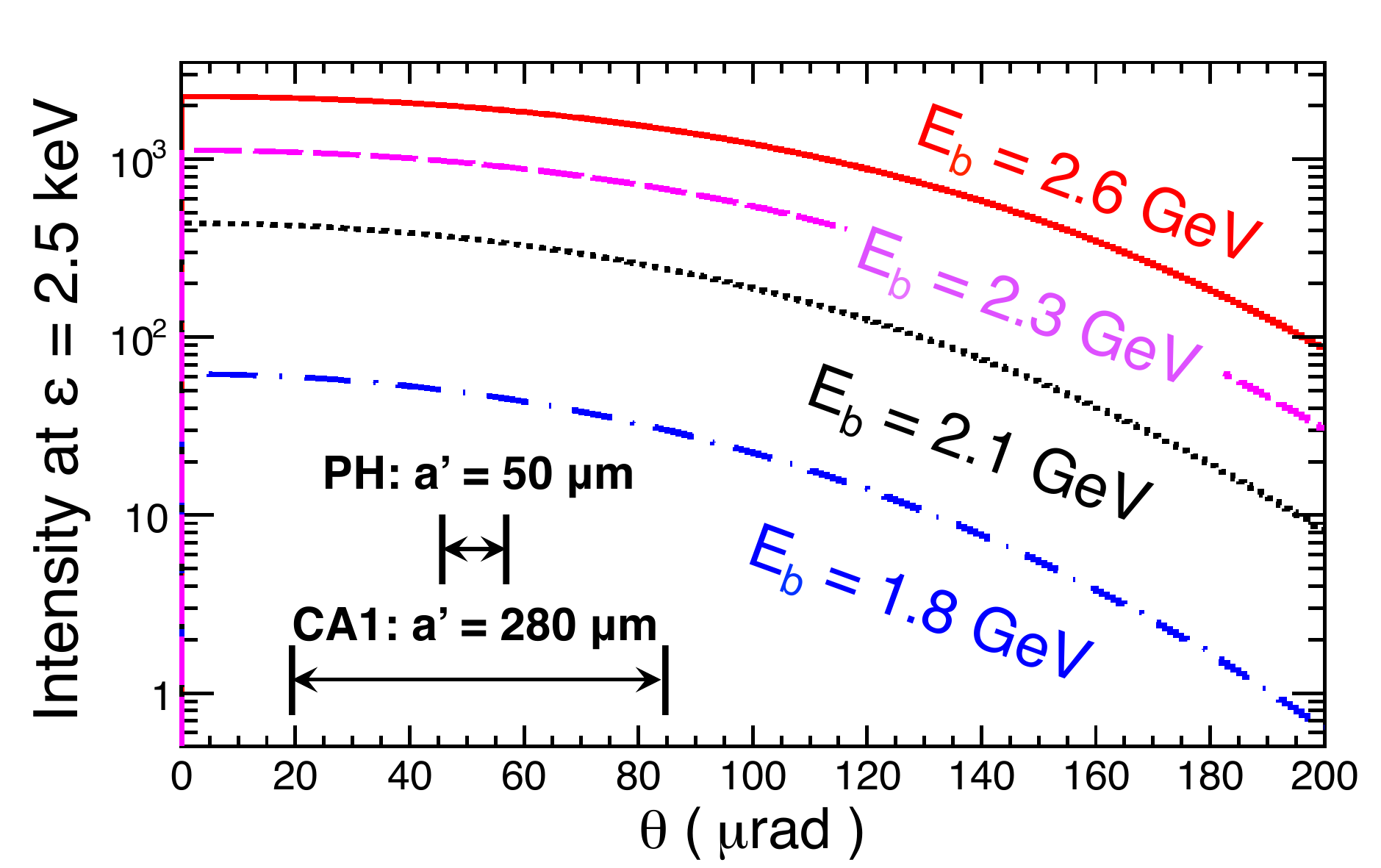}
   \caption{X-ray intensity from \Eq{eqn:power} 
            (in arbitrary units) 
            vs.~vertical angle $\theta$ at $\enr=2.5$\kev\ for synchrotron 
            radiation at several $e^\pm$ beam energies used at CESR-TA.
            Two (uncentered) spans are indicated, one each for 
            pinhole (PH) and coded
            aperture (CA1) optical elements (see \Tab{tab:optic}); 
            nominally these should aligned to be
            centered on $\theta=0$.
}
\label{fig:SRang}
\end{figure}

\begin{table}[hbt]
\small\selectfont
\caption{Parameters defining the CESR xBSM x-ray spectrum.
The critical energy \ecrit\ is defined in \Eq{eqn:ecrit},
and \sth\ in \Eq{eqn:sth}.}
\label{tab:spec}
\setlength{\tabcolsep}{1.50pc}
\begin{center}
\small\selectfont
\begin{tabular}{ccc}
\hline\hline
\rule[10pt]{-1mm}{0mm}
\Eb~(GeV) & \ecrit~(keV) & \sth(3\kev)~($\mu$rad) \\
\hline
\rule[10pt]{-1mm}{0mm}
 1.811 (``1.8'') & 0.416  &  61\\
 2.085 (``2.1'') & 0.634  &  65\\
 2.305 (``2.3'') & 0.858  &  68\\
 2.553 (``2.6'') & 1.166  &  72\\
\hline\hline
\end{tabular}
\end{center}
\end{table}

For CESR-TA, the hard-bend magnet at the xBSM source point has 
$\rho = 31.65$~m. \Tab{tab:spec} shows the critical
energies and \Fig{fig:SR} shows \Eq{eqn:power} evaluated at $\theta=0$
for the four $e^\pm$ beam energies from
1.8 to 2.6\gev\ which were used to characterize xBSM performance;
\ecrit\ ranges from $\sim$0.4-1.2\kev. 
Most of the detected x-ray intensity lies in the range 
$1<\enr/\ecrit<10$. Over this range one can
approximate~\cite{ref:Jackson} the differential radiated power near 
$\theta=0$ as
\beqa
\label{eqn:SRapprox}
\dfrac{dW}{d\theta}
&&\!\!\!\!= r\;\left|\;\Esyn\left(\enr >\ecrit,\theta=0;\gamma,\rho\right)\;\right|^2\non\\
&&\!\!\!\!\approx \dfrac{10}{3}\,\gamma^2 \;{\enr\over\ecrit}\; e^{-\enr/\ecrit}\; ,
\eeqa
which monotonically decreases by a factor of $\sim$1000
over the range $\enr/\ecrit=1-10$. \Eq{eqn:SRapprox}
gives a value $\sim$15-20\% smaller than \Eq{eqn:SR} at $\enr/\ecrit=1$
but converges to it as $\enr/\ecrit$ increases.

 For $\enr>\ecrit$, the angular dependence of the radiated power is 
approximately Gaussian with width 
\beq
\label{eqn:sth}
\sth \equiv {1\over\gamma}\sqrt{\ecrit\over{3\enr}}
= \sqrt{{\gamma\over 2}\,{{\hbar c}\over{\rho\enr}}}\quad .
\eeq
For $E_b=2.6$\gev\ and $\enr=3\ecrit$, 
$\theta_c=60~\mu$rad, which is nearly double the
maximum allowed ($\sim$35$\mu$rad for an optical element centered on
the beam) by the xBSM optics (apertures
of 50-300\mi), for a beam vertically centered on the optical
element. As shown in \Fig{fig:SRang}, variation of
radiated power with vertical angle is modest but not negligible at CESR-TA,
falling by $\sim$10\% (or less, depending on the optical element
and beam energy) from the center to the edge of the images.
  
\begin{figure}[b]
   \includegraphics*[width=\figwid]{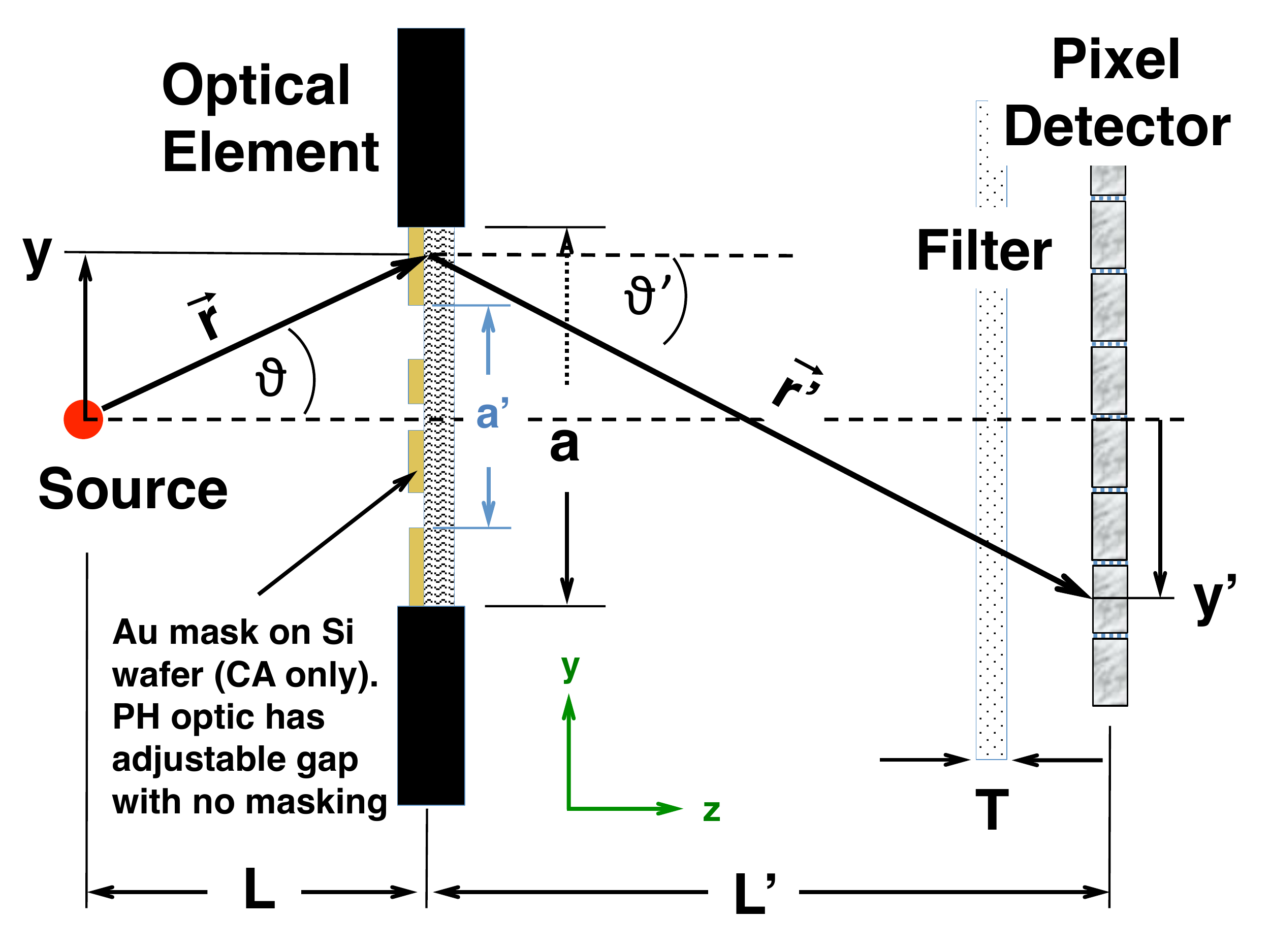}
   \caption{Simplified schematic of xBSM layout (not to scale).
   The distinction between $a$ and $\ap$ is that $a$ 
   is the total vertical extent of partial transmission
   through the mask material, and $\ap$ is the vertical
   extent of features (slits) in the mask; for the pinhole (PH)
   optic, which has no masking, $a=\ap$.}
\label{fig:geom}
\end{figure}

\begin{table}[t]
\small\selectfont
\caption{Geometrical parameters defining the CESR-TA xBSM 
         beamlines. Geometrical quantities are defined in \Fig{fig:geom}.
Distances assume the coded aperture optic; the pinhole optic is 25\mm\ closer
to the source point and hence has a magnification value about 1\% larger
than shown. The uncertainties on $L$ are from an optical survey; on $\lp$ from
the survey, known CESR orbit, and associated depth of field. }
\label{tab:geom}
\setlength{\tabcolsep}{1.00pc}
\begin{center}
\small\selectfont
\begin{tabular}{ccc}
\hline\hline
\rule[10pt]{-1mm}{0mm}
 Parameter & $e^-$ beamline & $e^+$ beamline \\
\hline
\rule[10pt]{-1mm}{0mm}
 $L$ & $4356.5\pm3.9$\mm & $4485.2\pm4.0$\mm \\
 $\lp$ & $10621.1\pm1.0$\mm& $10011.7\pm1.0$\mm\\
 $M\equiv\lp/L$ & $2.4380\pm0.0022$ &  $2.2322\pm0.0020$\\
 \ap & $\approx 50-300$\mi & same as $e^-$\\
 $a$ & $\approx 50-1000$\mi & same as $e^-$\\
 $2\theta_{\rm max}=\ap/L$ & $11-69~\mu$rad &  $11-67~\mu$rad \\
\hline\hline
\end{tabular}
\end{center}
\end{table}

\subsection{Point response function}
\label{sec:prf}

After passing through an optical element, x-rays from 
a point source 
form a diffraction pattern with single or multiple prominent peaks on the 
pixel detector. 
This pattern is the {\it point response function} 
(\prftt) \prf{y}, the expected x-ray intensity distribution
at the image plane
for zero beam size and a given x-ray spectrum, beam line geometry,
optical element, and filter in front of the detector. 
A simplified schematic of the geometry is shown in \Fig{fig:geom}, 
and the gross dimensions in \Tab{tab:geom}.
A non-zero vertical beam size creates an image with
broader peaks than that of a point source because the image is
a weighted superposition due to a continuum of vertically offset point sources. 
Extracting the vertical 
beam size from image data requires knowledge of the \prftt.

Diffraction of light from a point source through an optical element
to an image plane, with geometry as shown in \Fig{fig:geom}, can be
described by the Fresnel-Kirchoff formula~\cite{ref:Jackson}, which expresses the
complex amplitude for the wavefront at the image plane as an integral
over the optical element footprint. Simplifying the general formula to that
of a one-dimensional optic and image, we obtain:
\beqa
\Eimg(\,\yp,\,\enr\,) = 
-i\sqrt{k}\; \ds{\int}_{-a/2}^{+a/2} &&\!\!\!\!\!\!
\left( \Epll\upll + \Eprp\uprp \right)\;\;\times\non\\
&&\!\!\!\!\!\!\dfrac{e^{i\,k\,\left(\,r\,+\,\rp+\,(n-1)\,T\,\right)}}{\sqrt{r\,\rp}}\;\;\times\non\\
&&\!\!\!\!\!\!\dfrac{\cos\theta+\cos\thetap}{2}\;\; dy
\eeqa
where $a$, $r$, $\rp$, $\theta$, and $\thetap$ are defined in \Fig{fig:geom};
\Epll\ and \Eprp\ are given by \Eq{eqn:SRpol} and depend upon 
$\theta$ and therefore on $y$;
$n$ is the (complex) index of refraction, which contains effects 
of any partially opaque materials blocking the aperture, both a
phase shift (real) and absorption (imaginary), which have
$y$ and $\lambda$ dependences; and $T$ is
the material thickness (which may also vary with $y$).

Because x-ray wavelengths are small compared to aperture sizes
and both source-optic and optic-image separations, further simplifications
are possible. Typical apertures at CESR-TA range from 50-300\mi,
distances from 4-10~m, and wavelengths from 0.1-1\nm. Image offsets
$\yp$ are less than 1\mm. These parameter values result in the
well-known diffraction figure of merit 
$a^2/(L^{(')}\lambda)\approx 1$; this value is not small enough
for a far-field (Fraunhoffer diffraction) approximation, 
but appropriate for the near-field (Fresnel) diffraction.

We approximate
the source as a vertical line, with no horizontal or longitudinal extent.
Both bunch and detector dimensions (\Tabs{tab:cesr_params} 
and \ref{tab:det}) are very small compared to $L$ and $\lp$, resulting in 
the magnification $M\equiv \lp/L$ being nearly constant
over the bunch volume, justifying the approximation.

In Fresnel diffraction, the wavefronts must be considered
as cylindrical for a one-dimensional optic, not planar, and the
image remains sensitive to the optic-image separation.
Assuming the source is located at $y=0$,
\beqa
\cos\theta &\approx & 1\; ,\non\\
\cos\thetap &\approx & 1 \; ,\non\\
r+\rp &\approx & L + \lp + \dfrac{y^2}{2L} + \dfrac{\left(\yp-y\right)^2}{2\lp}\; ,\non\\
r\,\rp &\approx & L\,\lp\; ,
\eeqa
which lead to
\beqa
\label{eqn:kirchoff}
\!\!\!\!\!\!\Eimg(\,\yp,\, \enr\,)&&\!\!\!\!\!\!\!\!\!= -i\sqrt{k}\;\;\;
\dfrac{e^{ik(L+\lp)}}{\sqrt{L\,\lp}}
\times\non\\
&&\!\!\!\!\!\!\!\!\!\!\!\!\ds{\int}_{-a/2}^{+a/2} 
\left( \Epll\;\upll + \Eprp\;\uprp \right)\times\non\\
&&\!\!\!\!\!\!\!\!\!\!\!\!\exp\left[ik\;\left(\dfrac{y^2}{2L}+
\dfrac{\left(\yp-y\right)^2}{\lp}+(n-1)\,T\right)\right]\; dy\,;\qquad
\eeqa
\ie for a wave number $k$, 
the complex amplitude at any point \yp\ on the image plane
is calculated by integrating over the aperture, accounting at
each point $y$ for 
the synchrotron radiation amplitude in the absence of the optic,
the relative phase advance at that wavelength 
due to both path length and material, and the attenuation 
of intensity due to distance traveled and material traversed.
The \prftt\ is then obtained by integrating over all wavelengths
(energies):
\beqa
\label{eqn:prf}
\prf{\yp}\equiv 
\int_0^{+\infty} \srd\;\srf \left| \;
\Eimg(\,\yp,\, \enr\,)\; \right|^2\;d\enr\,
 .
\eeqa
where \srd\ specifies the spectral response of the detector,
empirically determined in \Sec{sec:specresponse}, and
\srf\ the spectral response of whatever partially transmitting
material are present in the filter and optical 
element. Accounting for the effects of materials in the path of the x-rays
is straightforward; however, one must account for 
the $y$-dependence of $T$ and the $\lambda$ dependence of $n$.
Designating each traversed material by the subscript $m$, the
complex index of refraction term is
\beqa
\label{eqn:atomicfactors}
(n-1)\,T\equiv \lambda^2\;\dfrac{r_eN_A}{2\pi}\sum_m \dfrac{\rho_m\, T_m}{A_m}
\left[\; f_1^m(\lambda) + i\, f_2^m(\lambda)\;\right]\;,
\eeqa
where $r_e$ is the electron radius, $N_A$ is Avogadro's number,
$\rho_m$ is the density of material $m$, $A_m$ is the atomic
number of material $m$,
and $f_1$ and $f_2$ are the {\it atomic scattering factors}
tabulated~\cite{ref:henke} as functions of x-ray energy.
The factor $f_2$ accounts for absorption of x-rays, and
can have both smoothly varying energy dependence and 
abrupt, discontinuous spikes in transmission corresponding to the 
various atomic shell energies.

\begin{figure}[tb]
   \includegraphics*[width=\figwid]{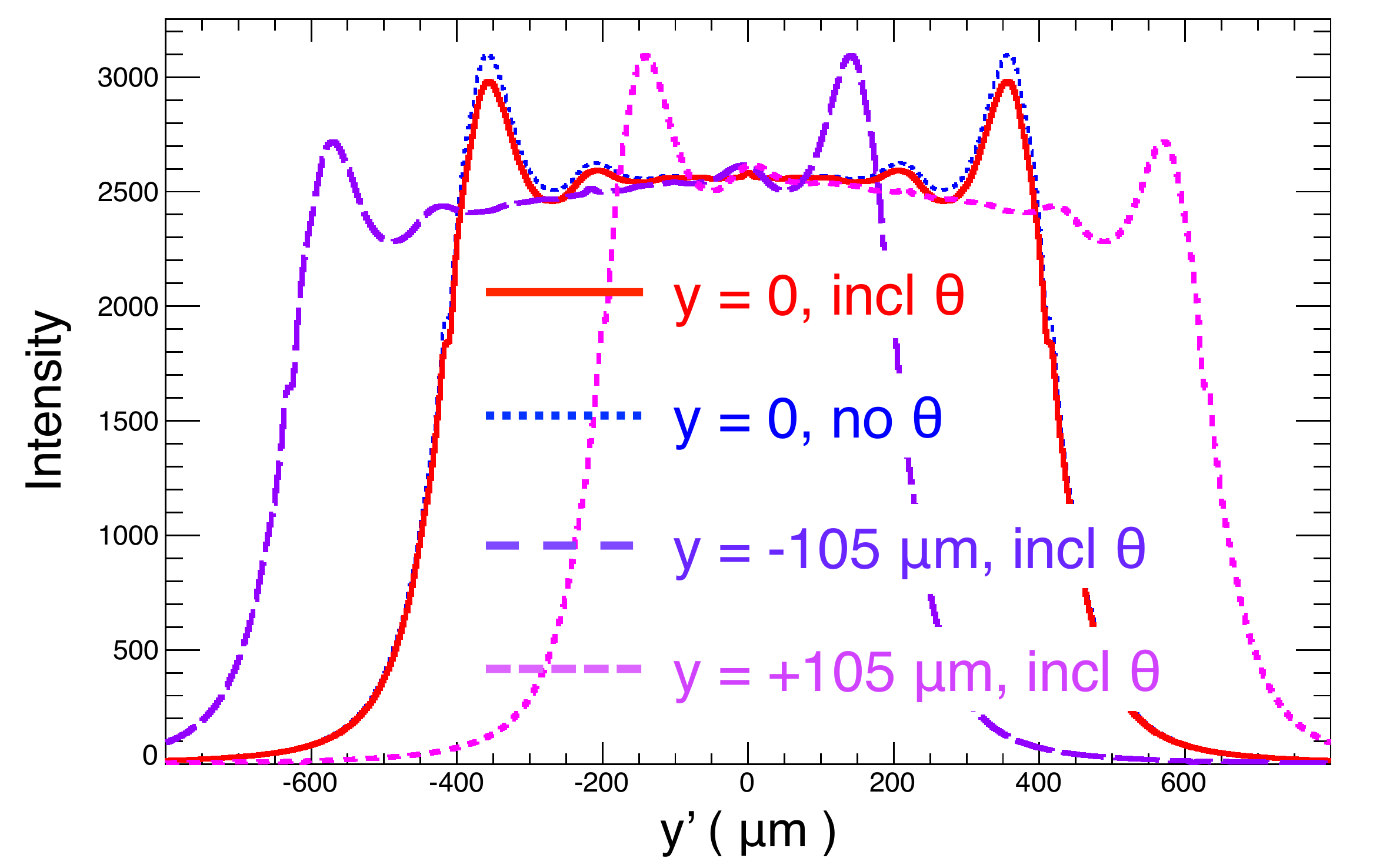}
   \caption{Predicted point response functions for
     a 300\mi\ pinhole optic at $E_b=2.1$\gev. 
     As indicated, 
     the curves correspond to the source point being centered on
     the slit or offset from center, either including or not including
     dependence of the synchrotron radiation intensity on
     the vertical angle $\theta$ out of the bend plane.}
\label{fig:prffour}
\end{figure}

\begin{figure}[h]
   \includegraphics*[width=\figwid]{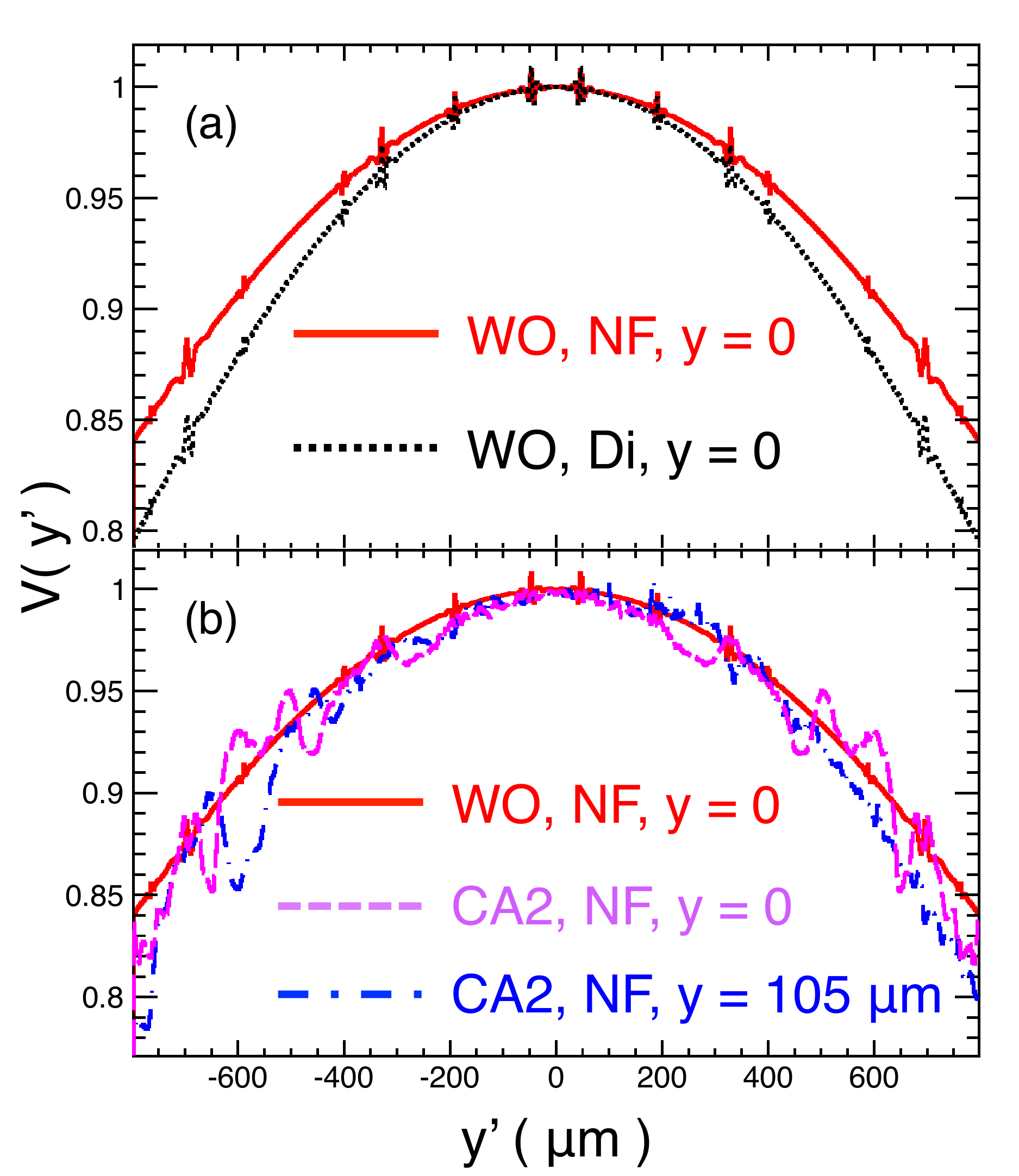}
   \caption{The angular correction function $\vee{\yp}$ (see \Eq{eqn:prfvee})
     at $\Eb=2.1$\gev\ for different combinations of an optical
     element and filter with vertical beam offsets. Optical elements (WO
     and CA2) are
     described in \Tab{tab:optic} and filters (NF and Di) in \Tab{tab:fil}.
The small and narrow (fwhm$\,\approx\,$10\mi) spikes in the WO-optic curves, 
which are expected to be smooth, are artifacts of the numerical 
approximations; broad (fwhm$>$20\mi) features are due to interference.}
\label{fig:vee}
\end{figure}

  It should be noted that \Eq{eqn:prf} implicitly assumes that
the source point is vertically centered on the optical element.
Generally this is not the case, and it matters because
of the $\theta$-dependence of \Eq{eqn:SR}. 
\Figure{fig:prffour} shows\footnote{\Figure{fig:prffour} and all subsequent
figures displaying a quantity dependent upon the x-ray spectrum
(such as a \prftt) apply the measured spectral response of the
detector described in \Sec{sec:specresponse} and effect of filters and 
materials in the x-ray beam using \Eq{eqn:atomicfactors} and 
the atomic scattering factors of 
\cite{ref:henke}.}  how $\theta$-dependence and beam position 
offset affect the box-like \prftt\ for a wide-slit pinhole optic.
Note that when the distribution shifts due to a beam position offset,
it is also slightly distorted relative to the shape with no offset.
While this dependence can be accommodated exactly, it is computationally
cumbersome to do so relative to what can be accomplished using a
factorization approximation on the \prftt\ for beam position 
$\off\ne0$. To first order, the \prftt\ can be expressed as the product
of two functions, one assuming a vertical beam profile that is uniform
in \off, and the second that imposes the vertical envelope of the 
synchrotron radiation field,
which might not be centered on the optical element. This second function
we designate as \vee{\yp}, which is the ratio of \Eq{eqn:prf} evaluated 
for a given optical element to that evaluated with vertical angle 
$\theta\equiv 0$ for the same optical element. 
(For pinholes, \vee{\yp} is obtained
from a wide-open optical element, whereas for coded apertures (CA),
where there is at least partial transmission to the entire
detector, \vee{\yp} is obtained for the CA itself.)
By construction, $\vee{0} = 1$, $\vee{\yp}=\vee{-\yp}$,
$\vee{\yp}$ is concave-down, and \prfv{\yp}{\off} 
is exact for the trivial case of a wide-open optical element. 
\Figs{fig:vee}(a) and (b) show \vee{\yp}
for a wide-open optic and a coded aperture; the curve for the 
diamond filter (relative to none) accounts for the stiffer spectrum with
correspondingly steeper \yp-dependence. Thus, the \prftt\ for
$\off\ne0$ is 
\beqa
\label{eqn:prfvee}
\prfv{\yp}{\off}\approx\dfrac{\prf{\yp + M\off}}{\vee{\yp + M\off}}\times
\vee{\yp - \off}
\eeqa
where  $M\equiv\lp/L$ is the {\it magnification}
(the ratio of the optic-image separation to source-optic distance),
the first term is the \off-independent \prftt, and the second imposes
the vertical beam profile.
For any pinhole optic with opening $\sim$50\mi, the approximation
is essentially exact for beam offsets up to $\sim$200\mi.
The approximation in \Eq{eqn:prfvee}\ is exact for \off=0, and
can be seen in \Fig{fig:vee}(b) to be valid for a coded aperture 
to within a few percent rms for beam energies and 
beam offsets encountered at CESR-TA. The importance and usefulness of
accounting for $\off\ne0$ is demonstrated in \Sec{sec:detoff}.

\subsection{Image processing}
\label{sec:image}

Once the \prftt\ \prf{\yp} is in hand, each measured image can be fit 
for the beam size, beam position, and intensity
that minimize $\chi^2$ between data points and
the smeared, offset, and amplified \prftt, assuming a Gaussian beam profile
and incoherent radiation from different vertical slices of the bunch.
Vertical beam position and detector offset for any given image are highly 
anti-correlated, so typically the detector is centered during instrument setup;
however the formalism below includes detector offset, with the understanding
that it cannot vary from one turn or bunch to the next in a given dataset.
The fitting function \fitf\ has dependent variable \yp\ and four 
parameters [beam intensity (\amp), 
vertical beam size (\siz), vertical beam offset from the center of the 
optic (\off), and vertical offset of the center of the detector from the center
of the optic (\dtr)]:
\beq
\label{eqn:defined}
\fitf \equiv \dfrac{\pfcn}{
\ds\int_{-\infty}^{+\infty}d\eta\,\ds\int_{\;0}^{+\infty} dk\; \left| \;
{\bf E}(\,\eta,\, k\,)\; \right|^2
}
\eeq
where
\beqa
\label{eqn:basicf}
\basicf \equiv & \dfrac{1}{\siz\sqrt{2\pi}}\times\non\\
& \ds\int_{-\infty}^{+\infty} 
&\!\!\!\!\!\!\!\!\exp{\left[-\dfrac{1}{2}\,\left({{\yp-\eta}\over{M\siz}}\right)^2\right]}
\times\non\\
& &\!\!\!\!\!\!\!\!\!\prfv{\yp - \eta}{\off + \eta/M}\;d\eta\;\qquad
\eeqa
and $\eta$ is an integration variable of the vertical coordinate
at the image plane.
The fitting function convolves a Gaussian shape approximating 
the beam profile with the \prftt,
translates it in the vertical dimension and amplifies it, thus 
extracting a best-fit beam size, beam position, and intensity of light
incident on the detector. The centered and 
smeared intermediate function, $\basicf$, is computed
from $\prf{\eta}$ by numerical integration for a discrete set of
evenly spaced beam sizes $\siz_j\equiv j\times\Delta \siz ,\ j=0,1,2 ...$, 
and source positions $\off_k\equiv\pm\, k\times\Delta\off ,\ k=0,1,2 ...$,
with values at intermediate beam sizes obtained from linear interpolation 
between the two nearest $\siz_j$ or $\off_k$, respectively. 
For our purposes we have found that there
are no changes in results for $\Delta\siz$ below 2\mi\ or $\Delta\off$=10\mi, 
and therefore use those values. The discrete set of {\it templates} 
$\basicjk\equiv\hfcn{\yp}{\siz_j}{\off_k}$ need
only be computed once, stored in files, and then be read into
memory for use in image fitting.

\subsection{Software tools}
\label{sec:soft}

Two general classes of software tools have been developed: one that
generates a point response function, and one that processes xBSM data.
The former class implements the optical element formalism of 
\Sec{sec:spectrum} and \ref{sec:prf},
numerically implementing \Eq{eqn:prf} using the known characteristics of
the beam, bending magnet, beamline geometry, optical elements,
filters, and detector spectral response, and the latter subsumes
all calibration, diagnostic, and analysis functions.
\PRFtt\ generation is implemented in a \CPP\ program using the CERN 
\ROOT~\cite{ref:root} package for diagnostic plots;
an \EXCEL\ version using more approximations is used for real-time feedback
and online analysis.
Two sets of code for data processing are maintained, one using
\MATLAB~\cite{ref:matlab} for online data acquisition, alignment, and
diagnostics during setup,
and a second in \CPP/\ROOT\ for more refined analysis.

\subsubsection{\PRFtt\ generation}
\label{sec:prfgen}

For each combination of beam energy, 
optical element, filter, and geometrical layout, a separate \prftt\ is needed
because it depends critically on the geometry and the x-ray spectrum.
A \prftt\ consists of a list of $[\yp,\prf{\yp}, \vee{\yp}]$ triplets 
at evenly spaced
\yp\ intervals covering not only the detector height but enough
beyond the detector edge to accommodate possible beam motion of 
up to $\sim 400$\mi. 
For each point, the integrals of 
\Eq{eqn:kirchoff} and \Eq{eqn:prf} must be performed. These proceed 
as follows:
\begin{itemize}
\item Parameters specifying the geometry, beam energy, bend radius,
optical element pattern, integration step sizes, mask and filter
materials and thicknesses, detector spectral response, etc., 
are assembled in a text 
{\it settings file} in which each parameter is identified by name and 
followed by its value.
\item A complex amplitude is accumulated at points on the image plane
separated by $d\yp=2.5$\mi\ for $\yp\in [-2000$, $2000]$\mi; each is
initialized to zero for each x-ray energy $\enr$ and suitably incremented
for each optic position $y$.
\item The integration of \Eq{eqn:kirchoff} is performed as a 
rectangle-rule sum taken in steps of $dy=0.1$\mi\ over all slits and masking 
strips of the optical element and at least 500\mi\ beyond the extreme 
slits at each end to include effects of continuous
partially transmitting masking material. The 
atomic scattering factors of \cite{ref:henke} are used
for evaluation of \Eq{eqn:atomicfactors}.
\item The integration of \Eq{eqn:prf} is performed as a simple sum
in steps of $d\enr=0.1$\kev\ in $\enr\in [0,20]$\kev, weighting each
entry by the detector spectral response \srd, measured in 
\Sec{sec:specresponse} (except, of course, during the procedures
of \Sec{sec:specresponse}, where $\srd\equiv 1$).
\item The angular correction function $\vee{\yp}$ is computed using
the optical element itself, except for a pinhole optic,
for which a a wide-open optical element is used.
\end{itemize}
For each of our coded apertures, \prftt\ 
generation requires $\sim$20 minutes of
single-threaded CPU time on an Intel\textregistered\ 
Xeon\textregistered\ CPU X5550 $@$ 2.67GHz (SPECint\_base2006 33,
SPECfp\_base2006 38).

\subsubsection{Data processing}

Both \MATLAB\ and \CPP/\ROOT\ data processing codes perform
similar basic functions:
\begin{itemize}
\item Read in the data text file, storing and making available the general 
  run information such as run number, beam current, beam energy, CESR bunch
  structure, pedestals, gains, range, optical element and filter selections, 
  etc. as well as the digitized pulse height for all detector channels for 
  all bunches and all turns in a fixed ordering.
\item Correct digitized data for individual channel pedestals and gains, 
  amplifier range, and bunch-to-bunch crosstalk (see \Sec{sec:cal}).
\item Define and fill image plots.
\item Fit a selection (\eg first hundred turns, every third turn, all turns,
  all bunches) of images to an appropriate \prftt\ for running 
  conditions for up to six parameters: beam size, beam vertical position, 
  intensity, an image component flat across the detector, magnification scaling,
  and a detector offset $\dtr$ (see \Eq{eqn:defined}). Generally
  only the first four parameters are floating and the last two are
  fixed to 1.0 and 0.0, respectively. The flat component 
  accounts for as much as $\pm10$\% percent of the image
  area\footnote{We take {\it image area} to mean {\it image intensity
      summed over all 32 detector channels}.} and 
  thought to be due to shifts in electronics response
  that are coherent across the detector. 
  Magnification scaling is only used to verify
  the validity of the distance ratios (see \Sec{sec:mag}).
  The value of $d$ is set to
  a nonzero value only for coded aperture data seen to be misaligned
  (see \Sec{sec:detoff}).

\item Prevent fitting if the area of the image is below a threshold 
  because in multi-bunch
  running it sometimes happens that some bunches are read out but are empty
  of current.
\item Calculate turn-averaged quantities of fit parameters 
  for each bunch, preventing bad-fit or
  no-fit non-statistical outliers from having undue weight.
\item As an alternate beam size measurement, construct and fit 
  the sum of all turns for each
  bunch to the \prftt, which includes beam motion effects but is nevertheless
  useful for comparison.
\item Store useful diagnostic plots and a summary of fit results 
  as text. If the run was taken for calibration purposes, store calibration
  results appropriately.
\end{itemize}

The \MATLAB\ 
programs are essential for alignment procedures and feedback on the
functioning of the electronics and proper filter and optical element selection,
all of which must be verified prior to taking production data. However,
to get quick feedback, generally only a small fraction of the data are 
examined (\eg 100 turns), or images from all turns are summed prior to 
being fit, because our \MATLAB\ tool cannot handle the load of fitting all
turns individually in a reasonable time. In addition, \prftt s in the 
\MATLAB\ program are limited
to multiple-Gaussian approximations to the binned \prftt s
from the \EXCEL\ tool mentioned earlier; 
template generation and template use are only implemented 
in the \CPP/\ROOT-based code.

The \CPP/\ROOT\ 
program is independently coded and performs all the above functions,
but has more flexibility. Specifications on desired analysis are made in text
{\it settings files}, including which bunches and turns are to be fit, fitting
options (which parameters are floating or fixed), the \prftt\ to be fit,
etc. Furthermore, more diagnostic information is created, such as
two-dimensional plots of fit parameter value vs.~beam current 
(or run number) and bunch number. Running it creates a text output summary 
{\tt  .out} file and a binary {\tt.root} file,
the latter of which stores plots, which can be viewed,
manipulated, and extracted with the interactive \ROOT\ object browser.
The \prftt\ can be specified as a multi-Gaussian function (for direct comparison
with the \MATLAB\ fits) or a binned version as described in \Sec{sec:prfgen}.

In addition, the \CPP/\ROOT\ analysis converts three of the fit parameter variations
in the time domain (turn number, each turn corresponding to $\sim$2.5~$\mu$sec)
into the frequency domain using a discrete fast Fourier transform
(FFT), computed using \FFTW~\cite{ref:fftw}. 
Resonances show up at frequencies of characteristic oscillations (such as 
horizontal and vertical tunes of the storage ring, 
which can vary slowly with time), or low frequencies associated
with A/C line voltage and power supplies. 

 To perform all the above functions, the \CPP/\ROOT\ data analysis
program requires
$\sim$7~msec per bunch per turn of
single-threaded CPU time on an Intel\textregistered\ 
Xeon\textregistered\ CPU X5550 $@$ 2.67GHz (SPECint\_base2006 33,
SPECfp\_base2006 38). The time is dominated by the $\chi^2$-minimization
fitting inside \ROOT's \MINUIT~\cite{ref:minuit} 
package.

\subsection{Optical element design}
\label{sec:design}

\begin{table}[hbt]
\small\selectfont
\caption{CESR-TA xBSM optic element parameters. Geometrical quantities are
  defined in \Fig{fig:geom}. Coded aperture patterns are shown in \Fig{fig:caphoto}.}
\label{tab:optic}
\setlength{\tabcolsep}{0.23pc}
\begin{center}
\small\selectfont
\begin{tabular}{llc}
\hline\hline
\rule[10pt]{-1mm}{0mm}
Category & Parameter & Value \\
\hline
\rule[10pt]{-1mm}{0mm}
WO        & $\ap\equiv a$ & 40~mm\\
(wide-open) & & \\
\hline
\rule[10pt]{-1mm}{0mm}
PH        & Tungsten blade $T$ & 2.5\mm\\
(pinhole) & Downstream taper & $2^\circ$\\
          & $a$ & 0-200\mi\\
          & \ap & $\equiv a$\\
\hline
\rule[10pt]{-1mm}{0mm}
CA1 & Si $T$ & 2.5\mi\\
(coded   & Au $T$ (2 chips) & $0.54\pm0.05$\mi\\
aperture)& Au $T$ (2 chips) & $0.69\pm0.05$\mi\\
         & $a$    & 1000\mi\\
         & \ap    & 280\mi\\
         & \# slits & 8 \\
         & Min/Max slit width & 10/40\mi\\
         & Transmitting fraction of \ap  & 54\% \\
         & Feature placement accuracy & 0.5\mi \\
         & Edge quality rms deviation & 0.1\mi \\
         & Pattern: S=slit, M=mask ($\mu$m) & \\
         & \; 20S-10M-20S-10M-40S- & \\
         & \;\qquad 30M-10S-10M-10S-10M- & \\
         & \;\qquad\qquad 30S-40M-10S-20M-10S & \\
\hline
\rule[10pt]{-1mm}{0mm}
CA2 & Si $T$ & 2.5\mi\\
         & Au $T$ & $0.62\pm0.05$\mi\\
         & $a$    & 1000\mi\\
         & \ap    & 296\mi\\
         & \# slits & 5 \\
         & Min/Max slit width & 10/68\mi\\
         & Transmitting fraction of \ap & 65\% \\
         & Feature placement accuracy & 0.5\mi \\
         & Edge quality rms deviation & 0.1\mi \\
         & Pattern: S=slit, M=mask ($\mu$m) & \\
         & \; 24S-10M-38S-42M-68S- & \\
         & \;\qquad 42M-38S-10M-24S & \\
\hline
\hline
\end{tabular}
\end{center}
\end{table}

Optical elements used at CESR-TA are listed in \Tab{tab:optic},
and include WO (wide-open), an adjustable vertical pinhole (PH), and two
coded apertures (CA1 and CA2). In this section we discuss
characteristics of each and relevant design considerations.

Design and quantitative evaluation of optical elements requires a 
figure of merit for beam-size resolving power. For the moment, we restrict
ourselves to an idealized quantity wherein systematic effects involved in
reconstructing beam size are ignored, so that the resulting function $Q(\siz)$
expresses the statistical power of a particular optical element at 
beam size \siz. We form a $\chi^2$-like quantity based on the fact that the
pulse height in each of the 32 pixels is proportional to the number of 
photo-electrons collected there and which will fluctuate according to counting 
statistics. Pixelizing the kernel of the point response
function \basicf\ defined in \Eq{eqn:basicf}:
\beqa
\pjs\equiv \int_{\rm pixel~j} \basicf\; d\yp
\eeqa
then
\beqa
\qs &\equiv & \qz\;\left(\dfrac{\siz}{\delta}\right)^2\times\non\\
&&\sum_{\rm pixels}\dfrac{\left[\pjs-\pjd - \dsd \right]^2}{\pjs+\pjd + 2\,\pz}
\eeqa
where $\delta$ is an arbitrary change in beam size 
(we use $\delta=8$\mi),
\dsd\ is the difference between value of \pjs\ 
averaged over all pixels ($j$) and the similarly averaged \pjd, 
 \pz\ is an estimate of the effective electronic pedestal noise, 
and \qz\ is an overall normalization factor.
The quantity \pz\ is inserted to prevent overweighting of differences between
small pulse heights, and here is chosen to be the same as
the peak value of \pjs\ for a 50\mi\ pinhole at \Eb=2.1\gev.
For optical elements that keep the primary image features
well contained on the detector for modest beam sizes, 
\dsd\ will be negligibly small; however, for very large beam sizes
or optic designs with image features close to the detector edges, 
it can become significantly nonzero. The resolving power \qs\
is an approximate indicator of the inverse 
statistical error squared per unit beam current for a single bunch,
single turn measurement at low beam current:  \qz\ has been chosen so that the 
expected statistical uncertainty
in beam size, $\Delta\siz$, at the CESR-TA xBSM can be expressed as
\beqa
\dfrac{\Delta\siz}{\siz}\approx \dfrac{10\%}{\sqrt{\qs\times I ({\rm mA})}}\;,
\eeqa
where $I$ is the beam current.
Note that for \qs=1 and the smallest beam current of interest,
0.1~mA, a barely acceptable measurement ($\sim$$3\sigma$
separation from zero) is obtained at all beam sizes;
\ie with this normalization, \qs\ values below unity
indicate a region of beam size where resolving power
is inadequate at the smallest beam currents. Our goal in optic
design, then, is to obtain the broadest possible regions
of beam size where \qs\ is not only larger than the alternatives,
but also greater than unity. Not surprisingly, the biggest challenge
is obtaining adequate sensitivity at small beam size,
where small changes in relative beam size alter the 
image shape relatively less than at larger beam size.

\begin{figure}[t]
   \includegraphics*[width=\figwid]{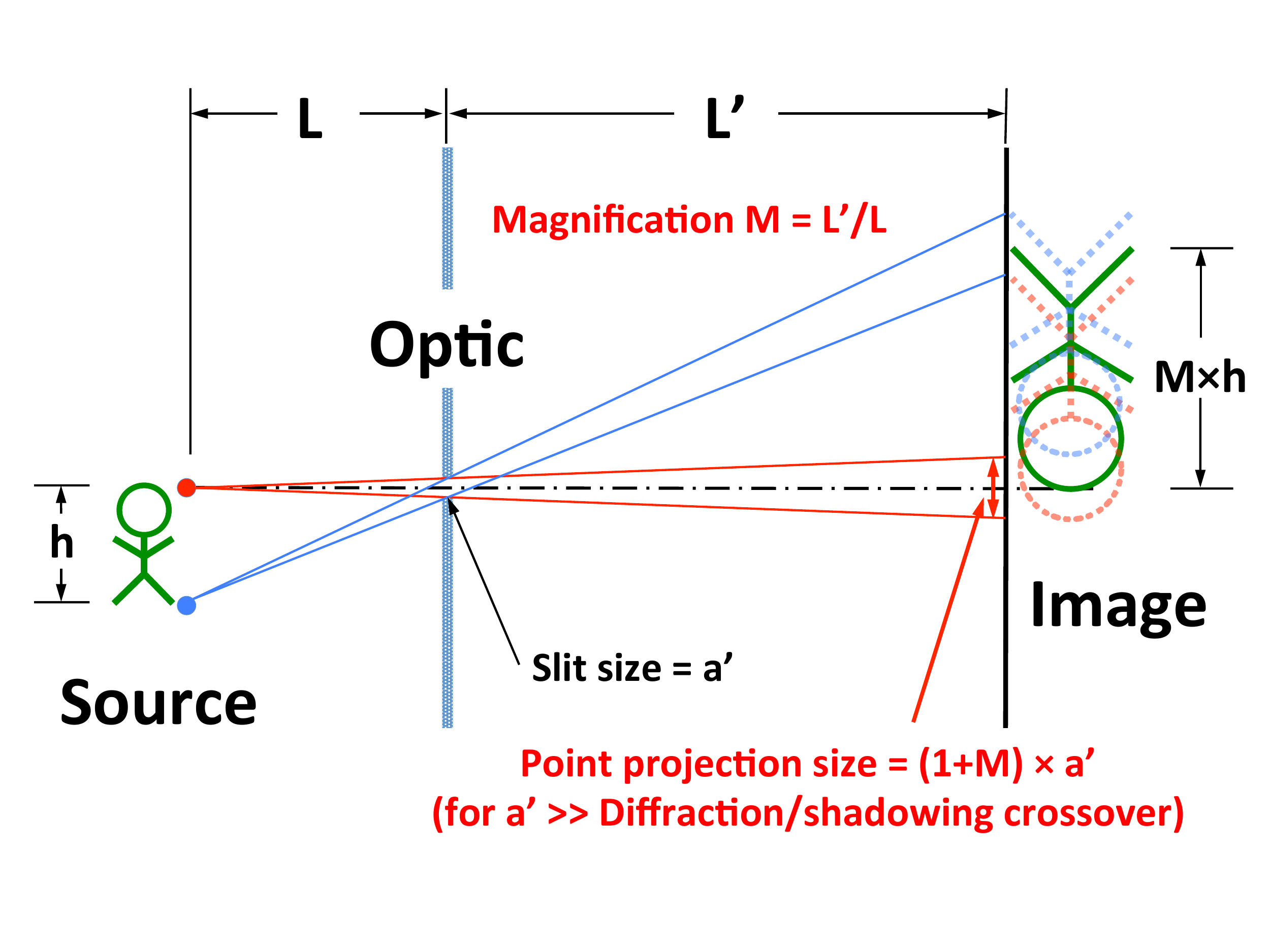}
   \caption{Simplified schematic of the basic principles of a pinhole camera.}
\label{fig:pinhole}
\end{figure}

\begin{figure}[t]
   \includegraphics*[width=\figwid]{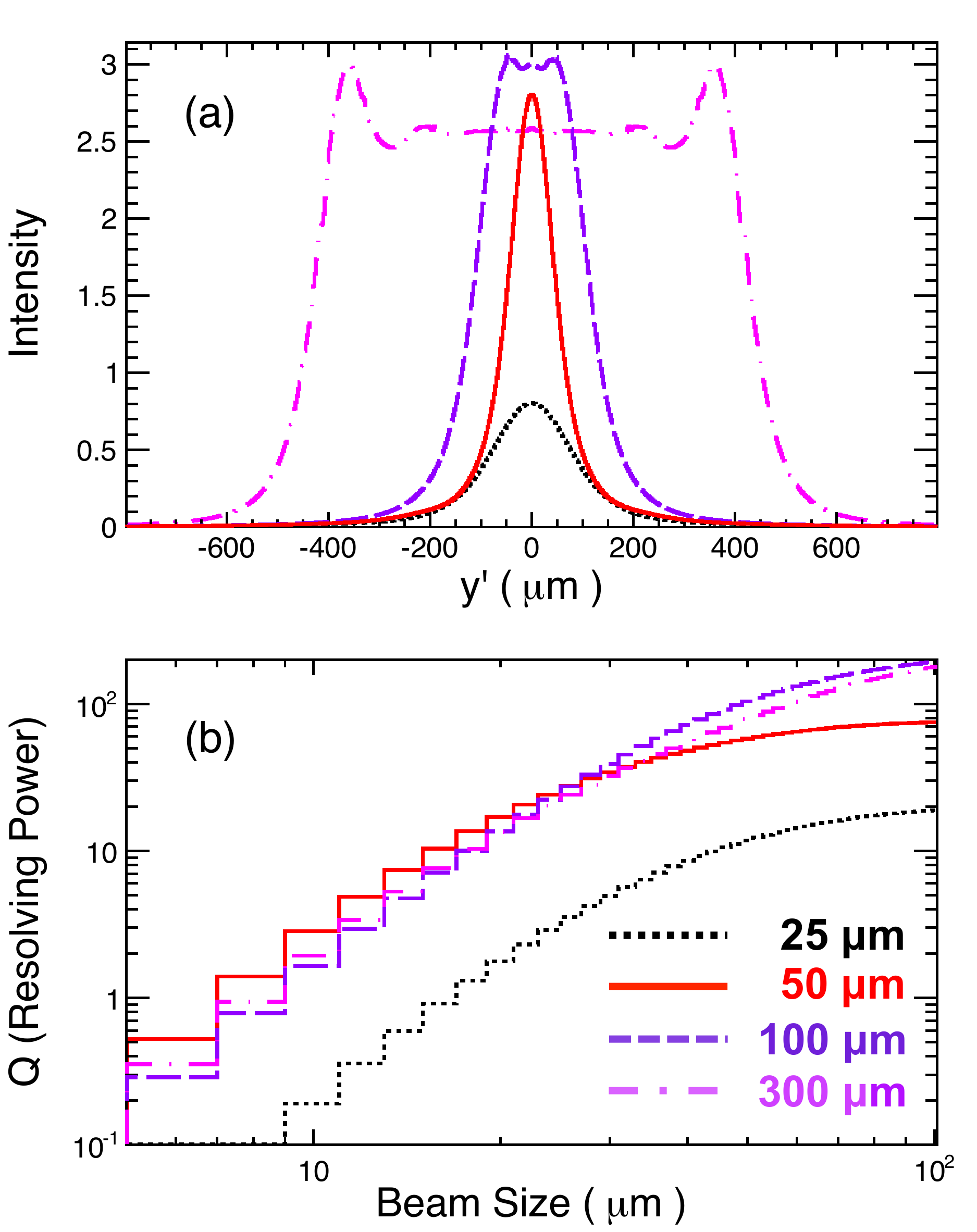}
   \caption{Predicted (a) point response functions 
     and (b) beam-size resolving-power curves for
     pinholes of indicated openings,
     all at $E_b=2.1$\gev.}
\label{fig:prfone}
\end{figure}

\begin{figure}[t]
   \includegraphics*[width=\figwid]{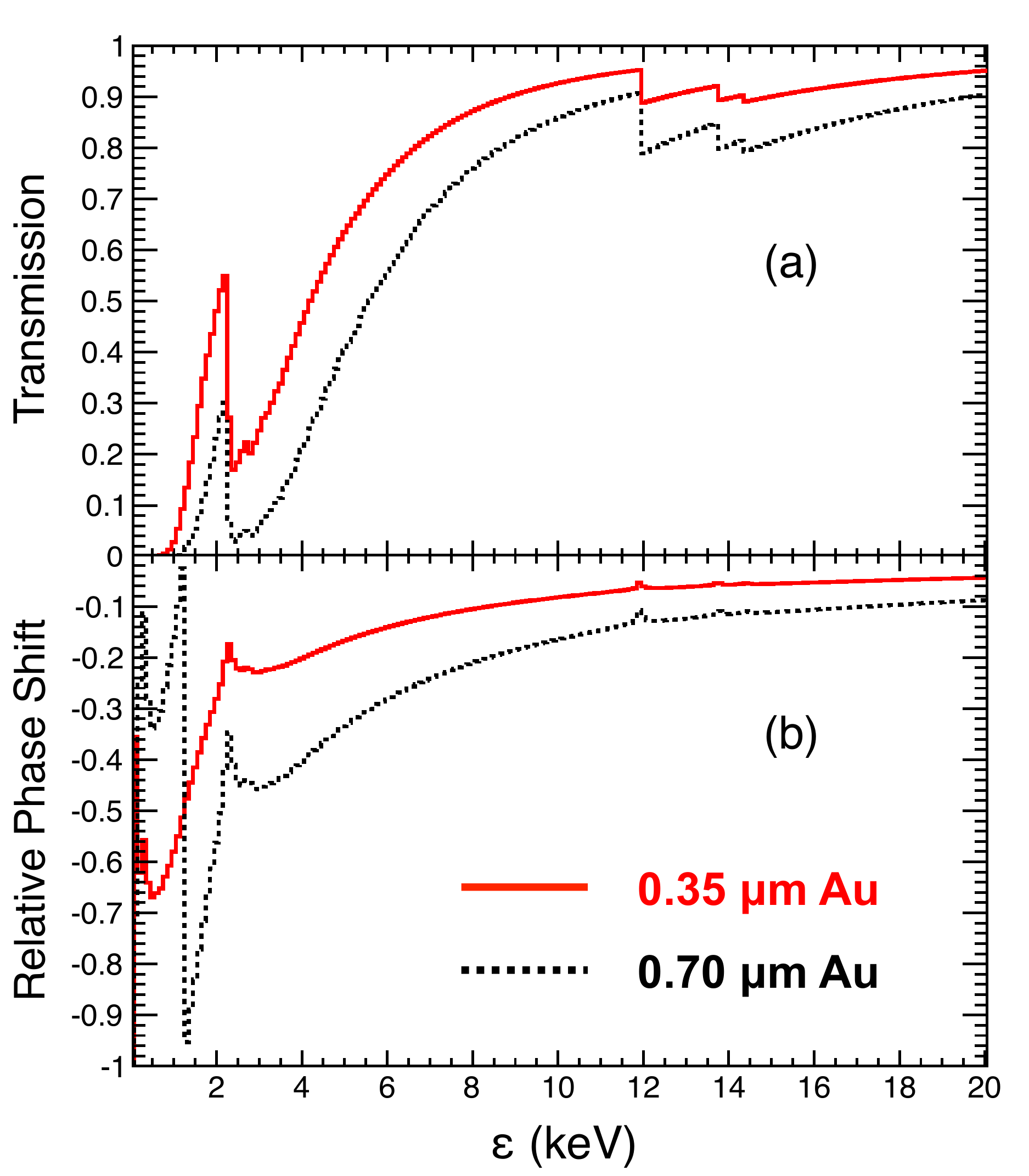}
   \caption{X-ray (a) transmission fractions, and (b) phase shifts 
     (in units of $2\pi$) as function of x-ray energy 
     for gold masking of thickness 0.35\mi\ (dashed) and 0.70\mi\ (solid). }
\label{fig:goldtrans}
\end{figure}

\begin{figure}[t]
   \includegraphics*[width=\figwid]{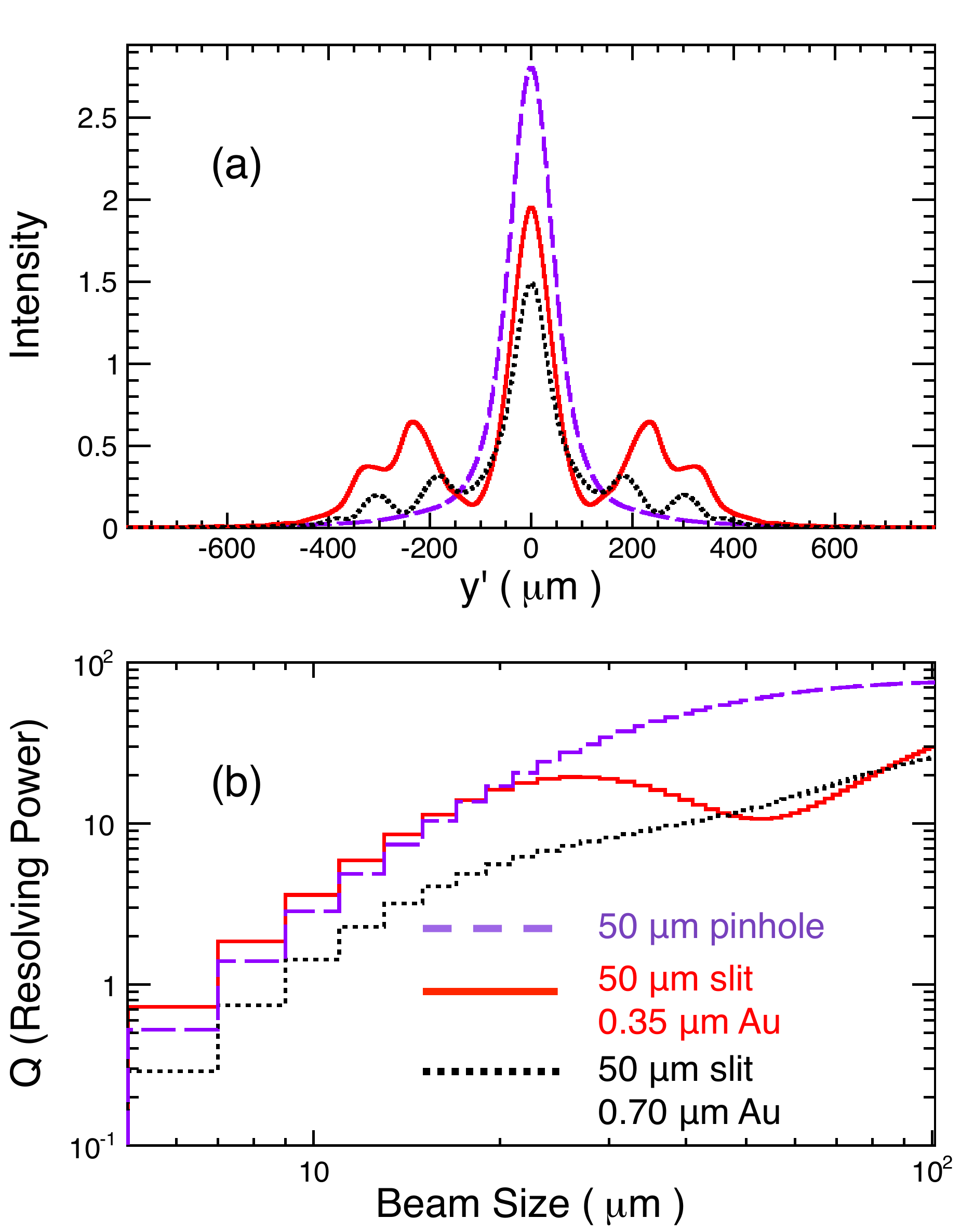}
   \caption{Predicted (a) point response functions, and
     (b) beam-size resolving-power curves for three different 
     optical elements at $E_b=2.1$\gev, as indicated.}
\label{fig:prftwo}
\end{figure}

\begin{figure}[t]
   \includegraphics*[width=\figwid]{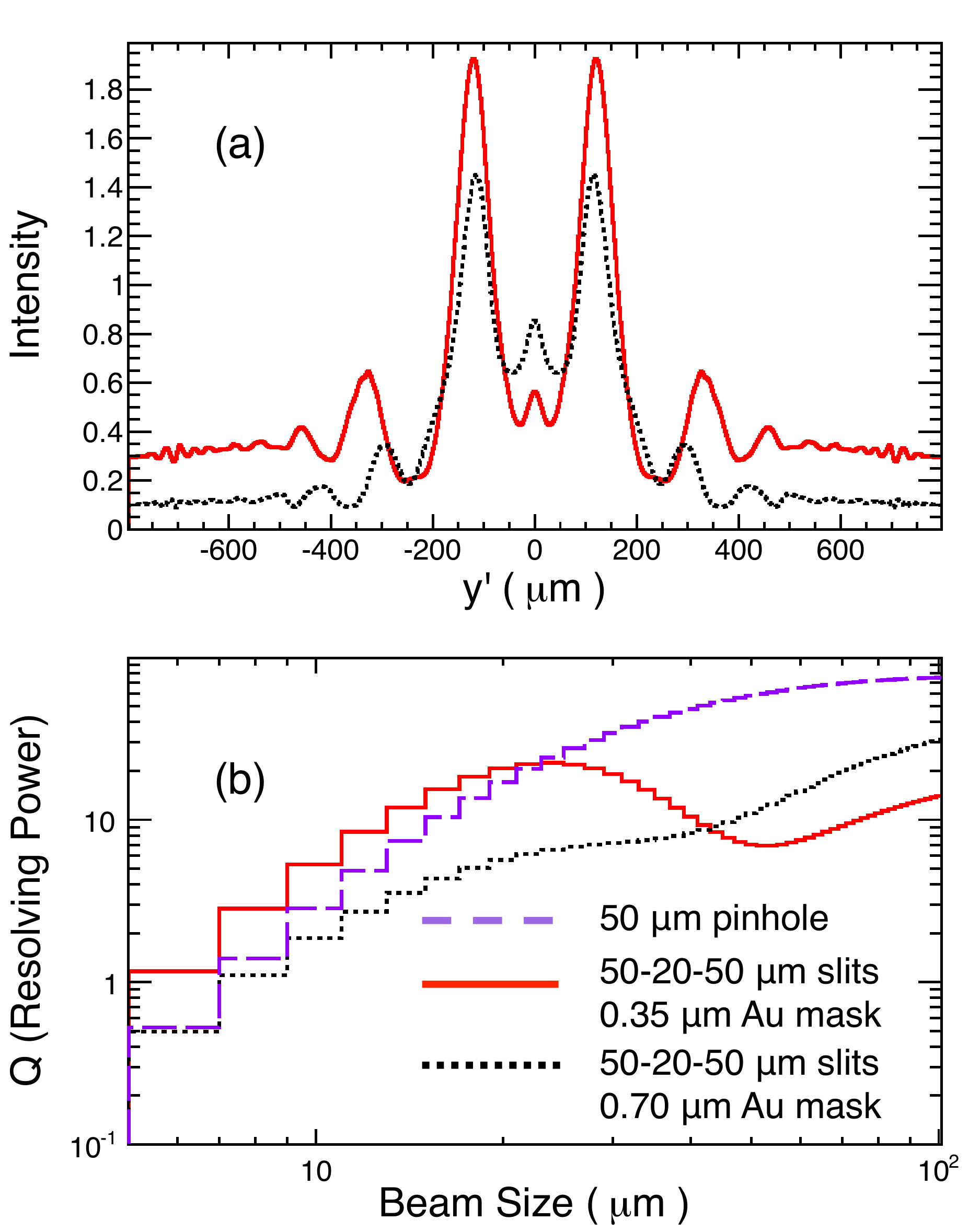}
   \caption{Predicted (a) point response functions, and
     (b) beam-size resolving-power curves for several
     optical elements at $E_b=2.1$\gev, as indicated.
     For comparison, (b) also
     shows the curve for a simple 50\mi\ pinhole, identical to that
     in \Figs{fig:prfone}(b) and \ref{fig:prftwo}(b).}
\label{fig:prfthree}
\end{figure}

\begin{figure}[t]
   \includegraphics*[width=\figwid]{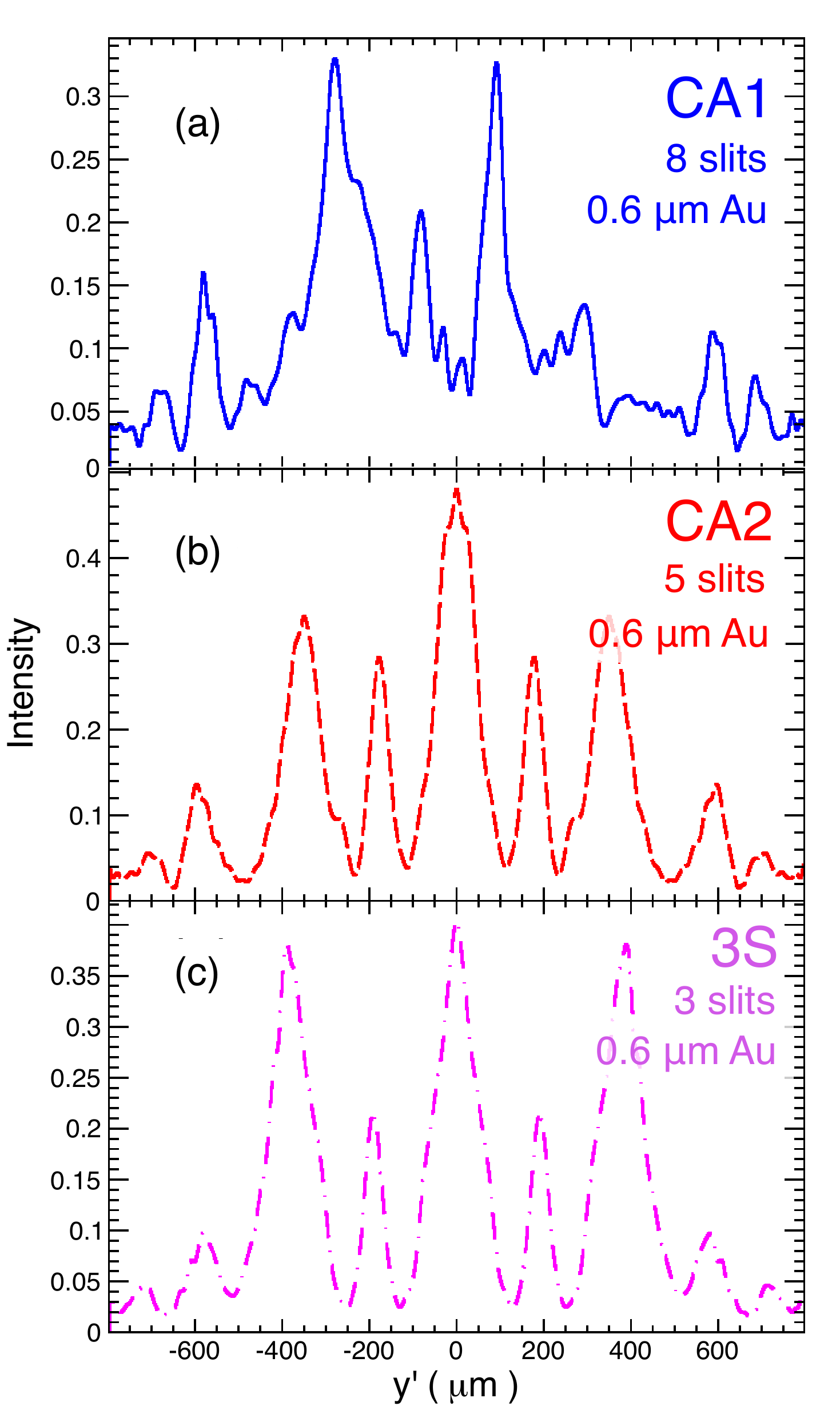}
   \caption{Predicted point response functions at $E_b=1.8$\gev,
     a 2.5\mi-thick Si substrate, and a 0.60\mi\ Au mask for
     (a) CA1 (8 slits: see \Tab{tab:optic}), 
     (b) CA2 (5 slits: see \Tab{tab:optic}), and (c) 3S (3 slits: see text).}
\label{fig:prfcaone}
\end{figure}

  Imaging of x-rays in the xBSM occurs according to the basic principles of a
pinhole camera, which are shown in \Fig{fig:pinhole}: ignoring diffraction,
each point at the source projects to a band on the image plane that is the 
size of the pinhole scaled by $(1+M)$, where $M$ is the magnification, 
whereas different points in the source plane separated by $h$ project in 
an inverted manner to a separation of $M\times h$ at the image plane. 
As the pinhole gap size \ap\ approaches the {\it diffraction crossover} 
(the aperture at which geometrical
anti-shadowing recedes and diffraction dominates) from above, 
the image width will decrease because diffraction has a small effect 
relative to the anti-shadow of the slit. At the diffraction crossover, the 
image from a single slit is narrowest; below the diffraction crossover, the 
width of the image grows. However, the area of the image scales linearly 
with \ap, regardless of the effect of diffraction on its shape; a 
consequence of these features is that the peak of the image lowers rapidly
as \ap\ shrinks below the diffraction crossover. These characteristics are 
evident in \Fig{fig:prfone}(a), which shows four point response functions 
for simple pinholes (slit sizes of 300, 100, 50, and 25\mi) for 
$\Eb=2.1$\gev\ and using a 4.4\mi\ diamond filter. In the absence of 
diffraction, all would be rectangular in shape, with widths at the image 
plane of $(1+M)$ times the gap sizes ($957$, $319$, $160$, and $80$\mi, 
respectively). In this case the diffraction crossover is about 50\mi.

\begin{figure}[t]
   \includegraphics*[width=\figwid]{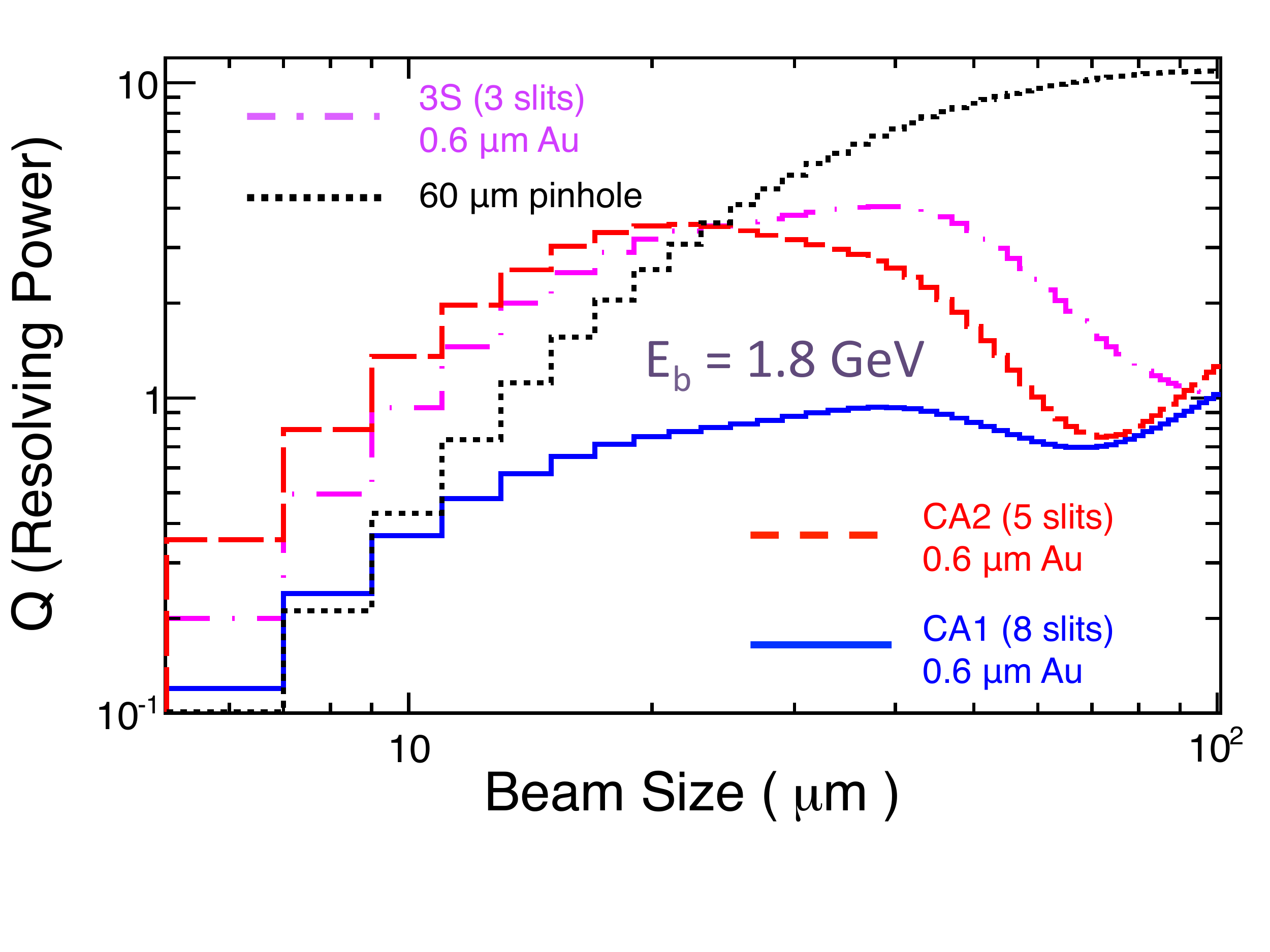}
   \caption{Predicted beam-size resolving-power curves at
     $E_b=1.8$\gev,
     for several optical elements (see \Tab{tab:optic} and text), as indicated.}
\label{fig:fomcacomp}
\end{figure}

The beam-size resolving power for four pinhole sizes is shown in 
\Fig{fig:prfone}(b), which highlights important principles applicable 
to optimizing effectiveness of optical elements: more light improves 
photoelectron-counting statistics, but for very small beam sizes, steep edges
and/or narrower peaks in the \prftt\ outweigh greater total energy deposition. 
For example, better resolution is possible above beam sizes of 
$\siz=40$\mi\ using a slit size of 300\mi\ 
compared to 50\mi, because above that size
neither has any steeply changing regions which offer high sensitivity, and 
the additional signal in the center of the distribution for the broader 
slit aids in distinguishing two nearby beam sizes. 
The crossover point for the 50\mi\ and 100\mi\ slits is even lower,
at about $\siz=25$\mi. The optimal choice depends upon 
the beam size range of interest.

We have also explored the use of multi-slit optical elements, 
known as coded apertures.
Coded aperture imaging is a technique well-developed
among x-ray astronomers~\cite{ref:dicke} which can,
in principle, improve on the spatial resolution of a pinhole camera, 
due, in part, to much greater x-ray collection efficiency.
A coded aperture has multiple slits, with the added feature that,
depending on the fabrication technique, the
mask material may be thin enough to partially transmit x-rays with 
a phase shift. Through interference, light passing through the mask
will affect the \prftt\ for good or ill, depending on slit and mask sizes and 
the x-ray spectrum. Our coded apertures use a gold
masking material of 0.5-0.7\mi\ thickness on top of a 2.5\mi\ thick silicon
substrate (which also absorbs some x-rays, but does so identically for
both slit and mask regions). 
The effective transmission and phase shift of gold masking 
(with thicknesses 0.35 and 0.70\mi)
are shown in \Fig{fig:goldtrans}(a) and (b), respectively.
In this example, the thinner masking is more effective than the
thicker choice  because it introduces a significant phase shift while 
preserving a larger fraction of the incident intensity for distribution among
the peaks. To illustrate the effect of partially transmitting
materials, \Fig{fig:prftwo} shows the \prftt s for several single slit
optical elements of the same slit size (50\mi) but different masking 
thicknesses (0.35\mi, 0.70\mi, and infinite). Note that
the 0.35-micron-thick gold has a large phase shift ($-\pi$ to $-\pi/2$) 
near the peak of the x-ray energy spectrum, in the 1-3\kev\ region.
The finite-thickness masking
introduces satellite diffraction peaks
into the \prftt s through interference of phase-shifted x-rays with
those that pass through slits.  The corresponding
resolving power of these three optical elements is shown in \Fig{fig:prftwo}(b),
where it is evident that the single-slit with
with thin-masking outperforms a single slit with thick masking 
below about $\siz=15$\mi.

Adding more slits can improve beam-size resolving power. Two identically-sized
double-slit (50-20-50\mi\ slit-mask-slit patterns) coded aperture \prftt s are 
shown in \Fig{fig:prfthree} for two thicknesses of gold masking. For the thinner
masking material the valley between the primary peaks is deeper, the peaks
themselves are higher, and the satellite peaks are more prominent. These 
features are seen to yield the expected effects in \Fig{fig:prfthree}(b), 
where the
resolving power comparisons reveal superior performance of the thinner masking
(relative to thicker masking) below $\siz=45$\mi. Notice, however, that the
single-slit 50\mi\ pinhole obtains better resolution (compared to the thin-mask
double-slit CA) above about $\siz=25$\mi, and is better at all beam sizes below
$\siz=100$\mi\ compared to the thick-mask coded aperture. Design of
an optical element must account for slit pattern and mask effects as well as
incident x-ray spectrum and target beam size; ignoring any of these
details can result in a complex coded aperture that does
not perform as well as a properly sized single slit with no 
masking material.

That the details of optical element design are subtle is exemplified by
comparison of two slit designs that were fabricated for this application
(\Tab{tab:optic}) as well as two alternatives. Coded
aperture CA1 is a Uniformly Redundant Array~\cite{ref:fencan} with
31 elements and has been extensively evaluated in run periods to date;
CA2 was designed with the conclusions of this section in mind, and 
will be used in our next run period. 
Comparisons are made at our smallest energy, as that is where
performance is inhibited most by lack of light at low beam current.
The \prftt s at $\Eb=1.8$\gev\ 
are shown in \Fig{fig:prfcaone}(a) and (b), 
respectively; beam-size resolving power is shown
in \Fig{fig:fomcacomp}. The key difference is wider and fewer slits in CA2
than for CA1; CA2 lets through more x-ray intensity, and has peaks with deeper
and wider valleys. An even simpler design, denoted as 3S,
features just three 60\mi\ slits separated from one another by 
60\mi\ of masking; its properties appear in \Figs{fig:prfcaone}(c) and 
\ref{fig:fomcacomp}.
Note that above $\siz=25$\mi, a simple 60\mi\ pinhole is superior in
statistical resolving power of beam size, because above that size the valleys
between the peaks in the coded apertures' \prftt s become populated, 
leaving the pinhole with
steeper edges with which to provide resolution.
CA1 has better resolving power than the pinhole below $\siz=9$\mi;
however, at this beam size, both have $\qs\ll 1$.
CA2 and 3S fare better than CA1 or the pinhole at moderate
or small beam sizes. It is interesting to note that
the slight superiority of CA2 over 3S at small
beam size is reversed at $\Eb=2.1$\gev, demonstrating
the sensitivity of performance to optic design and x-ray spectrum.

\section[Apparatus]{Apparatus}
\label{sec:app}

Two installations of the xBSM are maintained at CESR-TA: one for electrons,
and one for positrons. These are similar, differing only in the source-to-optic
and optic-to-detector distances ($L$ and $\lp$, respectively, 
in \Fig{fig:geom}), as shown in \Tab{tab:geom}. Each apparatus consists
of four components: beamline, optical elements, detector elements, and
readout. These are described below.
  
\subsection{Beamline}

\begin{figure*}[t]
\begin{center}
   \includegraphics*[width=7in]{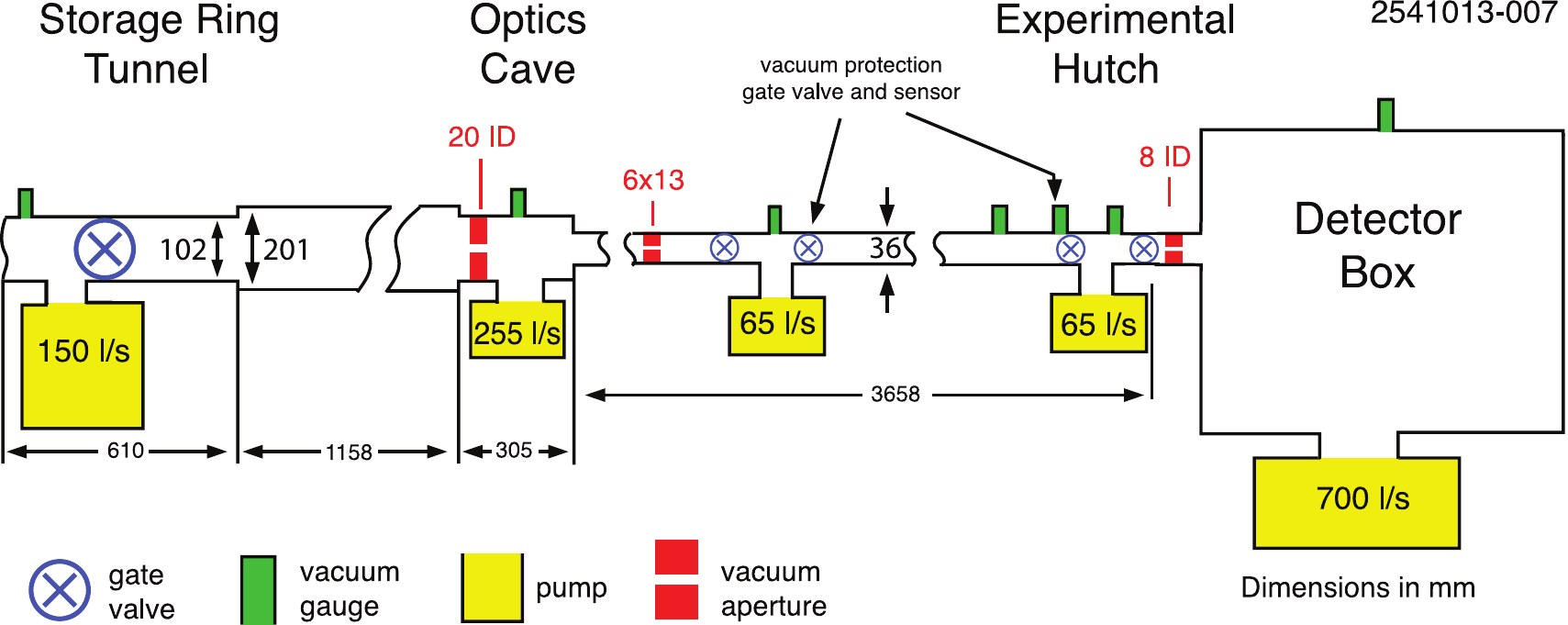}
   \caption{Layout of a CESR-TA xBSM beamline showing locations of pumps,
     impedance apertures, vacuum gauges, and relevant dimensions, in mm.}
\label{fig:beamline}
\end{center}
\end{figure*}

During each of two annual CESR-TA run periods, two CHESS~\cite{ref:chess} 
(Cornell High Energy Synchrotron Source) beamlines and hutches are 
re-purposed for xBSM use. A windowless evacuated x-ray beamline, 
approximately 14~m in length, connects the detector box to the CESR 
vacuum, as shown in \Fig{fig:beamline}. There is no material impeding 
the x-ray beam except for purposely introduced optical elements, defining 
apertures, and filters, all described below. Thus, the detector box volume 
is in contact with the CESR vacuum. While the CESR vacuum is maintained at 
approximately $2\times 10^{-9}$~Torr, the detector box contains electronic 
components, printed circuit boards, cables, and cooling lines that are not 
vacuum compatible. The detector box vacuum is maintained at approximately 
$10^{-6}$~Torr. The differential pressure is maintained without contaminating
 the CESR vacuum by a series of apertures in the beamline that 
increase the impedance and by the use of turbo pumps in each section between 
the apertures.  A fast gate valve, located 2~m upstream of the detector box, 
protects the CESR vacuum by closing if the pressure
measured at the vacuum protection sensor (see \Fig{fig:beamline}) exceeds 
$10^{-4}$~Torr. This trip level corresponds to $10^{-6}$~Torr at the optic
assembly due to the impedances and 
pumps. The windowless feature is essential for providing sufficient x-ray 
flux for single-shot capability at low beam energies and/or current 
(\ie having a viable beam size measurement from a single bunch on a single 
turn).

\subsection{Detector}

The detector is a linear array of InGaAs diodes with a 50\mi\ pitch and 400\mi\
horizontal width manufactured by Fermionics, Inc.~\cite{ref:fermionics}. 
Its parameters are summarized in \Tab{tab:det}.
The active InGaAs layer is 3.5\mi\ thick and is covered by a Si$_3$N$_4$ 
passivation layer 0.16\mi\ thick.  The time
response of the detector is sub-nanosecond (see \Sec{sec:readout} below).
The portion of a pixel sensitive to x-rays,
which was measured by sweeping a narrow vertically limiting aperture
immediately upstream of the detector,
is slightly smaller than the full pitch.
Hence there is no overlap in the response of adjacent pixels, 
which would otherwise alter the effective \prftt; neither is there an 
appreciable insensitive region of more than a few microns between 
adjacent pixels. While this detector is marketed for detection of
ultraviolet radiation (150~nm), we have found that it performs well 
in the x-ray (1~nm) range. A measurement of the detector spectral
response is detailed in \Sec{sec:specresponse}.
Other commercially available detector arrays 
tested did not meet the sub-ns time response requirement.

\begin{table}[t]
\small\selectfont
\caption{CESR-TA photodiode array characteristics. All
quantities marked {\it (spec)} are as specified by the manufacturer.
The active pixel width was measured in our apparatus. }
\label{tab:det}
\setlength{\tabcolsep}{1.35pc}
\begin{center}
\small\selectfont
\begin{tabular}{lc}
\hline\hline
\rule[10pt]{-1mm}{0mm}
Parameter & Value \\
\hline
\rule[10pt]{-1mm}{0mm}
\# channels, contiguous working  {\it (spec)}  &    64        \\
\# channels read out for beam-size images  &    32        \\
Pitch {\it (spec)}                               &   50\mi \\
Active pixel height {\it (measured)}       &  $45\pm3$\mi \\
Horizontal width {\it (spec)}                    &   400\mi \\
InGaAs thickness {\it (spec)}                     &   3.5\mi \\
InGaAs density                       &   5.5~g/cm$^3$ \\
Si$_3$N$_4$ thickness {\it (spec)}                 &   0.16\mi \\
Si$_3$N$_4$ density                   &   3.25~g/cm$^3$ \\
\hline\hline
\end{tabular}
\end{center}
\end{table}

\begin{table}[b]
\small\selectfont
\caption{CESR-TA attenuator and filter characteristics. The attenuation of any
limiting aperture alone is approximately the width relative to the active 
detector width of 400\mi.  }
\label{tab:fil}
\setlength{\tabcolsep}{0.95pc}
\begin{center}
\small\selectfont
\begin{tabular}{lcc}
\hline\hline
\rule[10pt]{-1mm}{0mm}
Material & Thickness & Horizontal aperture \\
\hline
\rule[10pt]{-1mm}{0mm}
 None & - & $1-2$\mm \\
 None & - & 171\mi   \\
 None & - & 105\mi   \\
 None & - & 35\mi  \\
 Diamond & $4.4\pm0.1$\mi & 105\mi \\
 Diamond & $4.4\pm0.1$\mi & 50\mi \\
 Diamond & $4.4\pm0.1$\mi & $1-2$\mm \\
 Molybdenum  & $1.91\pm0.06$\mi & $1-2$\mm  \\
 Aluminum & $7.2\pm0.7$\mi & $1-2$\mm  \\
\hline\hline
\end{tabular}
\end{center}
\end{table}

\begin{figure}[tb]
   \includegraphics*[width=\figwid]{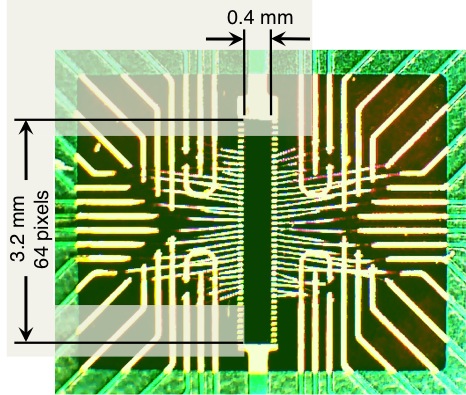}
   \caption{Annotated photograph of the CESR-TA xBSM photodiode array on its 
     mounting board.}
\label{fig:det}
\end{figure}

The detector, shown in \Fig{fig:det}, is attached to a custom-designed
printed circuit board. 
The common cathode of the detector array contacts the center trace through
conducting epoxy. Contacts of contiguous anodes alternate between the left and
right sides of the array. The printed pattern allows fabrication with 0.25~mm
center-to-center trace technology, and limits the bond length to 2.2~mm. 

The central 32 contiguous diodes of the array are used to measure
single-shot vertical profiles. Eight additional channels are connected
for diagnostic single-diode readout mode (\Sec{sec:align}). Another 3 diode
elements are attached to each of 
4 ground lines to reduce noise in the single-diode
readout. Thus, there are 45 connections to the printed circuits mounting board:
32 contiguous elements, the 8 single diode readout elements, 4 grounds, and the
common cathode. 

\begin{figure}[tb]
   \includegraphics*[width=\figwid]{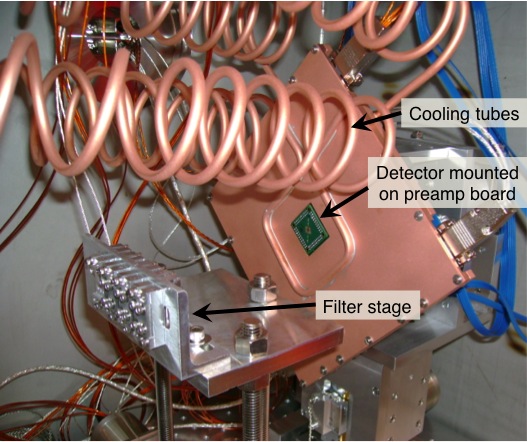}
   \caption{Annotated photograph of the CESR-TA xBSM detector mounted 
     on its preamp board and movable
     stage, inside the vacuum of the detector box. A water cooled copper jacket
     surrounds the preamp board. Cooling tubes and the filter stage area
     also visible.}
\label{fig:detmounted}
\end{figure}

The detector board is mounted on a preamp board, which, in turn, 
is attached to a stage with motor controlled horizontal and vertical
motion, as shown in \Fig{fig:detmounted}. 
This stage provides motion necessary for alignment (\Sec{sec:align}). 
Also visible in \Fig{fig:detmounted}, a set of filters and horizontally 
limiting apertures (attenuators) are mounted inside the detector box 
75\mm\ upstream of the detector itself. Their properties are
summarized in \Tab{tab:fil}. The filters
were selected to alter the x-ray spectrum incident on 
the detector in order to probe sensitivity of the detector and
beam size measurement to wavelength. The attenuators are used to 
protect the detector and prevent preamplifier saturation by
limiting x-ray flux in high current and/or high electron beam energy operation.
  
\begin{figure}[t]
   \includegraphics*[width=\figwid]{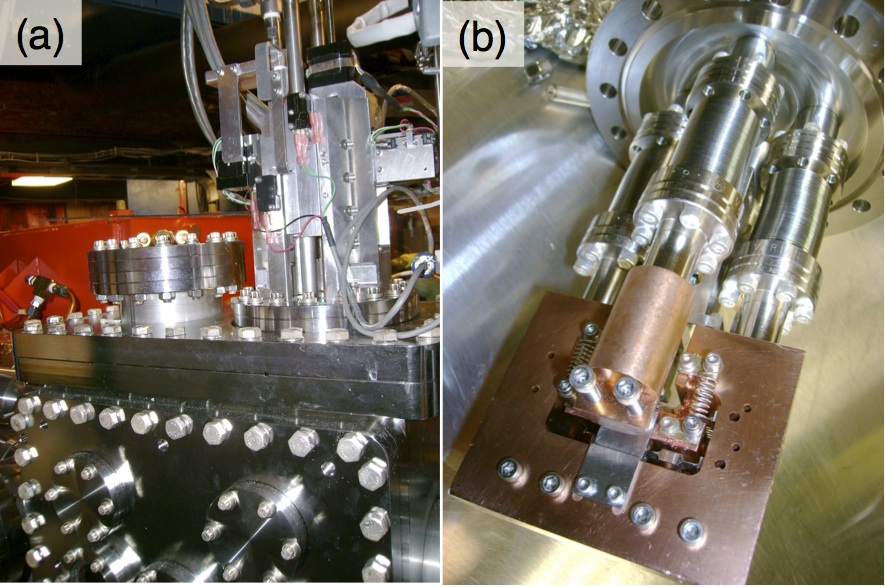}
   \caption{(a) The xBSM optics assembly mounted in the beamline. 
      (b) Motors visible on the top control the vertical locations 
          of two optics stages and the pinhole gap size.}
\label{fig:opticass}
\end{figure}

\subsection{Optical elements}
\label{sec:optics}

Optical elements are located in a dedicated enclosure in the beamline, shown
in \Fig{fig:opticass}(a), approximately 10~m upstream of the detector box 
and 4~m downstream of the source point. They are mounted on a 
stage, shown in \Fig{fig:opticass}(b).
The three struts provide vertical motion
and water cooling. 
Optical elements which have been tested include a pinhole, coded
apertures of different designs, and a two-dimensional Fresnel zone
plate. Each of these can be separately inserted into the beam with motor 
controls. 
  
\begin{figure}[tbh]
   \includegraphics*[width=\figwid]{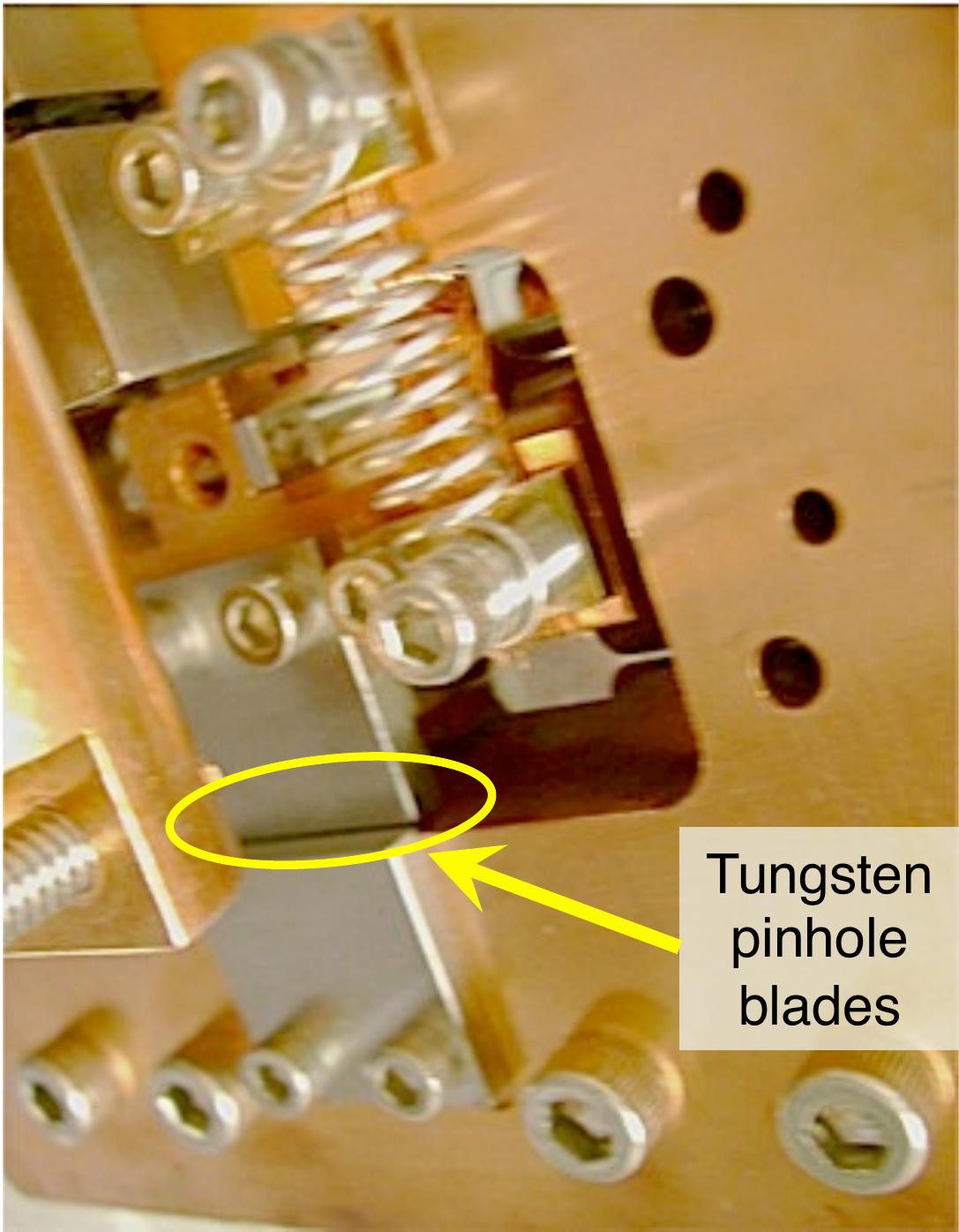}
   \caption{Annotated photograph of the tungsten blades forming the pinhole
     for the CESR-TA xBSM.}
\label{fig:pinholephoto}
\end{figure}
 
\begin{figure}[tbh]
   \includegraphics*[width=\figwid]{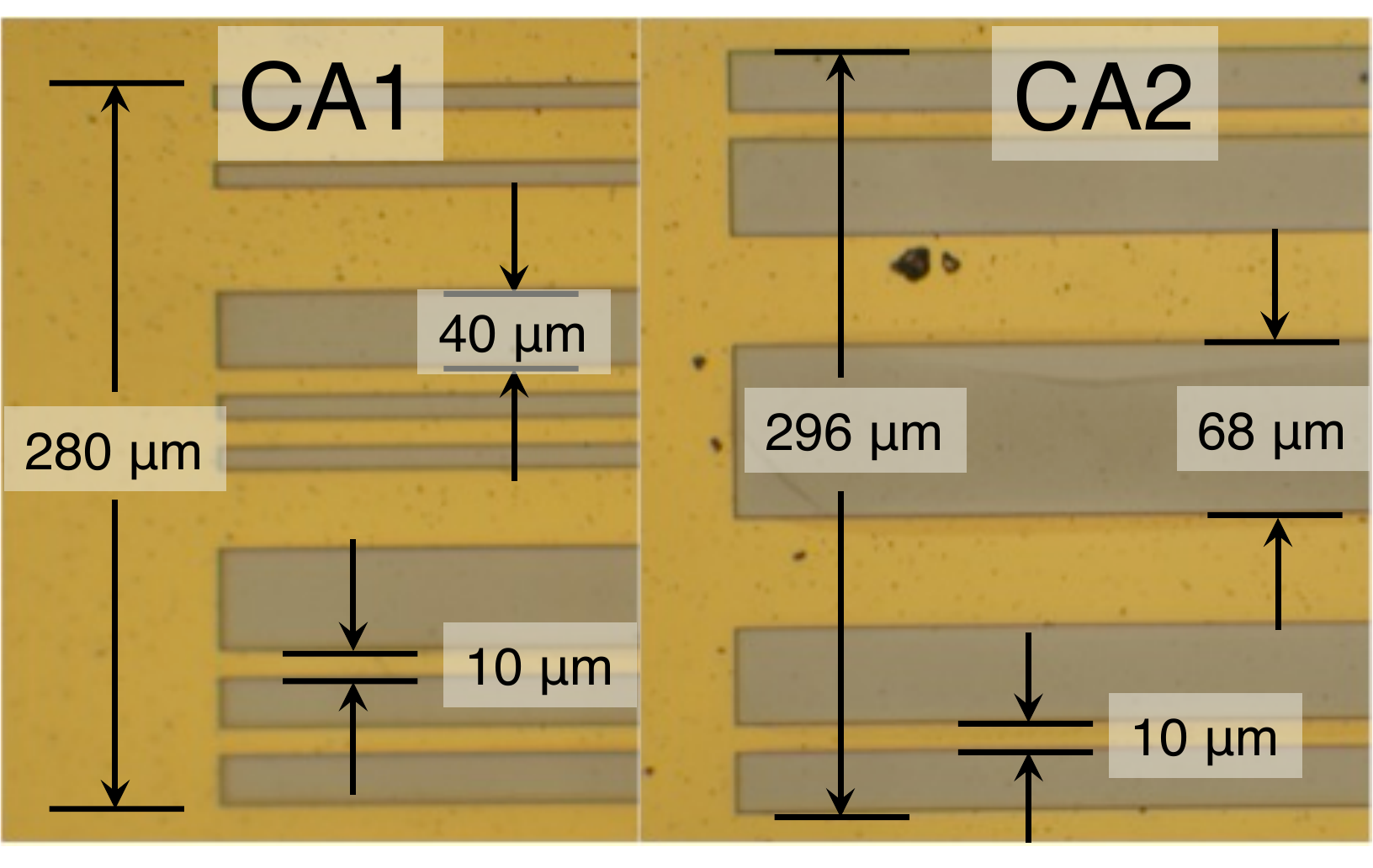}
   \caption{Photographs of portions of CESR-TA xBSM coded aperture 
            optical elements CA1 (left)
            and CA2 (right).
            Dark strips indicate transmission slits, while lighter
            areas represent the gold coating.
            The imperfections (black spots) are remnants of etching 
            resist with thickness $\sim$0.01\mi, which are essentially 
            transparent to x-rays.}
\label{fig:caphoto}
\end{figure}
 
\begin{figure}[tbhp]
   \includegraphics*[width=\figwid]{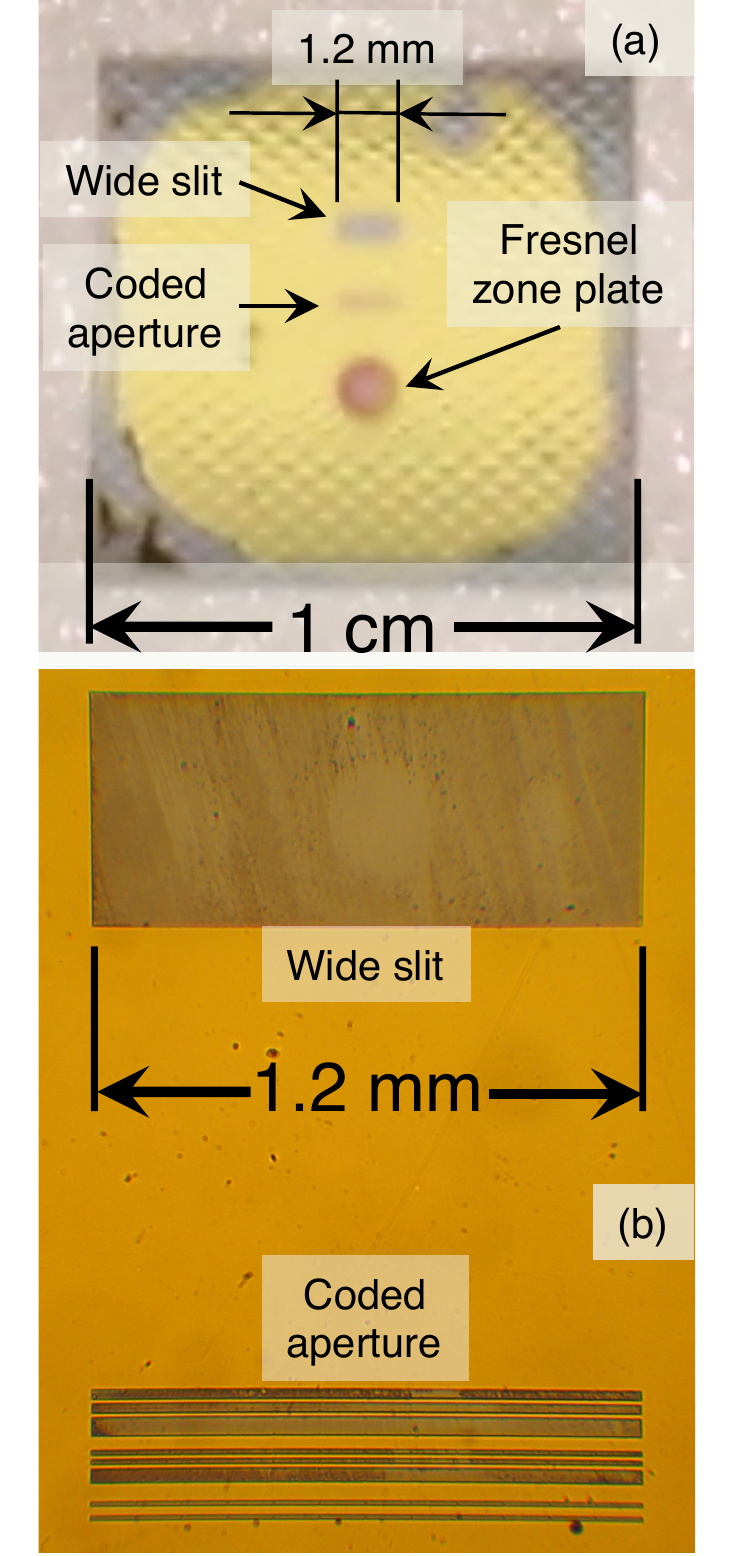}
   \caption{Annotated photographs of the silicon substrate optics chip. 
     (a) Three optical elements are 
     (barely) visible in the center of the 1-cm-square chip. 
     (b) The coded aperture and simple rectangular hole are visible with
     reflected light. For scale, the common width of the
     features is 1.2~mm.}
\label{fig:opticschip}
\end{figure}
   
\begin{figure}[tbhp]
   \includegraphics*[width=\figwid]{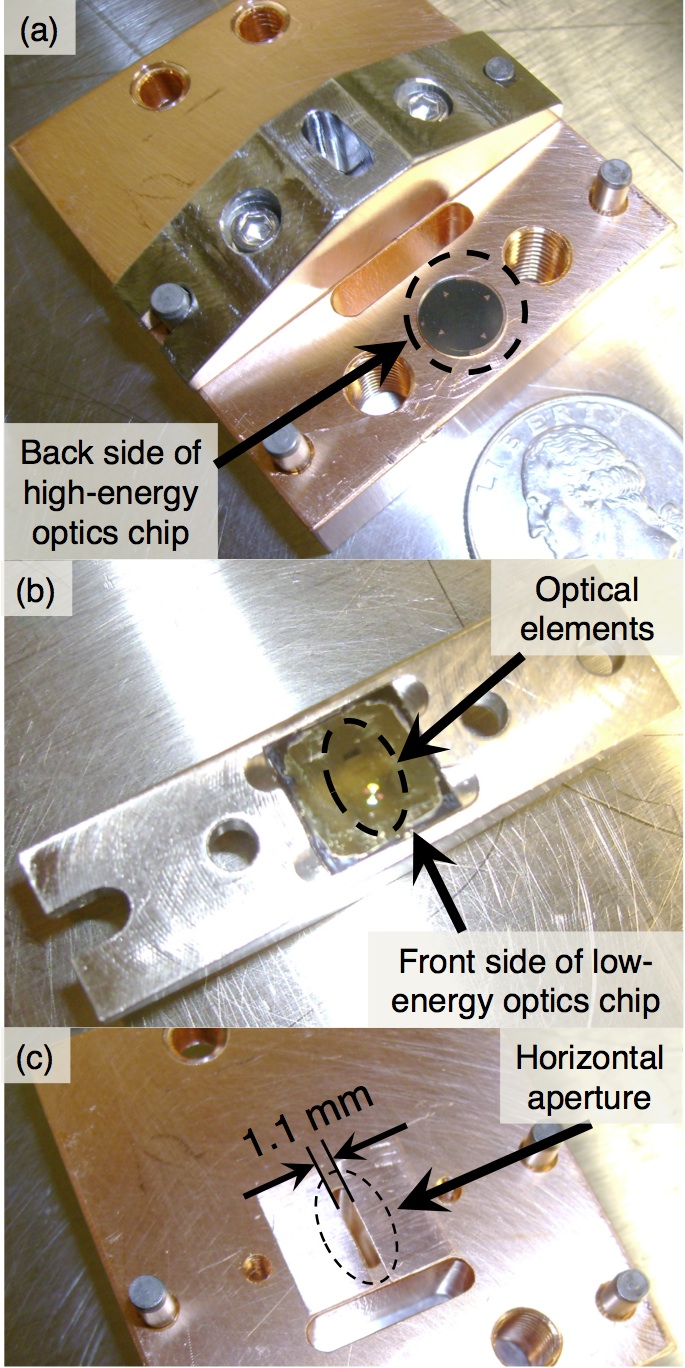}
   \caption{(a) The copper support block that attaches to one of the struts 
     in \Fig{fig:opticass}(b) holds 2 optics chips. Here, one chip is
     exposed while the 
     other is behind the stainless steel cover. (b) An optics chip is shown 
     mounted in the stainless steel cover. (c) For low energy chips, the 
     copper mounting block provides a 1.1\mm\ horizontally defining aperture.}
\label{fig:opticmounts}
\end{figure}

The one-dimensional pinhole is the simplest optic: a
vertically limiting slit. The adjustable slit consists of two tungsten blades, 
each 2.5~mm thick, as shown in \Fig{fig:pinholephoto}. The vertical
location of the upper blade is motor-controlled through one of the
struts: its position sets the slit opening. A 0.25~mm offset of the 
two blades along the beamline allows the opening to be completely
blocked without any contact between the blades. Faces of the tungsten
blades forming the slit are tapered away from the incident x-ray beam 
by an angle of 2$^\circ$ to prevent small angle reflections off the
faces from reaching the detector.
For the x-ray spectrum at typical operating beam energies, 1.8-2.6\gev, 
the gap size that results in the narrowest central peak in the 
image is in the range of 45-60\mi; travel of the upper blade allows
openings from zero up to 200\mi. A 500\mi-wide horizontal collimator 
sits just upstream of the pinhole to preclude x-rays reaching
the detector via any path not passing through the pinhole and to limit
the depth of field (\ie the longitudinal spread of the source).

A coded aperture, a 500\mi-wide slit, and a Fresnel zone plate 
are available.
These were acquired from Applied Nanotools, Inc.~\cite{ref:appnt}, and 
are etched with a proprietary process into a thin gold layer on a
2.5\mi-thick silicon substrate chip.
The two coded aperture designs we are using for low energy running, 
CA1 and CA2, appear in high resolution photographs in \Fig{fig:caphoto}, 
and have parameters summarized in \Tab{tab:optic}.
Optical measurements indicate that
the systematic placement of features is within 0.5\mi\ of the specifications. 
Edge quality of the etching is better than 0.1\mi\ rms deviation. 
The full optics chip which contains CA1 is shown in
\Fig{fig:opticschip}(a), and a view of the entire CA1 and wide slit in
\Fig{fig:opticschip}(b).

Optics chips are mounted on a copper support block which
provides precision alignment and thermal contact, 
as shown in \Fig{fig:opticmounts}. A 1.1\mm\ horizontally 
limiting aperture in the support block stops x-rays that are 
transmitted through the gold outside of the etched pattern. A stainless
steel cover plate, shown in \Fig{fig:opticmounts}, 
presses the optics chip into the 
copper, enabling precision alignment. The support block is, 
in turn, mounted on a separate vertical stage downstream 
of the pinhole (foreground of \Fig{fig:opticass}(b)). 

Coded apertures suitable for higher energy running
 have been tested at beam energies of 4\gev\ and
above. The mounting block shown in \Fig{fig:opticmounts}(a) 
shows one of these devices: a round 350\mi\ CVD (synthetic) diamond substrate 
with a pattern (not visible) etched into 8.7\mi-thick gold masking. This
device is discussed elsewhere~\cite{ref:arinaga,ref:hiopt}.

Performance of the pinhole and coded aperture are described in 
\Sec{sec:results}. The Fresnel zone plate was included in early optics 
chips as an option. This Fresnel zone plate is designed to focus 
2.5\kev\ x-rays in two dimensions. When used with a wide spectrum x-ray 
beam and a one-dimensional detector, however, beam size resolving power
is low. Use of an x-ray monochromator can address the wide spectrum,
but only by limiting the number of 
photons to about 1\% of those generated in the bend magnet, and 
so does not improve the resolving power enough for use in single-shot
operation.

\subsection{Readout \& data acquisition}
\label{sec:readout}

The xBSM electronic readout system has three main functions: 
$(i)$~to amplify each
pixel's photodiode signal, corresponding to 1-10$^4$ photoelectrons, 
synchronously with bunches passing the source point separated by as little 
as 4~ns; $(ii)$ to digitize the resulting voltage; and $(iii)$ assemble the 
digitized data 
for all channels from each turn and bunch into a data file. This challenge 
is met with a system designed fully in-house. Each of 32 independent readout 
channels consists of a transimpedance amplifier, a variable gain amplifier, 
an analog to digital converter, a field-programmable gate array (FPGA), 
a local buffer, and a programmable sample clock delay. 

\begin{figure}[b]
\begin{center}
   \includegraphics*[width=\figwid]{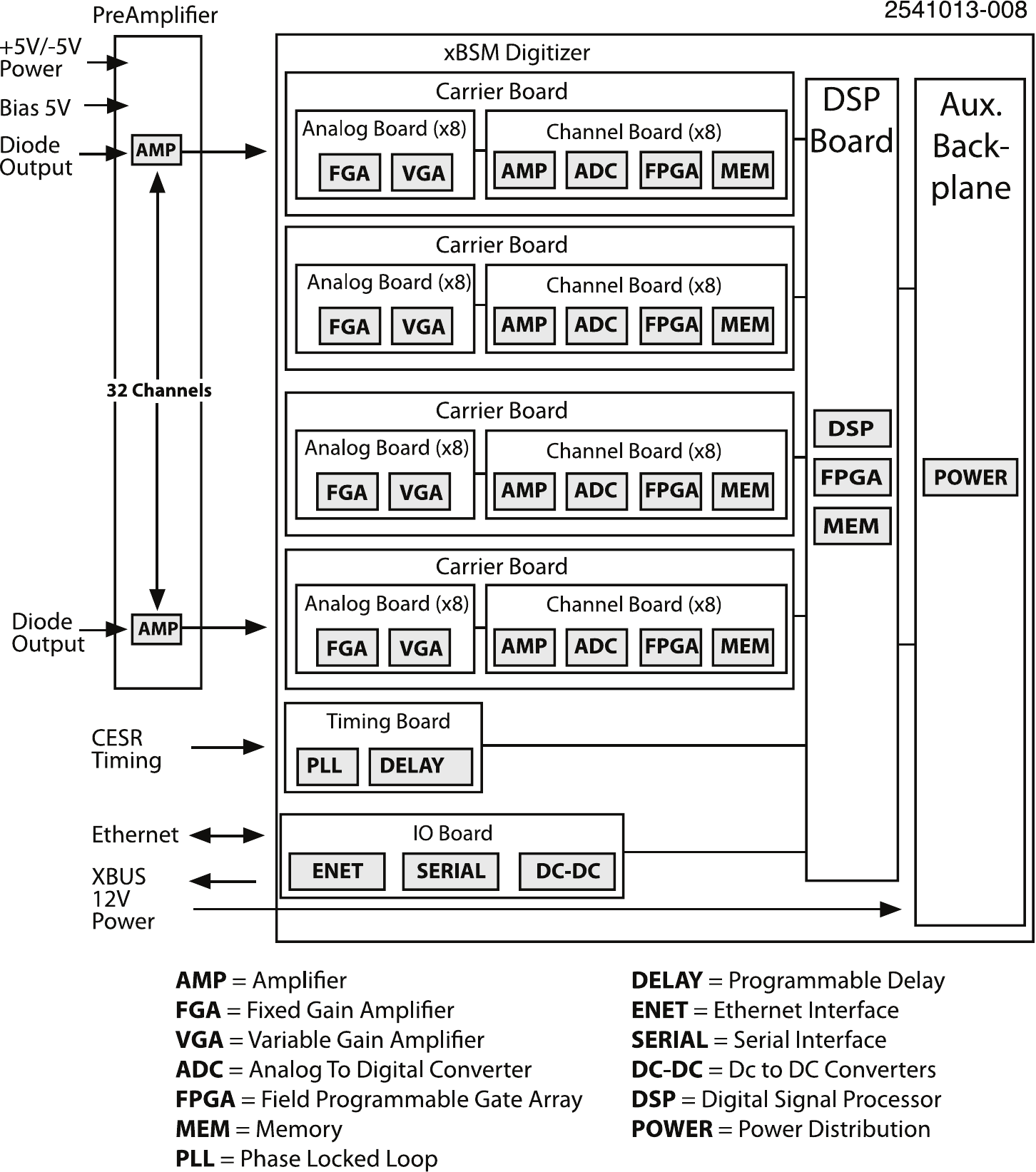}
   \caption{The CESR-TA xBSM electronics design.}
\label{fig:electronics}
\end{center}
\end{figure}
 
\begin{figure}[tbp]
   \includegraphics*[width=\figwid]{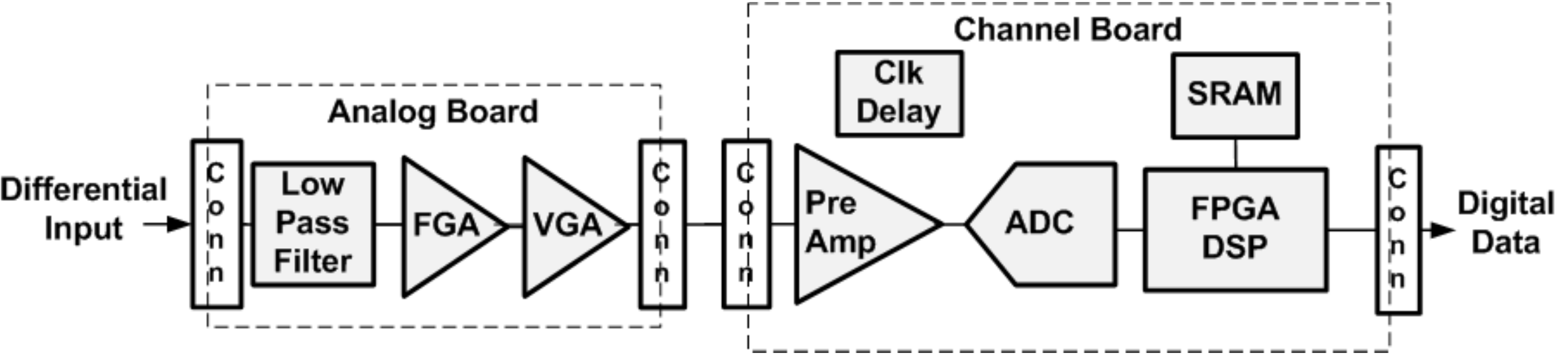}
   \caption{The CESR-TA xBSM function diagram of a single readout channel.}
\label{fig:function}
\end{figure}
 
Hardware components are modular, and are positioned
both inside and outside the evacuated detector box in custom housings. 
\Figure{fig:electronics} shows an overview of this system. Components 
inside the detector box are represented in major block on the
left. Signals from the diode detector are received, amplified and converted to
differential signals prior to being delivered out of the detector box 
in bundles of 8 channels of twin-axial cable. 
Components outside the detector box are represented
in the major block on the right. 
The bundle of 8 differential signals is received by a carrier board
(represented as a separate sub-block) 
that handles 8 channels of signal conditioning and digitization. 

\begin{figure*}[tbhp]
\begin{center}
   \includegraphics*[width=7in]{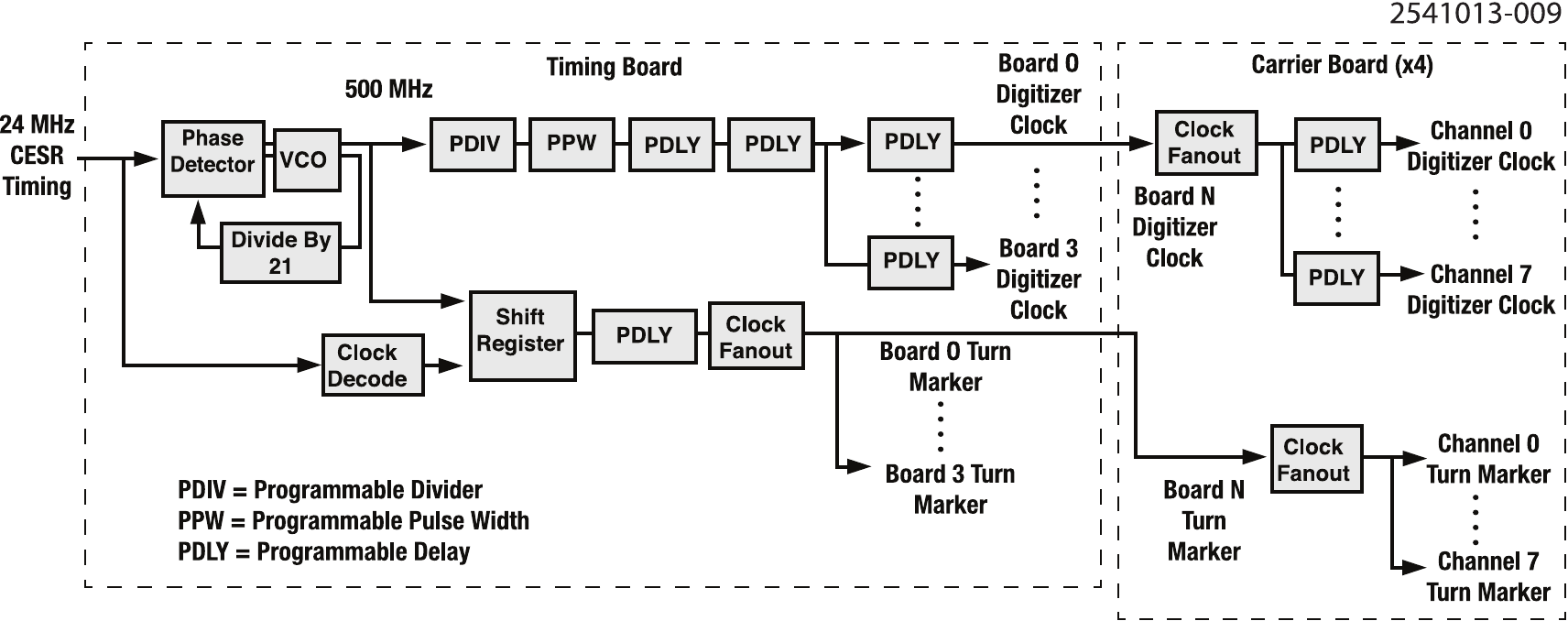}
   \caption{The CESR-TA xBSM signal chain timing.}
\label{fig:signalchain}
\end{center}
\end{figure*}

\Figure{fig:function} shows the functional blocks of 
each channel's signal processing. After a 500~MHz
low-pass filter, each channel provides 12~dB of fixed gain and $-4$~dB to
20~dB of digitally controllable gain in 24 steps of equal size. 
In ``automatic gain control'' mode, the readout software 
tests the signal and then sets the gain of all channels to the 
same, single highest 
value that does not saturate any single channel.  

A 10-bit analog-to-digital converter sampling at 300~MHz is utilized 
to provide the requisite timing performance. The data from the converter is
streamed through a local FPGA and is stored locally for each channel in an 
18~Mbit SRAM. This allows for $10^6$ samples to be captured on each channel
without data compression. Each channel receives its own sampling 
clock signal and has
an in-channel configurable delay to allow for channel-to-channel
synchronization. 

Synchronization to the CESR timing system occurs via a 24~MHz
encoded data signal. This signal, synchronized to the CESR RF system,
also contains revolution (turn) markers with a frequency of 0.4~MHz and
acquisition triggers. The same timing signal is also delivered to all of the
storage ring beam position monitors, so that synchronized turn-by-turn
measurements can be shared by both systems. 

A more detailed description of the internal timing control and distribution is
provided in \Fig{fig:signalchain}. An on-board voltage controlled 
oscillator generates a 500~MHz source clock synchronized to the 24~MHz 
encoded signal. This clock drives a divider with a programmable pulse
width in order to provide a fully configurable sampling clock. 
The sampling clock signal passes through a series of common programmable 
delays prior to being fanned out to each of the four carrier
boards. There are separate programmable delays at the carrier board level. 
The clock signal 
on the carrier board is then fanned out to the 8 sampling channels. Every
channel has its own programmable delay, allowing precise adjustment of the
sample point. 
The timing board also extracts the turn marker from the 24~MHz
encoded signal. The turn marker passes through its own set of programmable
delays and is fanned out to the 4 carrier boards where it is delivered to the
individual channels. Each channel can then be synchronized to the accelerator
revolution frequency. The digitizers run continuously. A bunch pattern
describing the time structure of filled RF buckets in CESR 
serves as a gating signal provided to all channels for
determining when the sample is to be stored in the local SRAM. 

\begin{figure}[t]
   \includegraphics*[width=\figwid]{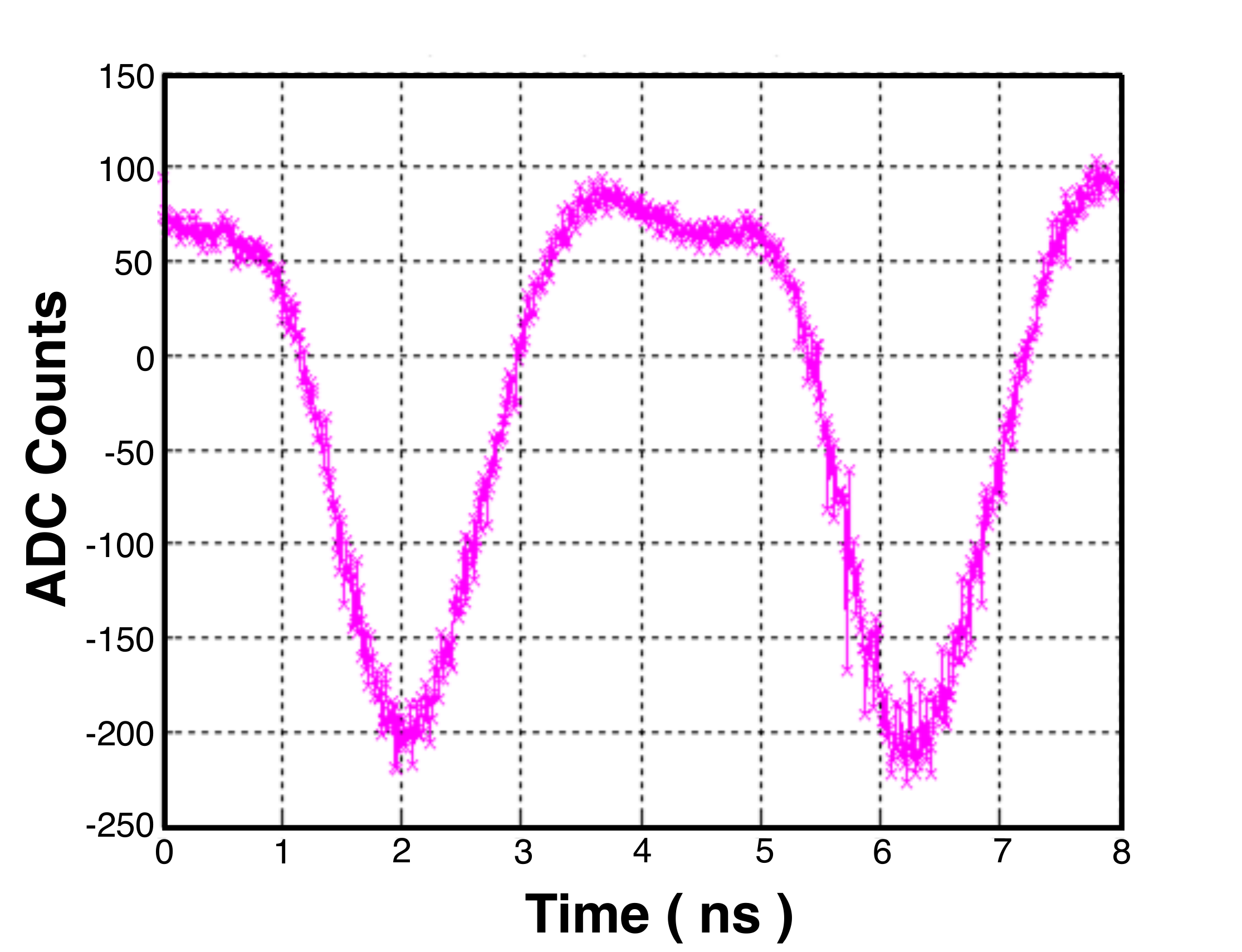}
   \caption{Pulse shape vs.~time of the CESR-TA xBSM photodiode readout.}
\label{fig:pulsevstime}
\end{figure}

During CESR-TA calibration procedures, 4-ns-spaced bunches can be measured
in detail. The digitizers can be operated in waveform-capture mode 
by shifting the sample time of subsequent measurements. 
\Figure{fig:pulsevstime} shows the
resulting digitized signal of two 4-ns-spaced bunches of electrons. The 
waveform-capture mode is used for the timing calibration, discussed further
in \Sec{sec:cal}, in which the programmable delay for each channel is 
adjusted in 10~ps increments to
center the sample point on the peak signal time.
Drift in these settings is small compared to the width of the 
optimal sampling section of the peak of the diode waveform. 
The main source of drift in the delay chips is thermal loading, 
but as the data acquisition hardware is temperature-regulated,
any remaining effect is small over the period of a day.  
The delays are tuned prior to each experimental shift. 
Variation in the RF phase due to beam loading has an
effectively negligible effect on the accuracy of the sampling point.

The digital data from each channel is collected on the digital signal 
processing (DSP) board. This board allows for on-instrument
processing and data manipulation. It also provides the link to the ethernet
interface which is contained on the board. All controls and data flow are
transmitted via an ethernet interface.  A diagnostic interface is provided 
via a serial fieldbus (XBUS) connection on the IO board. Power is delivered 
to the various boards via the auxiliary backplane.

Finally, digitized pulse heights are assembled into a text raw data file which
constitutes a {\it run}. Each file begins with a number of lines specifying
characteristics of the run, including its sequential number, a date and time
stamp, whether it contains
electron or positron data, motor control settings on optic, filter, and detector
elements, beam energy, beam current, bunch pattern, digital gain range setting,
and pedestal and gain values for the 32 channels of readout. These are followed
by a sequence of ordered lists, each of 32 numbers corresponding to the 32
contiguous pixel adc values. Single bunch runs generally span from 
100 to $64\times10^3$
turns, with 1024 turns most common. Multi-bunch runs typically also contain 
1024 turns per bunch for up to $\sim$50 bunches; the number of turns is limited
by the size of the SRAM as $\sim10^6$/(\# bunches). For multi-bunch running
the data file has turn-by-turn data for bunch \#1, followed by turn-by-turn 
data for bunch \#2, and so forth. Data files are maintained on 
a backed-up RAID array for subsequent offline analysis.

\section[Operations]{Operations}
\label{sec:ops}

Each CESR-TA run period typically has a duration of 9-21 days. 
Hence physical installation, data acquisition checkout,
alignments of the detector and optical elements, and setting
of the pinhole size are necessary for
each such run period. Alignment and calibrations can change slightly
during each run period, requiring re-tuning and re-calibration from day to day,
and the pinhole optic is set to a have a different gap size for each 
beam energy.

\subsection{Installation}
\label{sec:install}

Prior to each CESR-TA experimental run period, the xBSM hardware is 
installed into the CHESS hutches, one that is illuminated by x-rays 
from the positron beam and the other by x-rays from the electron beam.  
The hardware is broken into two main assemblies, the vacuum
controls and the detector box.  Each assembly is mounted on a rolling frame
and is placed into storage during CHESS user experimental run periods.  
The hardware 
is rolled into the hutch, placed on rigid stands, and bolted to the vacuum
chamber bringing the x-ray beam to the hutch.
The basic installation process places the hardware roughly into the
same position it was in at the conclusion of the previous CESR-TA run 
period. Both
assemblies have survey monuments permanently mounted on critical components
(flanges, box corners, beam pipes).  These monuments are used in conjunction
with survey monuments within the hutch to locate the hardware to within
~0.5\mm\ of the previously known good location, which
had been empirically determined to produce proper beam size measurements.
The xBSM alignment process is described further in \Sec{sec:align}.

After the hardware has been properly installed and positioned, 
the vacuum controls are used to
initiate the pump-down of the detector box.  A differential pumping scheme is
utilized which requires the detector box to be brought down
to $10^{-5}$~Torr before the upstream gate valves are opened to allow the 
transmission of x-rays from the source.
The initial pump-down takes approximately six hours due to an accumulation of
contaminants during storage.  Once the pressures have stabilized and the gate
valves are open, the system becomes ready for subsequent pre-operational
alignment and calibration.

\subsection{Detector \& Optic Alignment}
\label{sec:align}

Initially the detector box
and optical elements are positioned via an optical survey to an
accuracy of ~0.5 mm (\Sec{sec:install}). However, the 
$\theta$-dependence of the x-ray spectrum and intensity (\Eqs{eqn:power}, 
(\ref{eqn:SRpol}), and (\ref{eqn:sth}), as well as \Tab{tab:spec}) 
introduces sensitivity to variations in the $e^\pm$ orbit that 
typically occur in the storage ring during day-to-day tuning. 
The alignment procedure also
provides an initial check on the general beam quality. Prior to
alignment, the storage ring orbit is set to a nominal working point and
a low-emittance-tuning procedure is used to minimize vertical beam
size. The detector is reset to a standard working position
for the pinhole optic in both the vertical and horizontal directions.  
During the alignment,
minor corrections are made to both the electron orbit and the positions
of the detector and optical elements. 

The alignment procedure starts with scans wherein the detector position
is slowly moved through its horizontal and vertical degrees of freedom
while its position and synchronized response are periodically recorded. 
As described in \Sec{sec:app}, the detector is mounted on a 
stepping-motor controlled stage which allows vertical and horizontal motions
of $\pm10$\mm.  Completion time for each scan is $\sim$1~minute.
The current from a single diode in the array is monitored, with readings 
in the 0.1~$\mu$A range; the diode is biased with the cathode at ground 
and the anode connected to an ammeter input. Several such diodes are 
available, located just outside the array of 32 contiguous diodes 
connected to the digital readout, as shown in \Fig{fig:det}. 
Typically, we use a diode that is 0.925\mm\ from the center of that array.
The diode array position in the vertical direction 
is adjusted in order to center the single diode on the nominal detector 
position.

\begin{figure}[tb]
   \includegraphics*[width=\figwid]{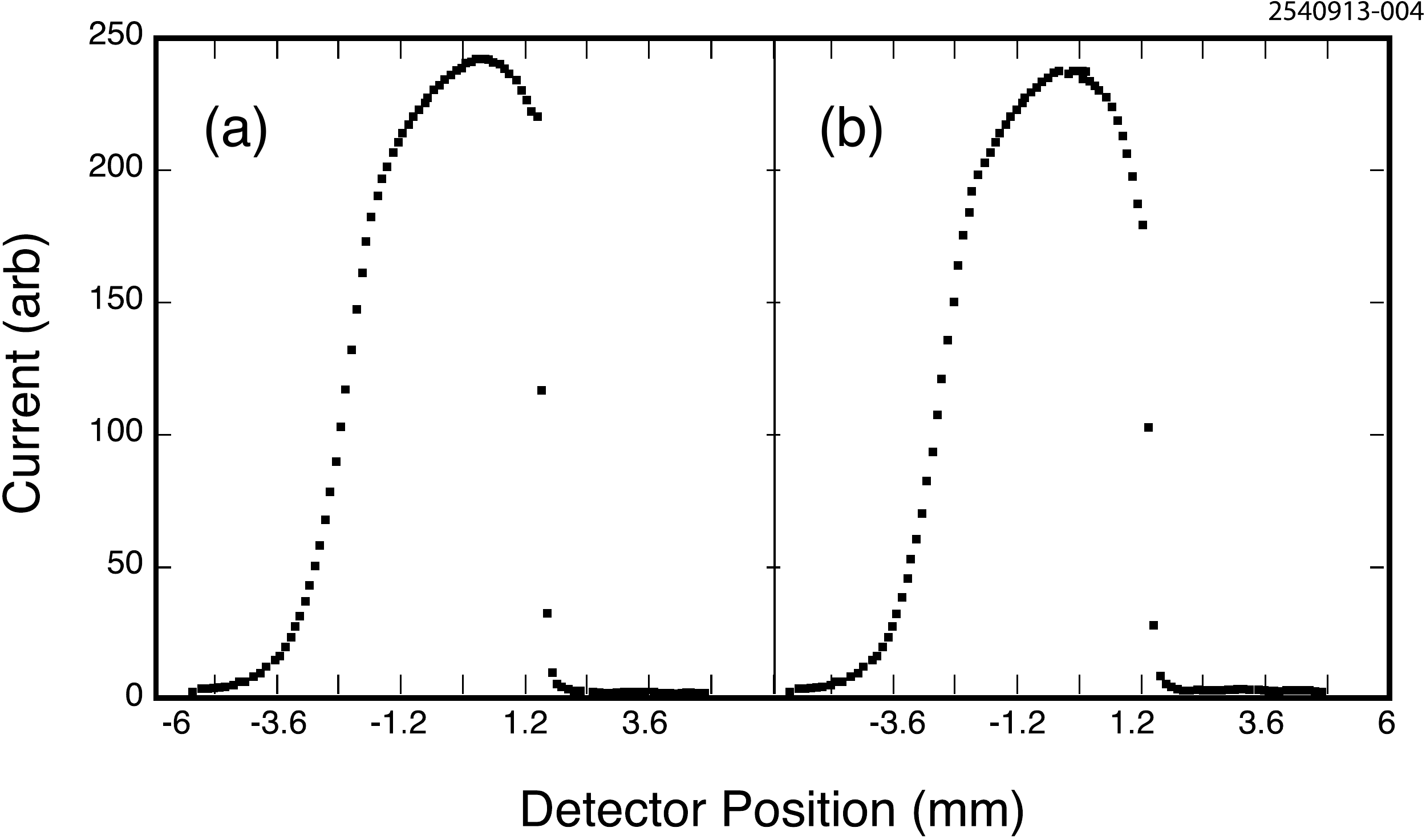}
   \caption{Single-diode current draw (arbitrary units)
   vs.~vertical position 
   of the detector stage as it is moved through an x-ray beam from 
   2.6\gev\ electrons with no optical element in place,
   both (a) before, and (b) after adjustment of the electron beam orbit.
   The vertical width of the x-ray beam observed here is 
   $\sigma\approx 1.1$\mm. The distribution in (a)
   is clipped above a position of 1.3\mm\ by the edge of an element in the beam
   line, whereas in $(b)$, the electron orbit has been adjusted to move the
   x-ray beam slightly away from the obstruction. }
\label{fig:align}
\end{figure}

\begin{figure}[tb]
   \includegraphics*[width=\figwid]{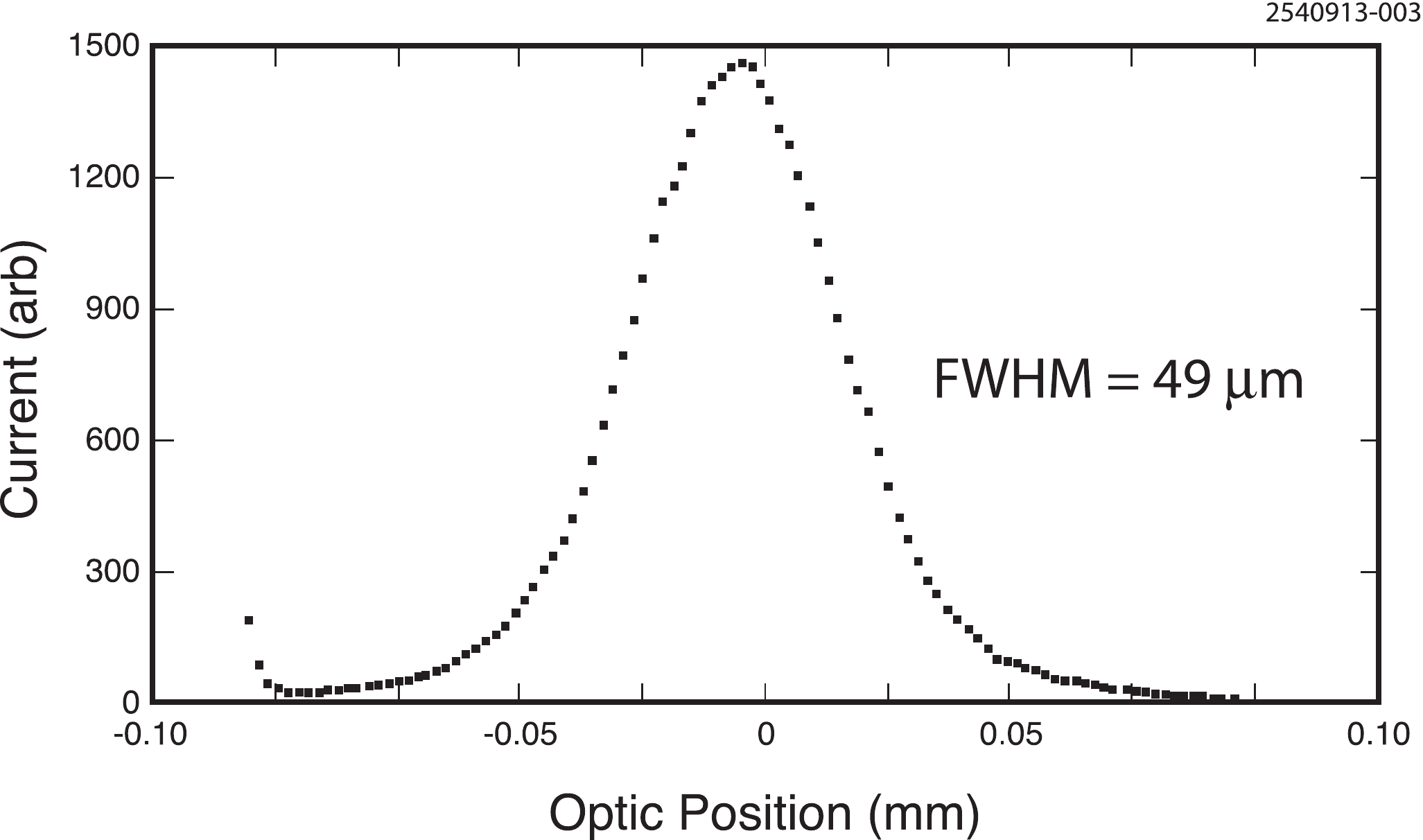}
   \caption{Single-diode current draw (arbitrary units) vs.~vertical position 
   of the optical element stage, using the pinhole optic with
   an opening of $\sim$40\mi. These data were taken just after
   the orbit adjustment shown in \Fig{fig:align}.}
\label{fig:aligntwo}
\end{figure}

In the first step of the alignment procedure, the detector is brought 
to a vertical reference location known to be in the nominal
plane of the storage ring. This is a preliminary position and may change
with subsequent tuning of the storage ring (see below).
Steering corrector magnets are used to move the horizontal $e^\pm$ orbit in the
storage ring to a reference orbit that is known to project light down the CHESS
beam line. The vertical orbit is corrected to be flat
(\ie in the plane of the storage ring) at the source point for the
x-ray beam line. The detector, read out in single-diode mode, is then exposed 
to the x-ray beam with no optical element in place.  

Typical initial vertical beam profiles are shown in \Fig{fig:align}, 
which feature widths consistent with that predicted by \Eq{eqn:sth} and 
shown in \Tab{tab:spec}. 
This example of an initial vertical beam profile has a 
single-diode current distribution that is truncated near one 
extreme of its vertical range, indicating that the x-ray beam 
has been missteered into an aperture.
This condition is corrected by fine-tuning the stored beam orbit.

The significant clipping visible in \Fig{fig:align}(a) 
was reduced to that shown in \Fig{fig:align}(b) with a subsequent orbit 
adjustment which moved the center of the x-ray profile
to a lower position, away from the aperture limitation.
The goal with such adjustments is to move the vertical position and 
direction of the x-ray beam such that the detector can be centered on
the beam envelope and any aperture clipping is beyond the edge of the
detector. In this example, the preliminary working position of the detector
center is $-0.26$\mm\ on the scale shown, while the center of the x-ray beam
envelope is at $-0.16$\mm. A small correction to the detector
position is made to center it on the beam.  
While the clipping still exists after the orbit adjustment in
this example, the beam profile is unaffected up to 1.3\mm\ from the peak,
which is more than the 0.8\mm\ half-width of the detector. After this
step, the $e^\pm$ beam orbit and detector vertical position (accounting for
the offset between the single diode and the center of the detector
array) are fixed for subsequent data acquisition. At this point, a
horizontal scan of the detector in single-diode-readout mode is performed 
to ensure that there is no upstream aperture limitation near the operating 
position; any clipping observed within the horizontal detector span indicates
a severe distortion relative to the reference orbit. After the root cause 
of any such distortion is found and addressed, the alignment procedure 
must be restarted.

In the next step, the optical elements are centered on the line between
the x-ray source point and the detector. The pinhole optic is mounted on
a stage that allows vertical motion. \Fig{fig:aligntwo} shows the result of
scanning the pinhole optic through the beam while logging the current in
the single diode. The location of the peak, in this case, $-0.005$~mm, is used
to locate the pinhole in subsequent data runs. The pinhole optic has a
0.5\mm\ horizontal aperture. The detector horizontal position is adjusted
to center the detector on the line defined by the x-ray source and the
optic aperture by conducting a single-diode scan of the detector  
in the horizontal direction.

The pinhole alignment provides a first measurement of the beam quality. 
The observed width of the distribution in \Fig{fig:aligntwo},
taken at the plane of the optical element, is proportional to that of 
the convolution of the \prftt\ with the magnified beam size, traced back 
from the detector plane to that of the PH optic. For \Fig{fig:aligntwo}, 
where fwhm~$\approx 49$\mi\ in the plane of the optic, this convolved rms 
width, referred to the source plane, is 
$\approx(1+1/M)\times{\rm fwhm}/2.36$, or $\approx30$\mi.
This is a common operating
condition for CESR-TA at $\Eb=2.1$\gev. Had this rms value been
much larger, it would have indicated that either the emittance 
or vertical beam motion were unacceptably large.
Such conditions would require retuning
the orbit and repeating this alignment procedure.

\begin{figure}[tb]
   \includegraphics*[width=\figwid]{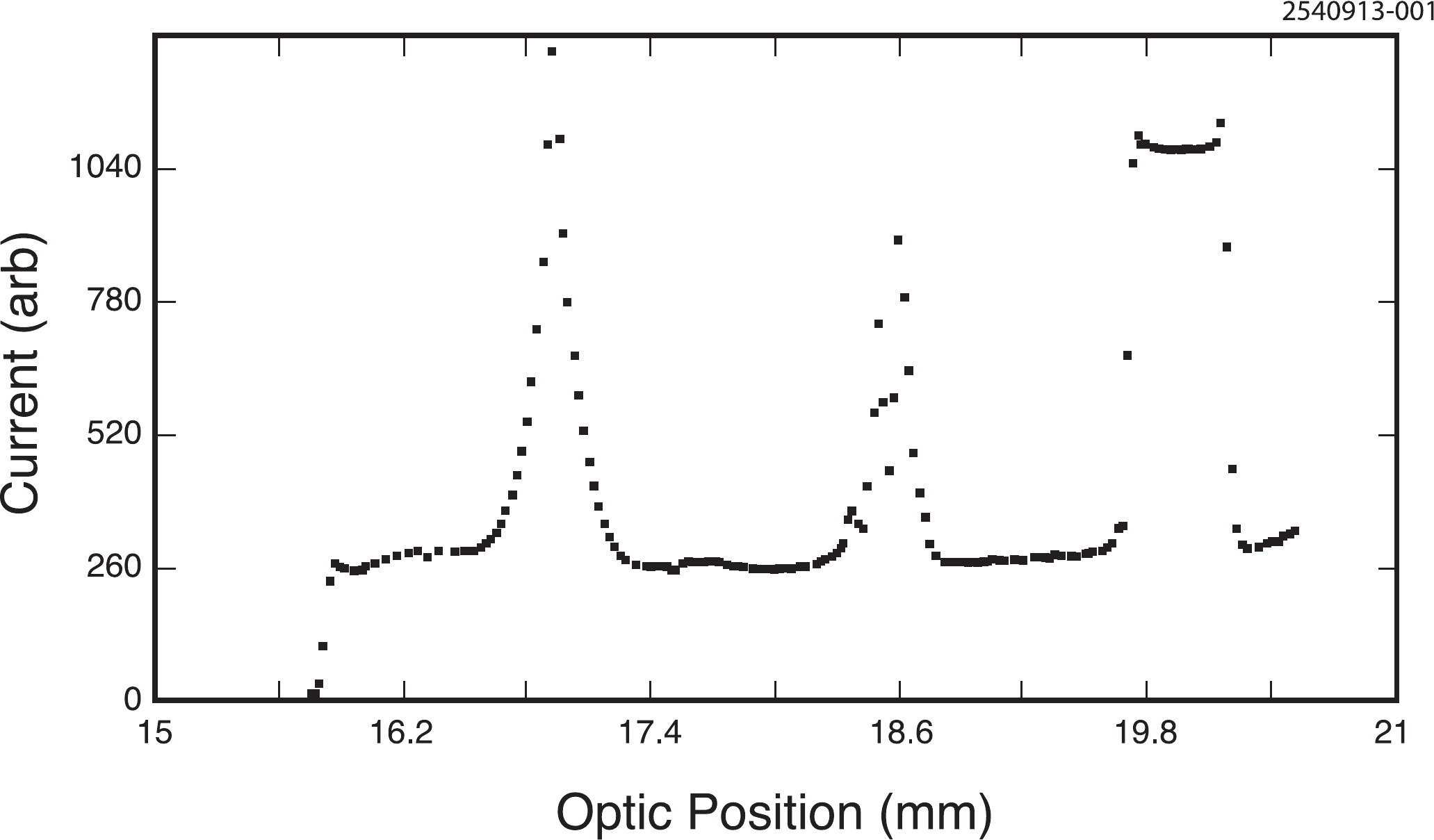}
   \caption{Single-diode current draw (arbitrary units) vs.~vertical position 
   of the optical element stage. These data were taken just after
   the scans of \Figs{fig:align} and \ref{fig:aligntwo}.
   Over this range, three elements of a gold-masked silicon optics
   chip are illuminated: a Fresnel zone plate is
   centered at $\sim$16.9\mm, a coded aperture at
   $\sim$18.5\mm, and a 0.5~mm-wide single slit at $\sim$20.0\mm.
   The non-zero intensity between the elements is a measure of
   transmission through the gold masking.}
\label{fig:alignthree}
\end{figure}

To align the etched-gold optical elements, the pinhole stage
is retracted and the optics chip moved into position. 
\Figure{fig:alignthree} shows
the result of a single-diode scan of the optics chip through the beam.  
The location of the center of
the coded aperture, in this case 18.5\mm, is used to locate the coded
aperture in subsequent data runs. As described in \Sec{sec:app}, these
optical elements have a horizontal aperture of 1.1\mm. As this aperture is
not aligned to the pinhole horizontal aperture, a separate horizontal
detector alignment is required. The horizontal position of the detector 
is adjusted by conducting a single diode scan while varying the horizontal 
location of the detector box. The detector box horizontal position is 
adjusted, rather than the horizontal position of the detector within the box, 
to preserve the alignment of the filter elements (described below) with 
respect to the detector.

\begin{figure}[tb]
   \includegraphics*[width=\figwid]{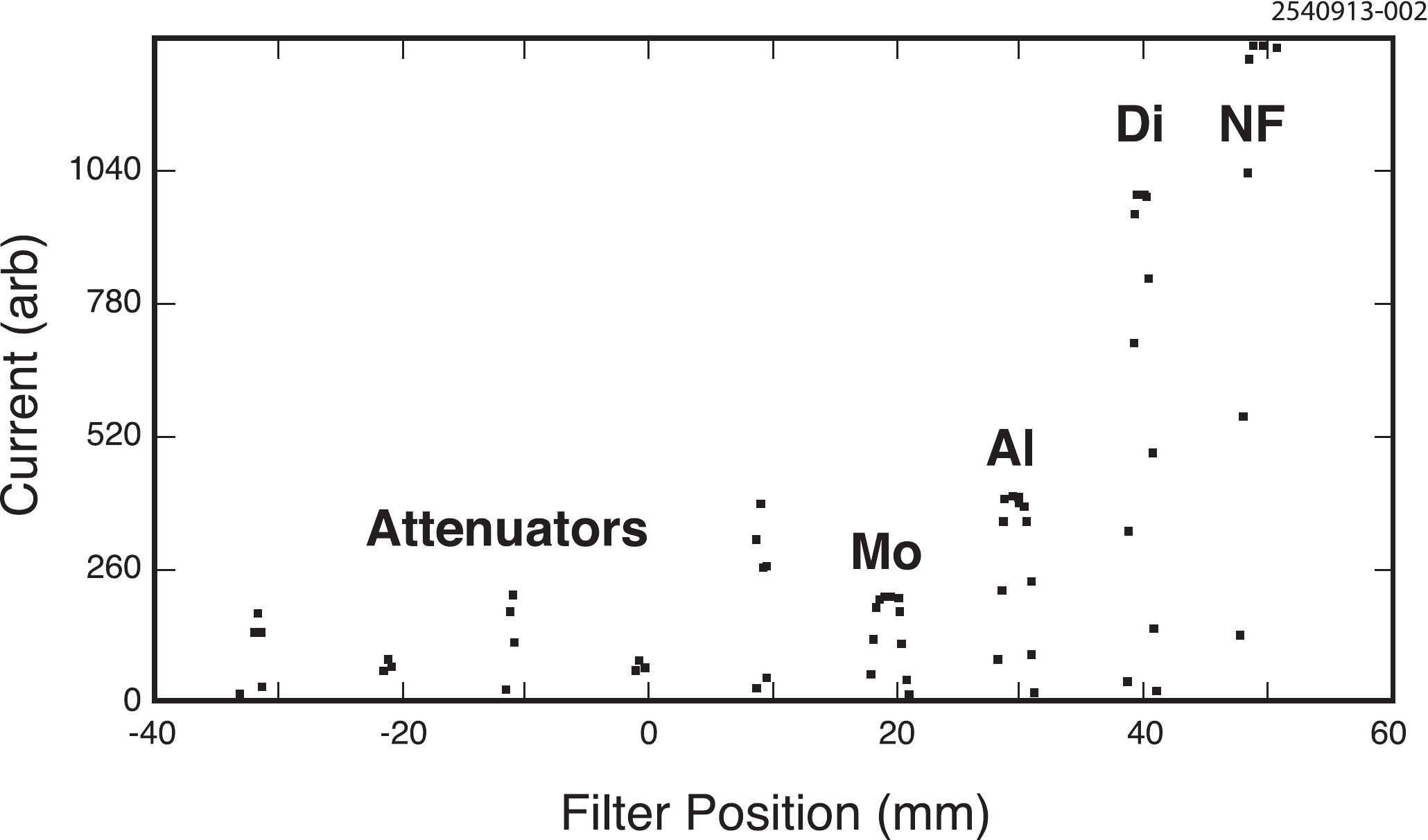}
   \caption{Single-diode
   current draw (arbitrary units) 
   vs.~vertical position of the filter stage, just upstream of the detector.
   These data were taken just after the scans of \Figs{fig:align}, 
   \ref{fig:aligntwo}, and \ref{fig:alignthree}. The locations of 
   several filters and attenuators are indicated.}
\label{fig:alignfour}
\end{figure}

Filter elements, used in the x-ray spectrum analysis described in
\Sec{sec:specresponse}, are located on a motor-drive stage in front of the
detector, as shown in \Fig{fig:detmounted}. 
A single-diode scan of the filter element stage is performed to 
locate the operating positions of the filters. \Figure{fig:alignfour}
shows the stage location for various filters and horizontally limiting
apertures. 

Final alignment is accomplished using the digital detector readout of the
full 32 channels by adjusting the vertical detector location to center the
diode array on the x-ray beam. To center the detector on the x-ray beam
envelope and to center the image on the detector, minor adjustments, on
the order of 100\mi, can be made to the detector and optic vertical
positions and/or the beam orbit. 

\subsection{Beam-induced coded aperture damage}
\label{sec:damage}

\begin{figure}[t]
   \includegraphics*[width=\figwid]{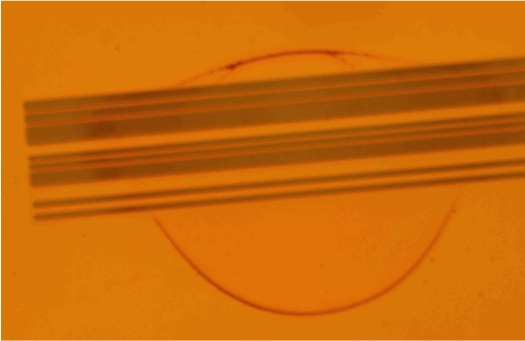}
   \caption{Photograph of a CA1 optical element after accompanying data
showed an alteration in image shape. Excessive x-ray power on the optic
has damaged the gold masking.}
\label{fig:damage}
\end{figure}

The sub-micron gold masking on coded apertures can be damaged by
exposure to power above a certain level. Care must
be taken to limit flux by use of attenuators (horizontally limiting
apertures) and adequate heat sinks for the gold masking. 
In one instance where adequate precautions
were not taken, such damage did occur to one CA1 element.
It was noticed that image patterns changed, and later
the optic was removed from service. A photograph appears in 
\Fig{fig:damage}, and shows evidence of alteration of the masking
by overexposure to the beam. Image data taken before and after the damage
occurred indicate that a significant fraction of gold masking was removed 
from the chip in the central, etched region of the pattern; 
the darkened arcs in the photograph,
taken with reflected light, may or may not correspond to such regions.
The damage may be due to raising
the temperature of the optics chip above the silicon-gold eutectic
point, 363$^\circ$C. It is estimated that the beam exposure causing
this damage was $\sim$22~mW. 
The damaged CA1 element had thinner gold masking (0.53\mi) than any
other optic chips we have used; thicker gold improves heat
transfer to the mounting (\Fig{fig:opticmounts}),
enabling higher operating power without damage.

\subsection{Calibrations}
\label{sec:cal}

After the instrument has been properly aligned and the detector and optics 
are deemed to be in a healthy state, the data acquisition system is
brought online and configured. The xBSM requires tight 
synchronization with the storage ring master timing controls in order to 
take bunch-by-bunch and turn-by-turn measurements. Timing controls within 
the data acquisition box are used to separately align each of the 32 channels 
to a reference bunch (train~1, bunch~1) in the storage ring. Typically, 
the time domain synchronization can be accomplished with an accuracy
of $\sim$50~ps. Once this synchronization is established, all timing control
settings are saved to disk, making the instrument ready to capture detector 
images of individual bunches in the storage ring.

Digitized readout of each pixel's pulse height on a particular
bunch and turn requires four different types of electronic
calibration corrections: pedestal, gain, preamplifier
range, and
bunch-to-bunch crosstalk. Under normal conditions,
all channels are set in common to one of the 25 ranges for each run.
Each range provides
$\pm512$ counts of readout; the ranges cover a dynamic range of 
$\sim$15, in logarithmic steps (nominal 1~dB per step). 
Pedestals, typically a few counts, apply to each
channel in each range, although they tend not to vary with range.
Gains provide a multiplicative scaling for each channel,
which again can vary with different range index settings but tend not to.
Finally, multi-bunch running requires correction for
the crosstalk between a signal in a particular channel in a particular bunch
and that same channel in subsequent bunches.
This is due to time structure of the amplified signal,
which includes an over-damped return to the baseline that needs
$\sim$10~ns in order to settle.
Bunch-to-bunch crosstalk is
largest for 4\ns\ bunch spacing, the smallest spacing
available at CESR-TA.

\begin{figure}[b]
   \includegraphics*[width=\figwid]{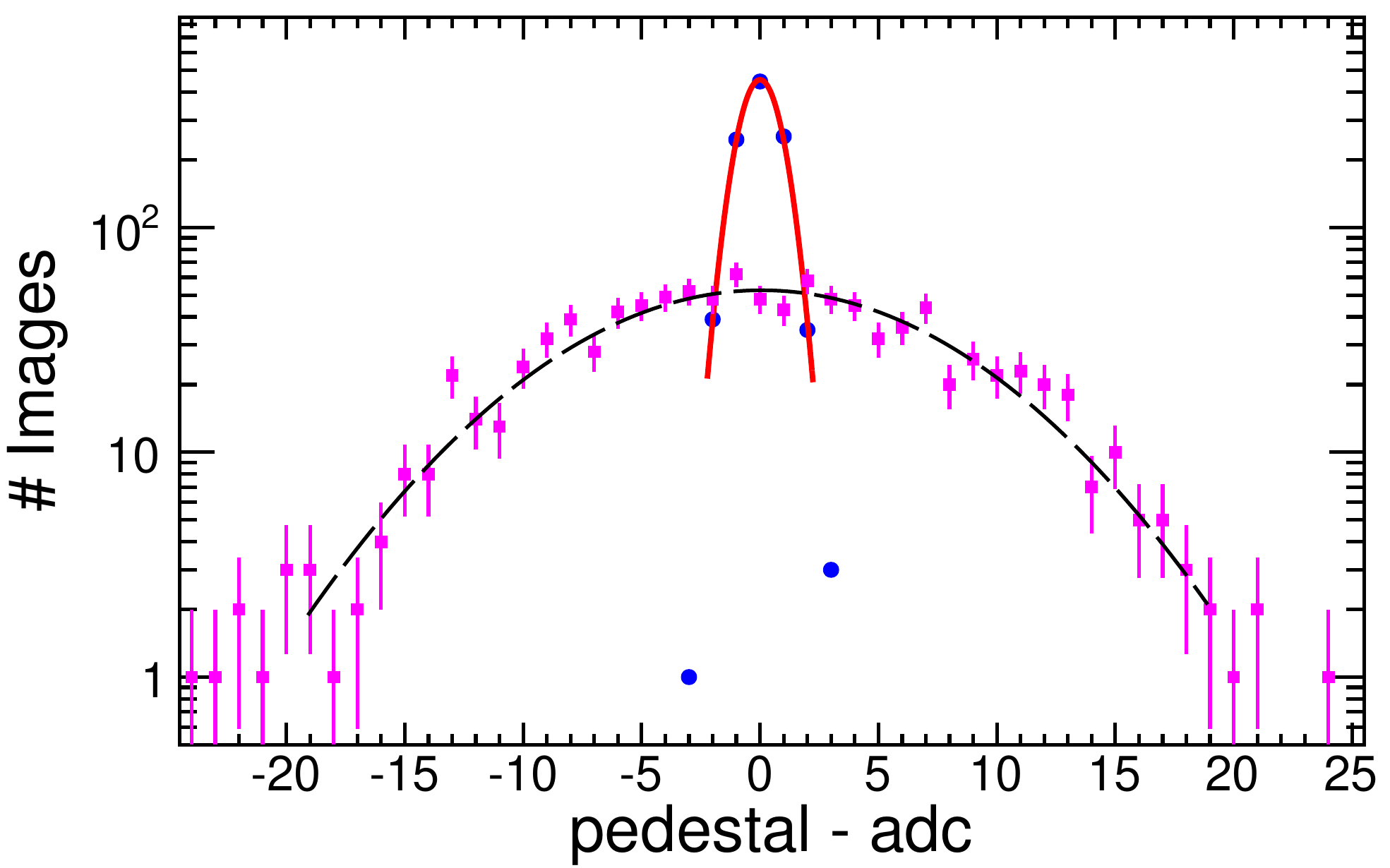}
   \caption{Pedestal distributions for channel \#1 on range index \#1 (circles
     and solid line Gaussian fit) and range index \#25 
     (squares and dashed line) 
     in pedestal runs.}
\label{fig:ped}
\end{figure}

\begin{figure}[tb]
   \includegraphics*[width=\figwid]{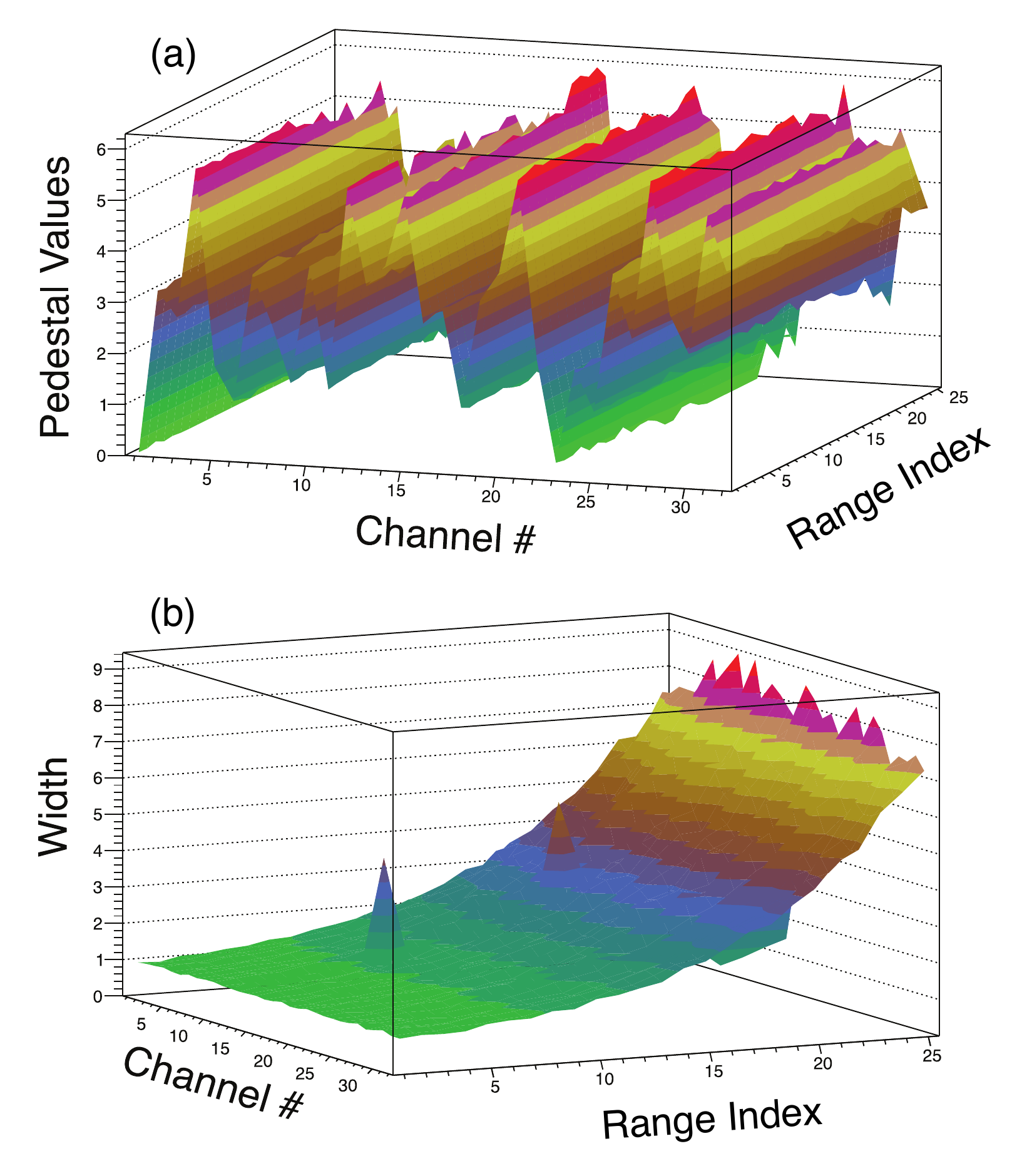}
   \caption{(a) Mean pedestal values and (b) their Gaussian widths, each as a function of channel number and range.}
\label{fig:pedrng}
\end{figure}

\begin{figure}[tb]
   \includegraphics*[width=\figwid]{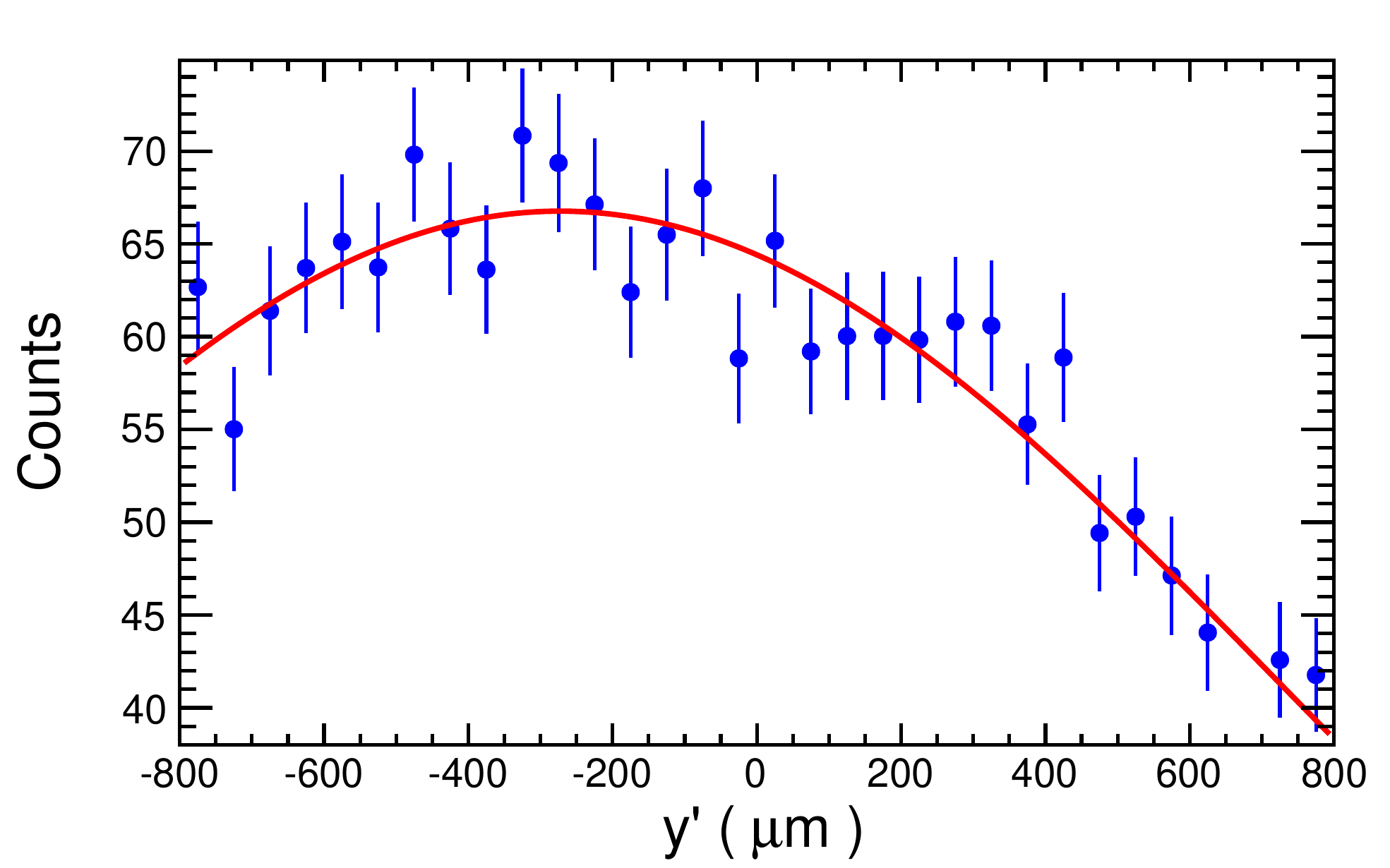}
   \caption{Typical detector image from a gain calibration
     run for a wide-open 
     optical element and preamplifier range index \#1.
     The solid line is a Gaussian fit to the data, which have
     already undergone a few iterations of calibration.}
\label{fig:gain}
\end{figure}

\begin{figure}[tb]
   \includegraphics*[width=\figwid]{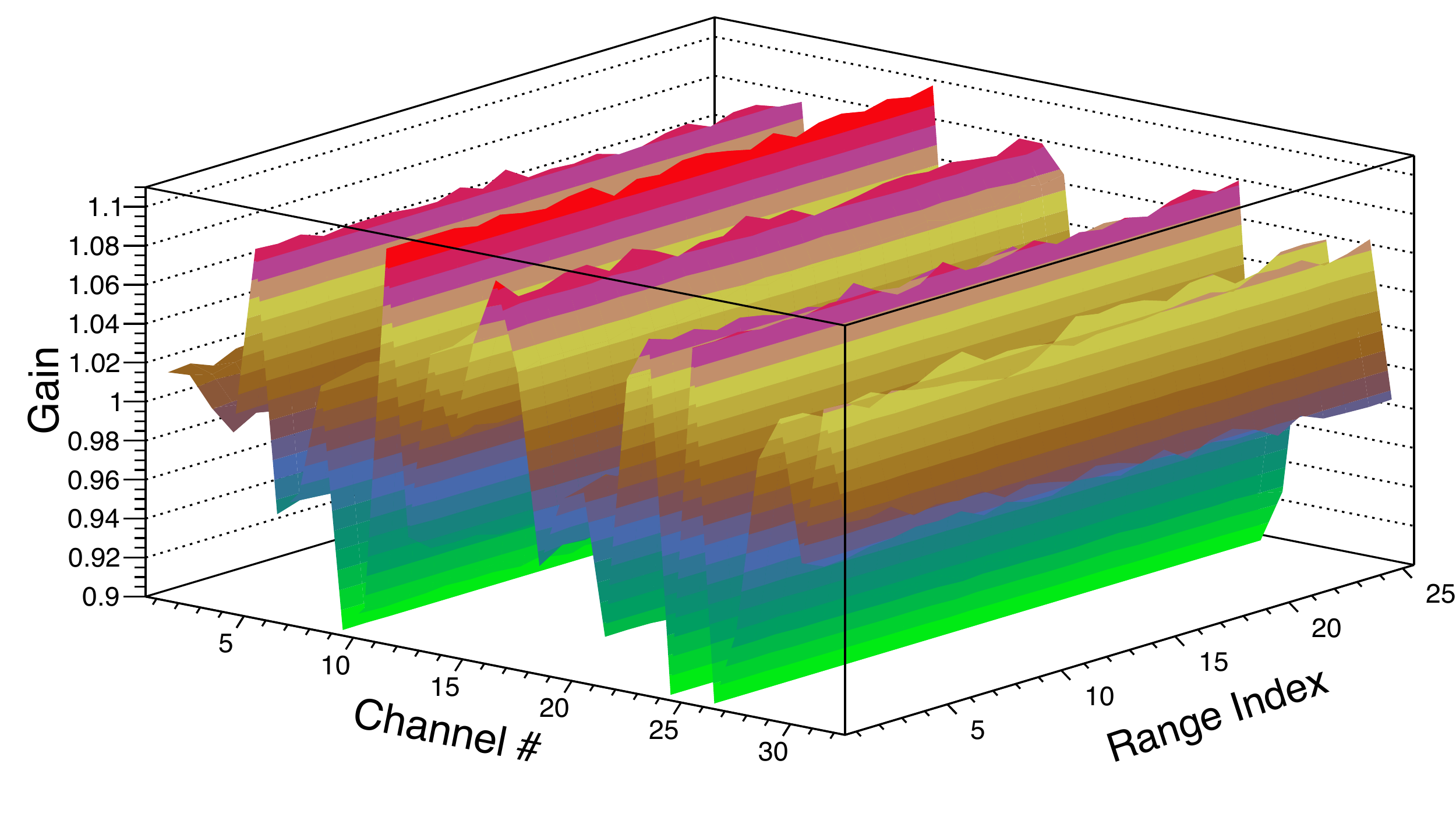}
   \caption{Gain values for all channels and ranges.}
\label{fig:gainrng}
\end{figure}

Mean pedestal values and associated Gaussian widths 
(which are used in image error bars) 
are obtained by reading out the detector with
circulating beam, but with a closed beamstop, preventing 
x-rays from reaching the detector. A typical distribution 
of pedestals observed in detector images
and Gaussian fits to them are shown in \Fig{fig:ped}. 
\Figures{fig:pedrng}(a) and \ref{fig:pedrng}(b) show
pedestal and width values, respectively, 
vs.~channel and preamplifier range index.  Gain and range
calibrations both require single-bunch data with a wide open
optical element, which exposes all 32 channels to substantial
radiation that varies smoothly across the detector
(see \Eq{eqn:sth}). For each range, gains are 
computed from a fit of the images to a Gaussian;
a typical detector image and Gaussian
fit is shown in \Fig{fig:gain} and gain values vs.~channel
and range index in \Fig{fig:gainrng}. The patterns and approximate
reflection symmetry about the middle of the detector in both
pedestal and gain values vs.~channel number reflect the
symmetry and characteristics of the physical
layout on the readout board. Variation with preamplifier range
is small relative to channel-to-channel gain variation.
During the gain calibration, images are reviewed for basic structure with an 
eye towards identifying individual channels which are behaving badly. 
Such bad channels have typically taken the form of a constant
or stuck value, a consistently low or high value or intermittently 
inaccurate values. Bad channels are identified and diagnosed at the 
hardware level until all channels are performing reasonably.
If electronics components are replaced, channel calibrations 
are performed again, starting with pedestals.

\begin{figure}[tb]
   \includegraphics*[width=\figwid]{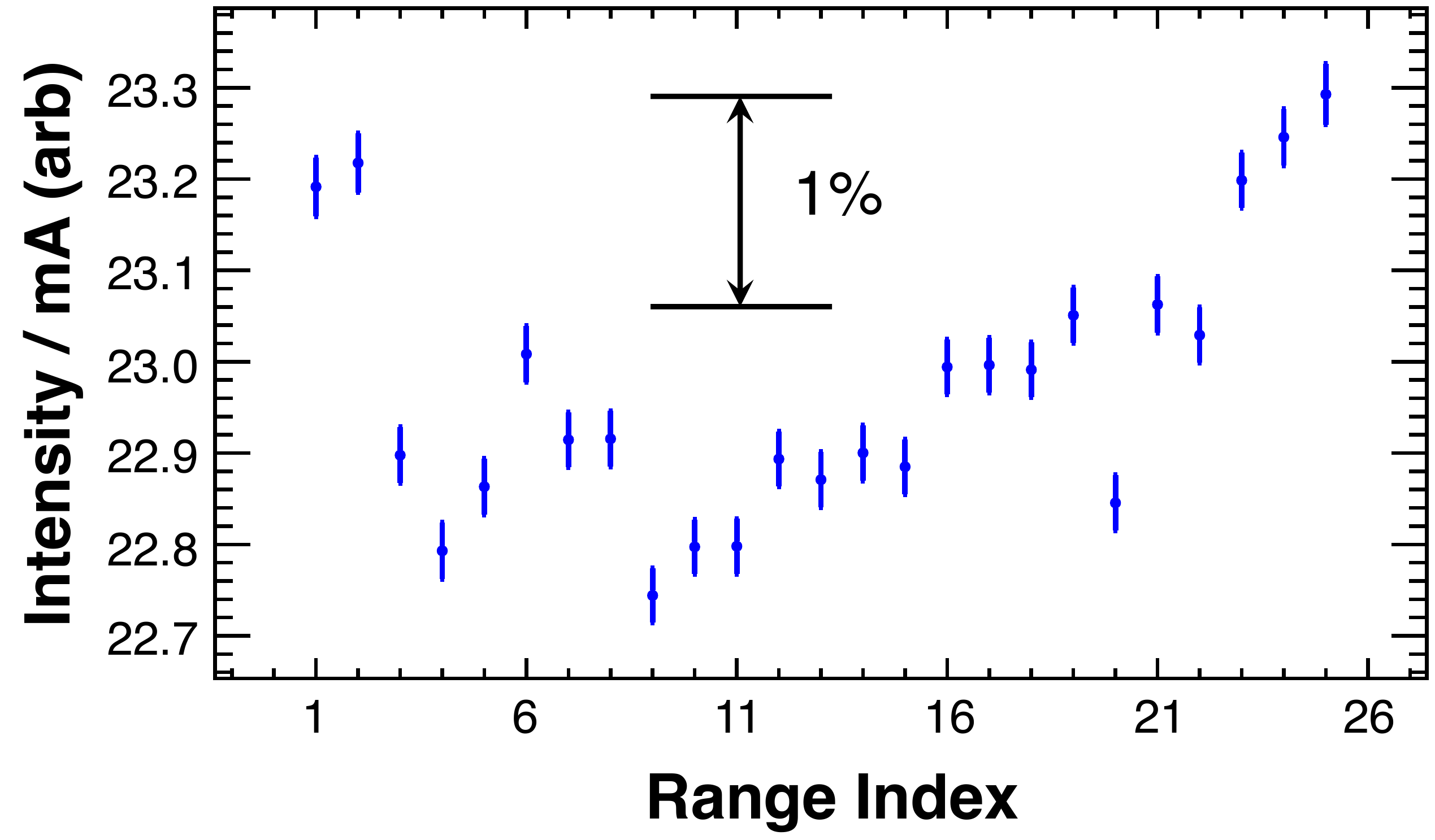}
   \caption{Peak intensity per mA of beam current vs.~preamplifier 
     range index for the D-line xBSM installation. Ideally all data
     points should be consistent with the same fixed value; their
     spread shows the $\sim$1\% variation across the full dynamic range.}
\label{fig:range}
\end{figure}

\begin{figure}[tb]
   \includegraphics*[width=\figwid]{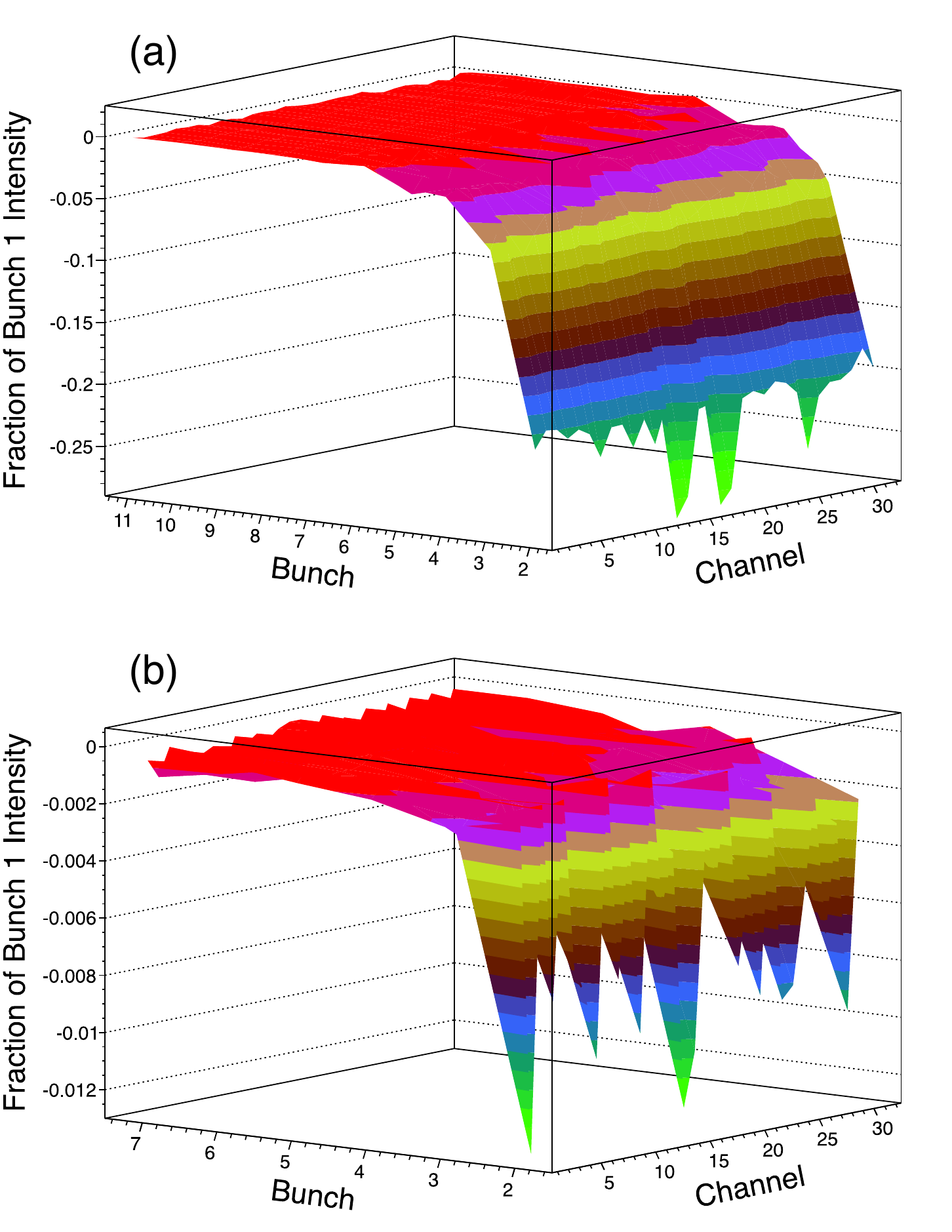}
   \caption{For a multi-bunch fill with bunch separations of
    (a) 4~ns, or (b) 14~ns
             and
            current only in bunch \#1,
            the pulse height in each detector channel 
            measured for succeeding bunches as a fraction
            of the corresponding value in bunch \#1, 
            plotted vs.~channel and bunch number.}
\label{fig:xtalkfour}
\end{figure}

Range calibration is completed using
WO-optic data in all ranges when the single bunch
current is large enough to give significant
pulse height in all ranges but not so large
as to cause saturation on the highest gain range;
all ranges are forced by the range correction to yield
the same number of corrected counts.
For amplifiers on both beamline installations
the final range calibration finds 
that each step in index indicates a gain change of 11.9\%
instead of the nominal 12.2\%, with a maximum 
deviation of any individual range from that average
step size of $\sim$1.3\%, as shown in \Fig{fig:range}.

The bunch-to-bunch crosstalk correction also requires
a special type of data collection, one in which
CESR is configured to run and trigger data collection 
with multiple bunches but
only the first is filled with electrons or
positrons. This dataset is accumulated with the wide-open optic. Each channel's
correction for a particular follow-up bunch 
is the fraction of the first bunch's pulse height
that is recorded there, after correcting for
pedestal, gain, and range. Bunch-to-bunch crosstalk 
corrections are then applied to true multi-bunch data
by adjusting pulse height in a particular channel
in a particular bunch to compensate for crosstalk from the
preceding bunches' pulse heights. As shown in 
\Fig{fig:xtalkfour},
the correction can be in the 20-30\% range for
the next bunch in 4~ns bunch-separation data,
and significant residual signals persist for several more
bunches, whereas for 14~ns bunch-separation data
the correction even for the first succeeding bunch
is $\sim$1\%.

After the timing values have been tuned and calibrations performed, 
 the instrument is deemed ready for experimenters.

\subsection{Setting the pinhole opening}

The \prftt\ for any optical element depends on the gap size,
and the optimal gap size for a pinhole depends upon the x-ray spectrum.
As the x-ray spectrum varies with the beam energy, the pinhole opening
must be reset for each change of \Eb, or upon initial setup each
run period. The motors controlled actuator determining the pinhole opening,
described in \Sec{sec:optics}, has digital
setpoints that are nearly linear in the gap size. However, the
zero-opening setpoint can vary between run periods, so confirmation of the
actual opening is performed using images from the data at different
actuator settings. The desired setting is that which minimizes the
fwhm of the measured image. While the optimal gap size is known from
modeling studies of resolving power to be about ten microns
wider than where the minimum width occurs, the minimum image width
setting is more robust in fitted beam size 
with respect to small changes in pinhole gap size,
and hence a reliable calibration point. 
Further information on pinhole opening studies appear in \Sec{sec:models}.

\section[Results]{Results}
\label{sec:results}

\subsection{Detector spectral response}
\label{sec:specresponse}

\begin{figure}[t]
   \includegraphics*[width=\figwid]{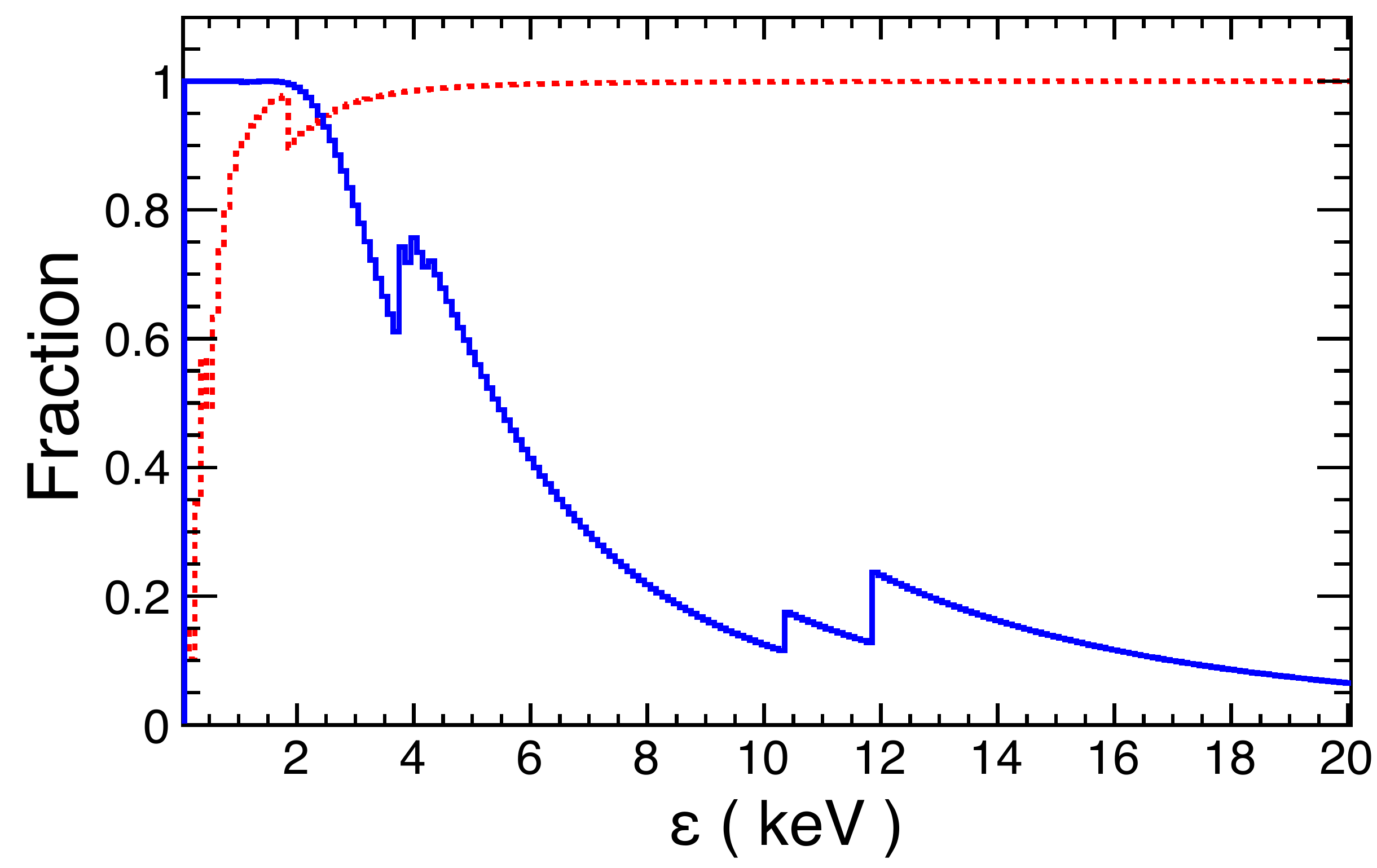}
   \caption{Predicted x-ray absorption of the InGaAs detector 
     (solid histogram) 
     and transmission for the Si$_3$N$_4$ passivation layer 
     (dashed histogram) as a function of energy.}
\label{fig:dettran}
\end{figure}

Precision fitting of images requires an
accurate \prftt. To do so, in addition to knowledge of the
geometry, optical element specification, beam energy, and filtering, 
one needs to know the effective
spectral response of the detector, at least to the extent that it
is not uniform in wavelength. This includes the energy-dependent
attenuation
of any coatings in front of the active material as well as
the absorption of the diode itself, including the degree to
which absorption of x-rays translates into pulse height.
As a nominal starting place we can account for the 0.16\mi\
of Si$_3$N$_4$ specified as the passivation layer and
assume 100\% efficiency of transmitting energy absorbed
in the InGaAs into detected photoelectrons, as shown in \Fig{fig:dettran}.
Note the two general features that inactive material will
inhibit the lowest energy x-rays from reaching the active part,
and that the ability of the active portion to absorb incident
radiation decreases steadily above 2\kev (aside from modest
dips and spikes from photoelectric absorption edges).

\begin{figure}[t]
   \includegraphics*[width=\figwid]{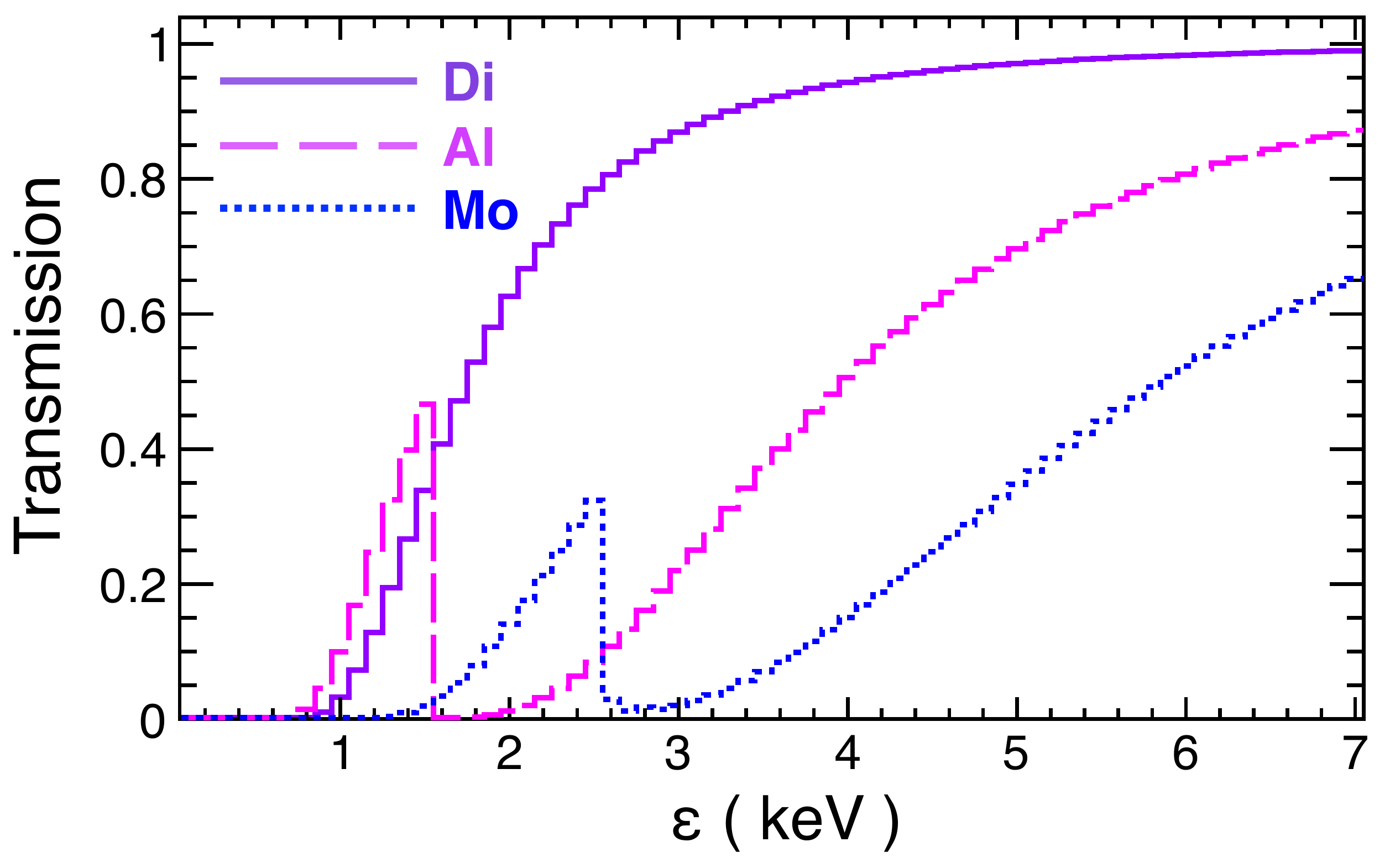}
   \caption{Filter transmission curves as a function of x-ray energy 
     for the diamond (solid), 
     aluminum (dashed), and molybdenum (dotted)
     filters, using thicknesses specified in \Tab{tab:geom}.}
\label{fig:filtran}
\end{figure}

\begin{figure}[t]
   \includegraphics*[width=\figwid]{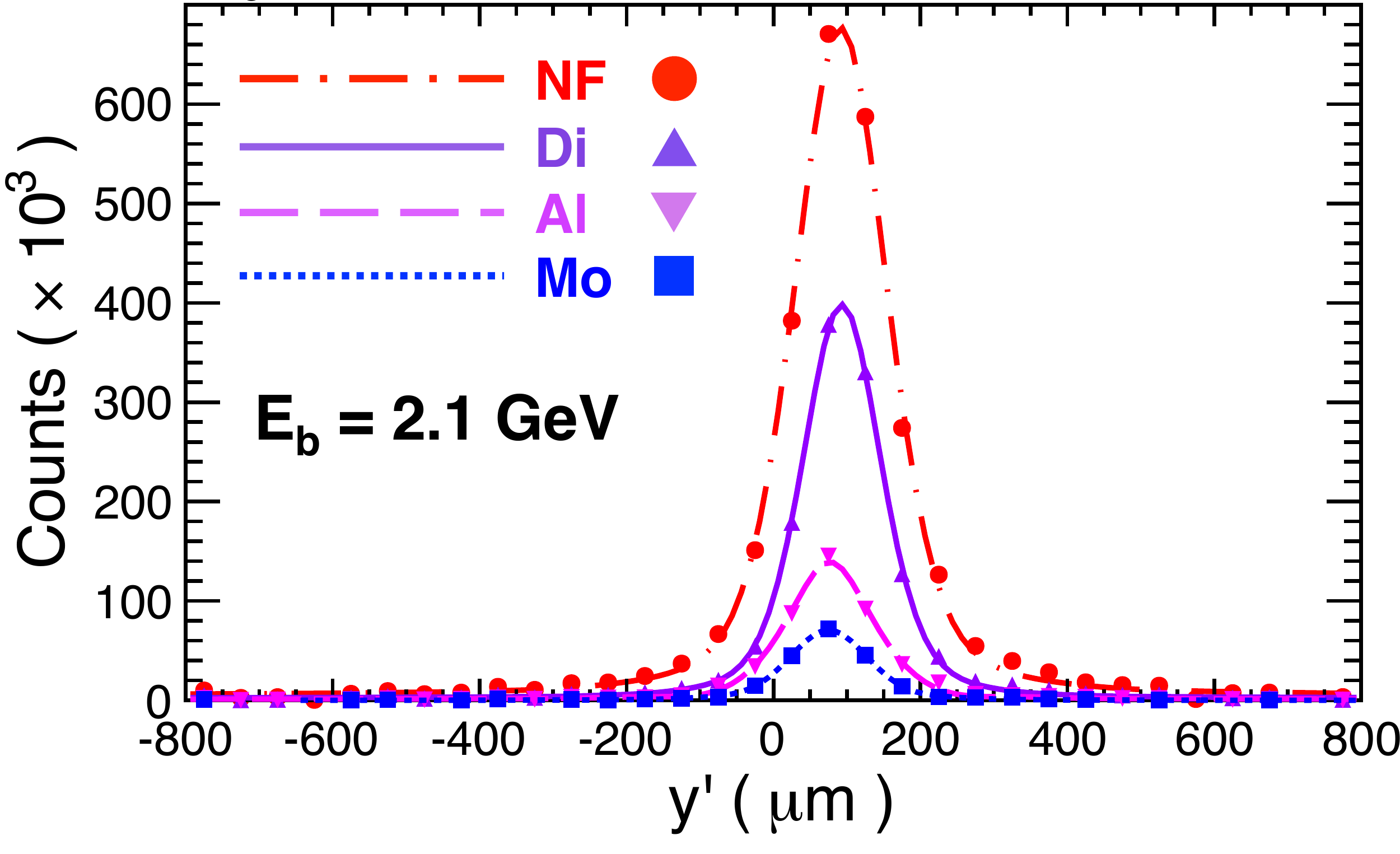}
   \caption{Detector images taken with the pinhole optic and 
            corresponding fits at $E_b=2.1$\gev\ with different filters in
            place, as indicated. 
            The images shown represent sums over 1024 turns. }
\label{fig:FilterScanMd}
\end{figure}

We determine the effective spectral response empirically because the
idealized assumptions do not account for the conversion of absorbed
energy into detected charge, and because specified layer thicknesses may
be inaccurate.
A laborious but effective technique to do so has been developed.
It relies upon the fact that our three different filters probe
different portions of the spectrum.
We compare the image areas per unit beam current in pinhole 
images with filters upstream of the detector to those without any filter
present. Gathering such data at four 
different beam energies enables a detailed description over the
range of important wavelengths. The method has sensitivity for two reasons:
first, because the underlying spectrum shifts upward for increasing beam 
energies (as shown in \Fig{fig:SR}), thus highlighting different
parts of the x-ray spectrum; and second, because each of the filters 
highlights a different energy range (as shown in \Fig{fig:filtran}).
We use ratios instead of absolute intensities to
avoid systematic issues associated with retuning the x-ray
beam trajectory at different beam energies.

\begin{figure}[tb]
   \includegraphics*[width=\figwid]{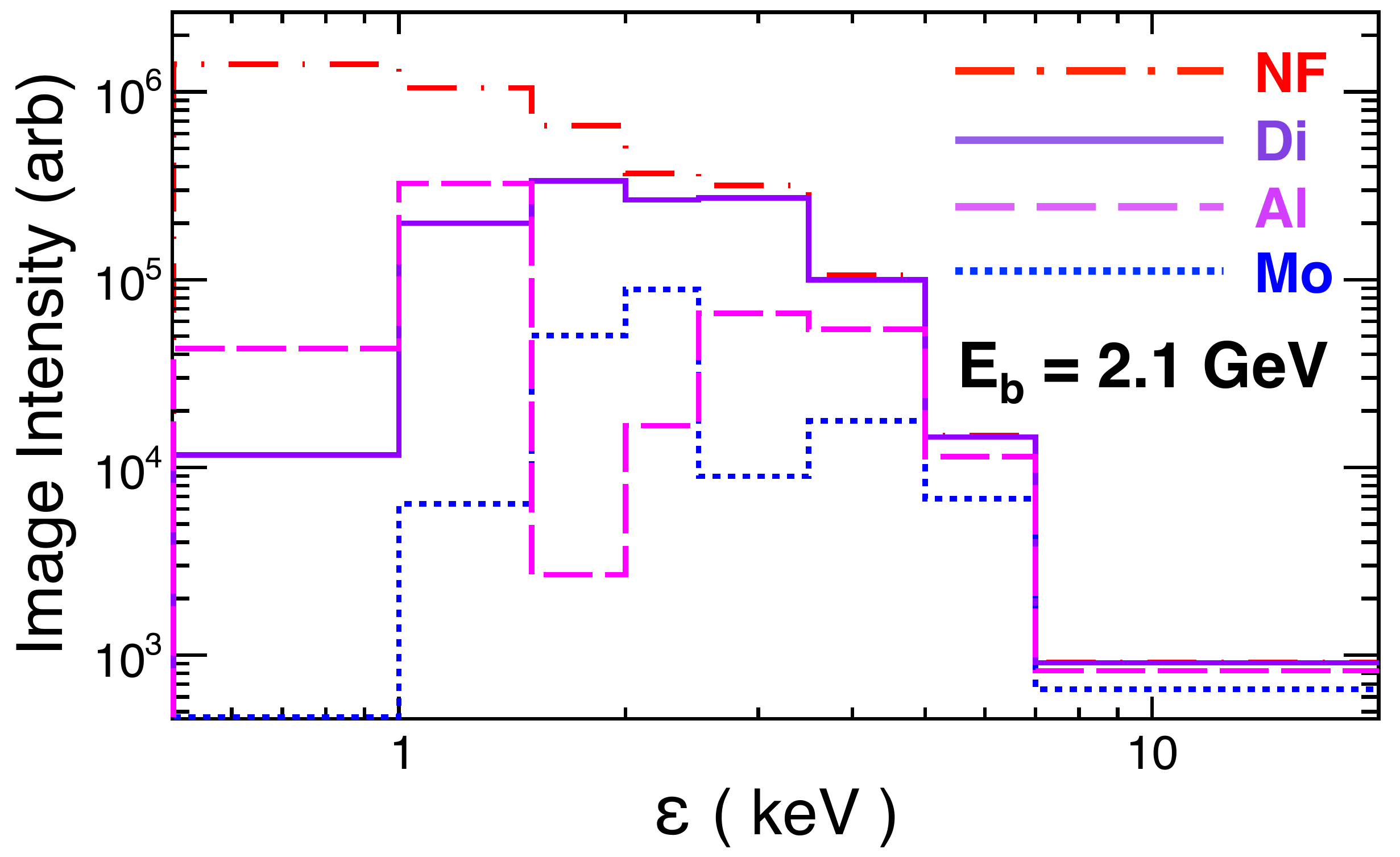}
   \caption{Contributions to \prftt\ areas from different x-ray 
            energy bins
            for $E_b$=2.1\gev\ 
            and different filters, as indicated.}
\label{fig:areaMd}
\end{figure}

\begin{figure}[tb]
   \includegraphics*[width=\figwid]{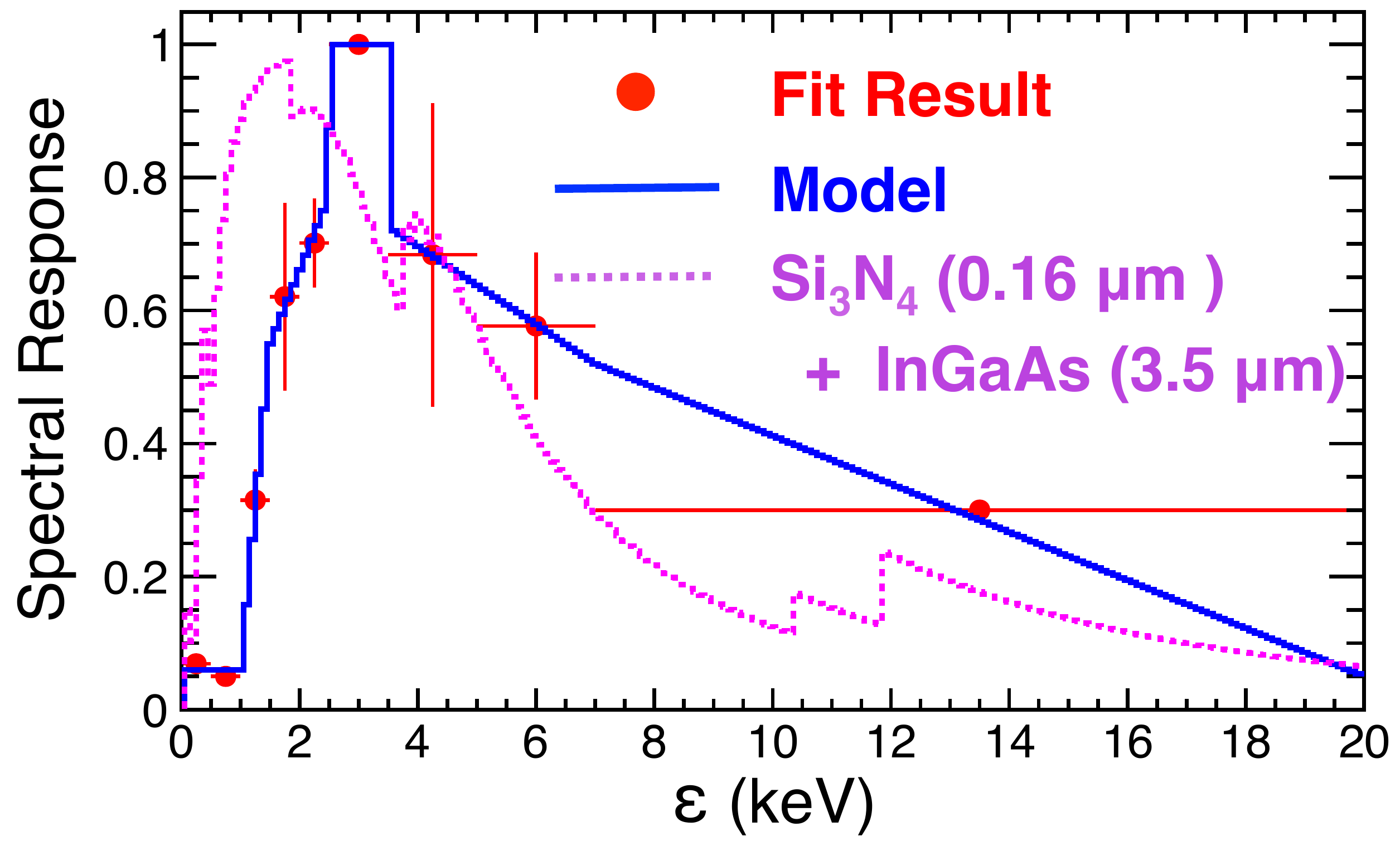}
   \caption{Fit results for the relative detector response as a function of
   x-ray energy (solid circles with error bars), a piecewise-linear model
   used for constructing \prftt s (solid line), and the curve predicted 
   (dotted line) 
   using the nominal materials and thicknesses in the detector.} 
\label{fig:detresponse}
\end{figure}

  Data accumulation and analysis proceed as follows:
\begin{itemize}
\item At each of the four beam energies, with the diamond filter in place, 
      set the pinhole opening to the value which
      minimizes the fwhm of the image. Maintain this gap for all four
      different filter runs at each beam energy.
\item Accumulate at least 1024 turns of single-bunch images
      with each of four filter configurations: no filter
      (\NF), diamond filter (\Di), molybdenum filter 
      (\Mo), and aluminum filter (\Al). The beam current 
      is recorded with each dataset for use in the next step.
\item Using a given spectral response\footnote{Initially, assume the
      nominal materials and a flat response; later, use
      the result of the previous iteration. The analysis converges after a
      single iteration.}, build \prftt s for each beam energy
      and filter. Then fit the measured images from each dataset to the 
      appropriate \prftt\ 
      for beam size, beam position, intensity, and a floating
      flat background term to account for uniform pedestal shifts.
      Tabulate the average intensity per turn per unit current for each
      beam energy and filter.
\item Form three ratios of intensities/turn/mA for each beam energy: 
      \Di/\NF , \Mo/\NF , and \Al/\NF. 
      Assign systematic uncertainties from variations
      observed from data taken at different times with different
      currents, which amount to 1-3\% for each numerator and denominator.
\item Perform a $\chi^2$-minimization 
      fit of the twelve intensity ratios to 11 or fewer
      floating parameters, each of which determines the relative strength
      of a different photon energy bin within the range 0-20\kev.
      In the fit, each strength parameter multiplies a corresponding
      energy bin of the spectrum obtained from \Eq{eqn:power}.
      At least one photon energy bin's strength must remain fixed
      in order to set a scale for the others; this procedure only determines
      a shape of the spectral response. We choose the nine energy bins 
      shown by the points with error bars in \Fig{fig:detresponse}, 
      holding the 2.5-3.5\kev\
      bin's strength fixed at a value of 1.0. (X-ray energies above 
      7\kev\ only contribute modestly at $E_b=2.6$\gev, and are not 
      significant at lower $E_b$.)
\end{itemize}

Complete datasets at four beam energies were accumulated
on CESR's D-line ($e^+$) xBSM station. Results are given
for the second iteration of analysis. Representative
images and fits are shown in \Fig{fig:FilterScanMd}.
Contributions to the areas from different x-ray energy bins are shown in
\Fig{fig:areaMd}.

The fit is not guaranteed to give physical results, and could potentially
chase down an artifact where $\chi^2$ is slightly smaller than
the ``right'' answer. These issues are dealt with by constraining
the energy bin strengths to all be positive, and to require that
these strengths fall as x-ray energy increases above 4~\kev, as must
occur due to the absorption capability of the InGaAs detector.
The initial unconstrained fit gives a negative result for the
0.5-1\kev\ bin and a value greater than 1 for the 7-20\kev\ bin;
both are unphysical and have large uncertainties. Hence we fix
the value of the 0.5-1\kev\ bin to be 0.05, a small but reasonable
value consistent within $1\sigma$ of the initial result.
We fix the value of the 7-20\kev\ bin to be 0.3 so as to be consistent
with a physical response that falls with increasing energy;
it will be shown later in \Sec{sec:specvar} that even large variations
in this value have virtually no effect on beam size.

\begin{figure}[tb]
   \includegraphics*[width=\figwid]{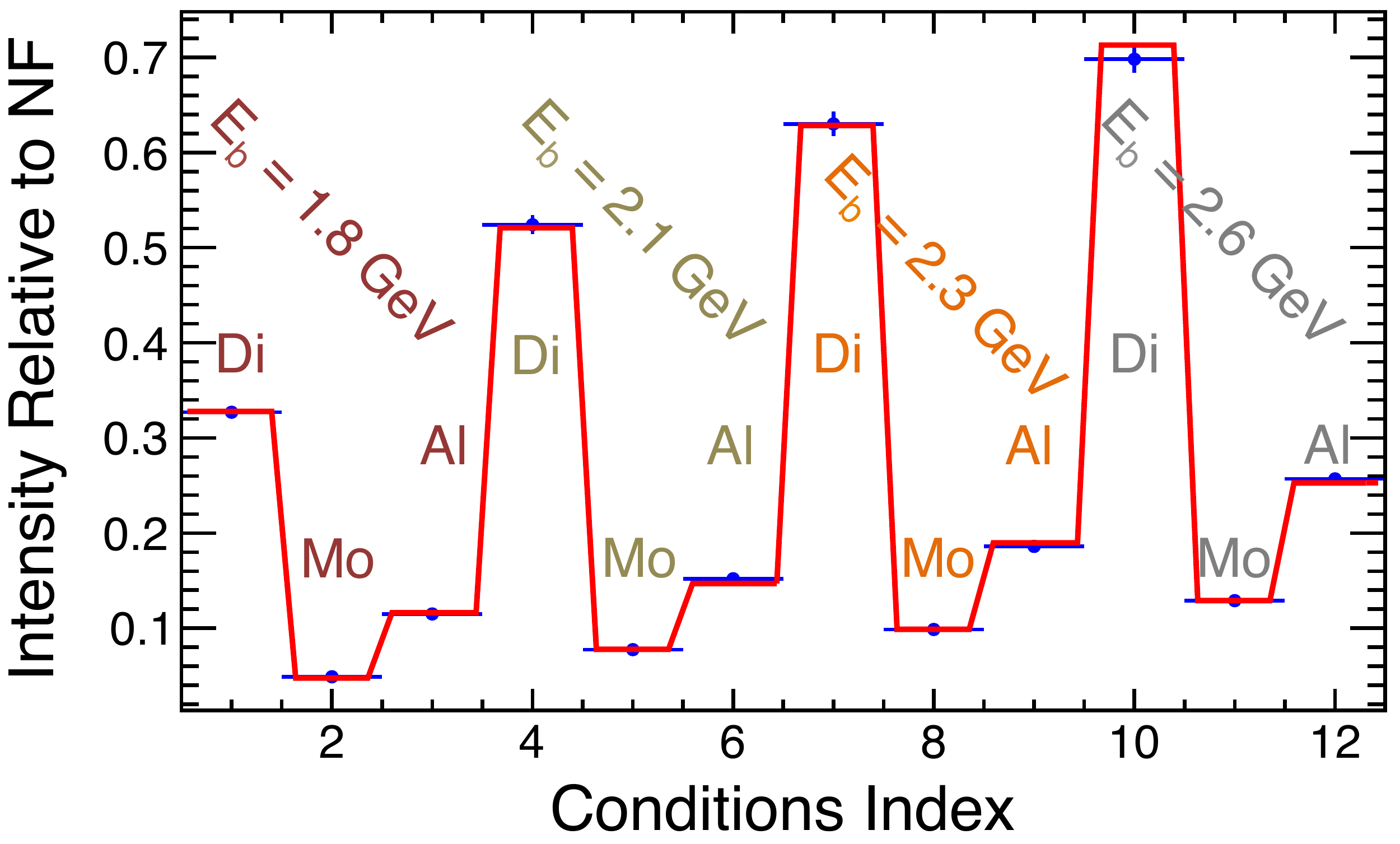}
   \caption{Filter scan fit of intensity relative to the \NF\ data 
   vs.~the conditions index, which 
   cycles through beam energy/filter combinations as indicated.
   Solid circles represent the data and the 
   histogram the best fit.}
\label{fig:filterscanfit}
\end{figure}

\begin{figure}[tb]
   \includegraphics*[width=\figwid]{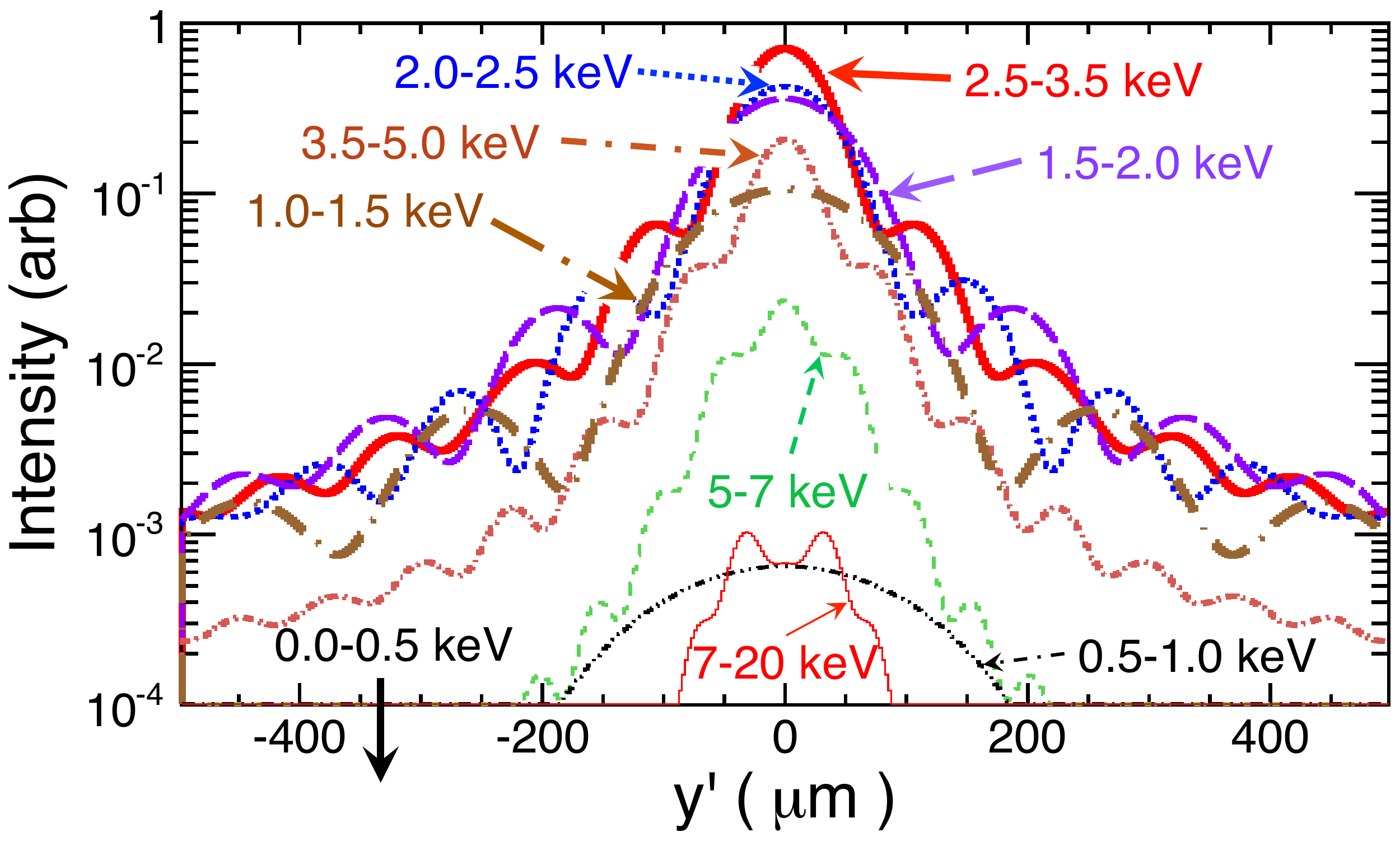}
   \caption{Contributions segregated by x-ray energy bin, as indicated,
            to the pinhole-optic \prftt\ for $E_b$=2.1\gev\ with
            the diamond filter assuming the detector response of the
            piecewise-linear model of \Fig{fig:detresponse}.}
\label{fig:prfbyenr2p1}
\end{figure}

Results of the fit to the data as described 
are shown in \Figs{fig:filterscanfit}
and \ref{fig:detresponse}. The fit has a reasonable
$\chi^2=4.9$ for six degrees of freedom. The histogram in
\Fig{fig:detresponse} represents the piecewise-linear model
of detector response (\ie the spectral response function \srd\ in 
\Eq{eqn:prf}) used in constructing point response functions.
Contributions to a pinhole \prftt\ from different x-ray energy 
bins at $E_b=2.1$\gev, assuming the piecewise-linear detector response
and the diamond filter, are shown in \Fig{fig:prfbyenr2p1}.
It is evident that the shape of this \prftt\ is determined
dominantly from x-ray energies in [1.5-5.0]\kev, and therefore
mostly depends upon just three of the bins other than 
the normalization bin, 2.5-3.5\kev,
two lower and one higher. This remains true at all beam energies from 
1.8-2.6\gev, making the physicality assumptions in the filter scan
fit at very low and very high x-ray energies of little but cosmetic import.

Estimating systematic uncertainties in the \prftt s from the
error bars of the fit requires some care due to substantial
bin-to-bin fluctuations. For example, fixing
the strength of the 2.0-2.5\kev\ bin to a value $1\sigma$ 
(10\%, relative) larger than the final fit 
results in the best-fit value
for 1.5-2.0 rising by 15\% (relative) 
and 3.5-5.0\kev\ by about 26\% (relative).
Such correlations are fortuitous for stability
of beam size with respect to spectrum uncertainty because
the higher energy bin contributes a narrower component
to the \prftt, and the lower energy bins a broader component,
so that overall, in this example, 
the \prftt\ shape is only very slightly altered.
Quantitative estimates of systematic uncertainty on beam size
from spectral response uncertainty is explored in \Sec{sec:specvar}.

\subsection{Beam size extraction}
\label{sec:size}

Each image in a data run is subject to a $\chi^2$-minimization fit 
to a template appropriate for the beamline geometry, optical element, 
beam energy, and filters in use for that run, with the fit parameters
being the amplitude \amp, the beam offset \off, the beam size \siz,
and a flat background term. 
To properly weight each of the 32 pixels in the fit, each point
must be given reasonable uncertainties. We assign the uncertainty
to be the quadrature sum of three terms: one for the least-count effects
on the gain range being used, one for the electronic noise corresponding
to the average pedestal width, and one to roughly account for the
statistical variation in the number of photoelectrons collected
for that pixel. The first is trivial to assign as a single count
and the second is derivable from the pedestal width measurements described
in \Sec{sec:cal}. The third term is obtained from data taken with
a wide-open optic, as shown in \Fig{fig:gain}; deviations from
the Gaussian fit are taken as attributable to the three sources
enumerated here, and the statistical portion is taken as that
which remains after the other two are accounted for.

\begin{figure}[b]
   \includegraphics*[width=\figwid]{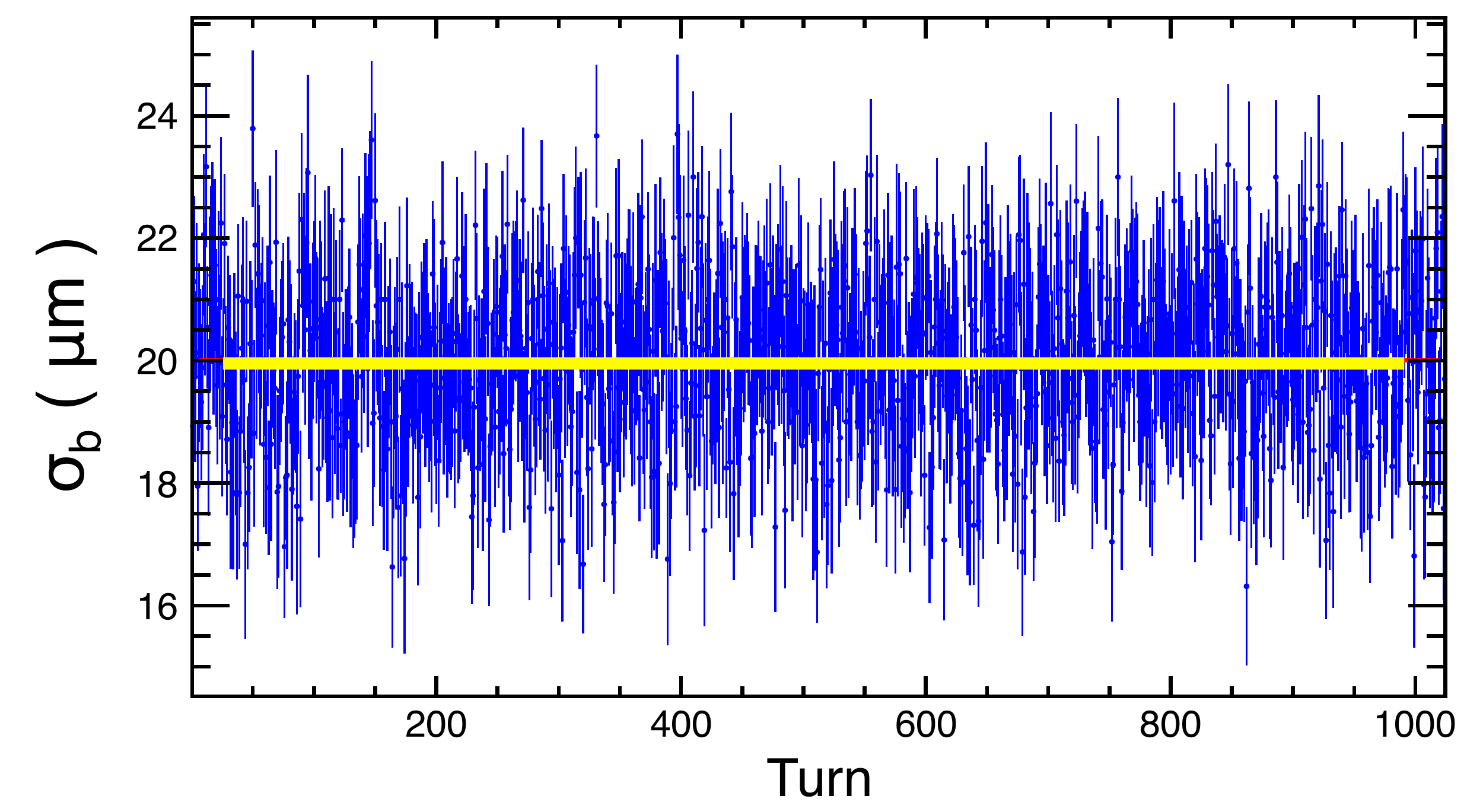}
   \caption{Beam size vs.~turn for a run of 1024 turns. The solid line represents
   \sizta, the weighted average, excluding outliers, of the 1024 measurements.}
\label{fig:bmsizshort}
\end{figure}

We have developed two methods for extracting a beam size that 
should be applicable to the entire run (typically $N$=1024-4096 turns) 
for each bunch. Each method has
its advantages and disadvantages. The first technique calculates
a {\it turn-averaged} beam size \sizta\ by taking a weighted
average of the $N$ individual turn image fits, 
using the statistical uncertainty from
each turn's individual fit (inflated by the factor 
$\sqrt{\chi^2/{\rm dof}}$ if that factor exceeds unity). 
Because failed fits on a small number of turns
can give results which skew a weighted mean, outliers beyond three
standard deviations from the mean
are excluded in two iterations involving recalculation of the standard
deviation after each. This method is quite robust when there is sufficient
pulse height on each individual turn so that the turn fits are
reliable. An example of a successful \sizta\ extraction is shown in 
\Fig{fig:bmsizshort}. Outlier rejection is applied to all fitted parameters
when reporting turn-averaged quantities.

The turn-average method can break down at very low current when the number of 
bad fits, and hence outliers, becomes large. We have also observed
that, at low current and using the pinhole optic, this method gives
a beam size that is biased low by up to several percent. This effect is
not fully understood, but is thought to be attributable to cases
where most pixel signals correspond to fewer than 20 photoelectrons, which will obey
Poisson rather than Gaussian statistics, and which will be more
strongly affected by least-count digitization effects. This effect 
was observed by comparing the beam size for a given moderate current run 
with no attenuator with that obtained just after 
attenuator insertion (which limited the horizontal detector exposure, thereby
reducing flux onto the detector). The effect appears in the attenuated data 
near 0.5~mA/bunch and grows as current decreases.

\begin{figure}[tb]
   \includegraphics*[width=\figwid]{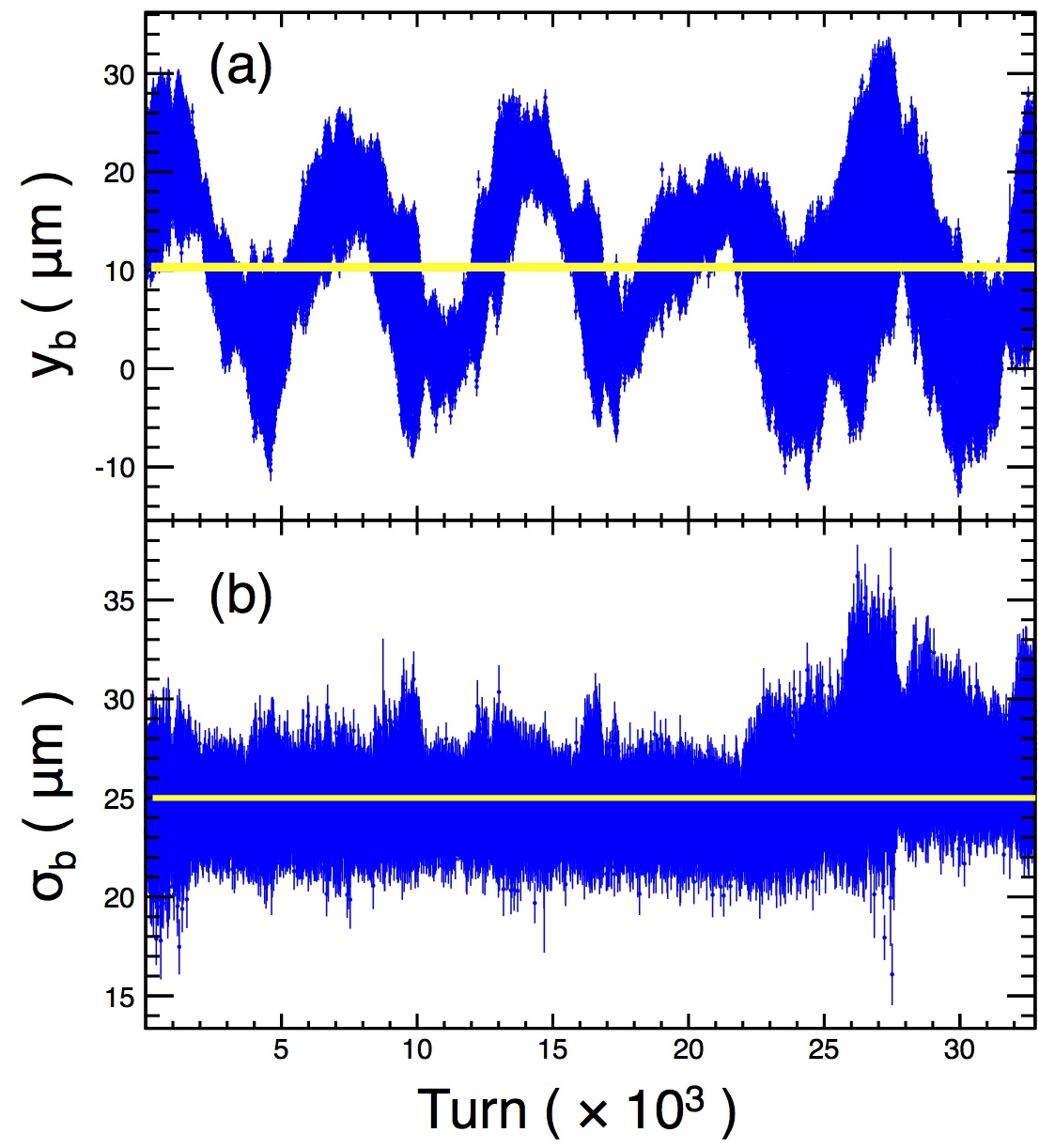}
   \caption{(a) Beam position and (b) beam size vs.~turn for a 
   single bunch run of 32768 turns. Each solid horizontal line represents
   the weighted average over all turns. excluding outliers.
   There is visible variation, beyond that expected from statistics alone,
   in both beam position and beam size as a 
   function of turn.}
\label{fig:sizlongrun}
\end{figure}

  Hence we developed a second method of computing beam size from 
turn-by-turn data;
it yields a quantity we call the {\it corrected image-summed} beam size, 
\sizcts. Here we treat all the turns for each bunch as a collective that has
several relevant properties: 
\begin{itemize}
\item  \sizts, the image-summed beam size obtained by simply adding all
turn images together before fitting to a template;
\item $\sigma_{\off}$, the rms beam motion for the run, obtained from the 
turn-by-turn \off\ values; and 
\item $\delta\off$, 
the median statistical error, on the turn-by-turn values of \off. 
\end{itemize}
In the limit of zero beam motion, the image-summed beam size 
\sizts\ matches the turn-averaged beam size \sizts\ at moderate or 
high currents. However, frequently there can be substantial vertical 
beam motion, so that a correction to the naive image-summed beam size is 
necessary. The correction first subtracts the rms beam position 
$\sigma_{\off}$ in quadrature, but then adds back the median statistical 
error on beam position $\delta\off$ (again, in quadrature) so as not to 
bias the result by a purely statistical nonzero value of $\sigma_{\off}$. 
Summarizing:
\beq
\sizcts^2 \equiv \sizts^2 - \sigma_{\off}^2 + (\delta \off)^2\; .
\eeq
For moderate to high current runs, $\sizcts=\sizta$ to within
a small fraction of a micron, regardless of beam motion.
At low energy and/or low current, however, \sizcts\
is slightly larger than \sizta, although only enough to 
reduce but not eliminate the low-current bias. In the remainder of this
section, the beam size quoted will be the 
turn-averaged  value, \sizta, because we will not be
showing very low current data.

We have also observed that values of either type of beam size
can vary by $\sim$1\mi, 
outside their statistical errors for 1024-turn (or fewer) measurements,
taken just seconds apart on the same CESR fill. One source
of such variation is feed-through of low frequency
noise (\eg AC line voltage) into
CESR bending and focusing magnets, which operate
on tightly regulated DC current. This becomes apparent for some
running conditions if one takes runs of, \eg 32768 turns, as seen
in \Fig{fig:sizlongrun}. When comparing beam size measurements
of the same CESR fill using data taken with 1024 (or fewer) turns,
several runs (if available) need to be combined in order to average 
over such effects;
additional data was not always taken, so statistical uncertainties 
in 1024-turn beam sizes are
not representative of deviation from the average over longer time scales.

\subsection{Optical element models}
\label{sec:models}

Measured beam size depends upon the \prftt, and the \prftt\ depends
upon the model, where by ``model'' we mean the
assumptions made when computing the \prftt. As described in
\Sec{sec:prf}, we choose to make all computations in the $yz$ plane,
ignoring the horizontal, due to the symmetry of the setup.
Geometrical quantities, shown in \Tabs{tab:geom}-\ref{tab:fil} 
are reasonably well-known from 
direct measurement and/or specifications of manufacturers. 
Effects of filters, optic substrates, and masking are taken from
the nominal compositions and thicknesses and material properties
catalogued in \cite{ref:henke}. 
The spectrum from \Sec{sec:specresponse} is used,
along with all the geometrical and material information, in \Eq{eqn:kirchoff}.
A step size of 0.1\mi\ is used along the optical element height,
and \prftt\ values are computed every 2.5\mi\ along the image plane;
both of these step sizes are finer than necessary.
The coded aperture \prftt\ is completely determined from these parameters.

\begin{figure}[tb]
   \includegraphics*[width=\figwid]{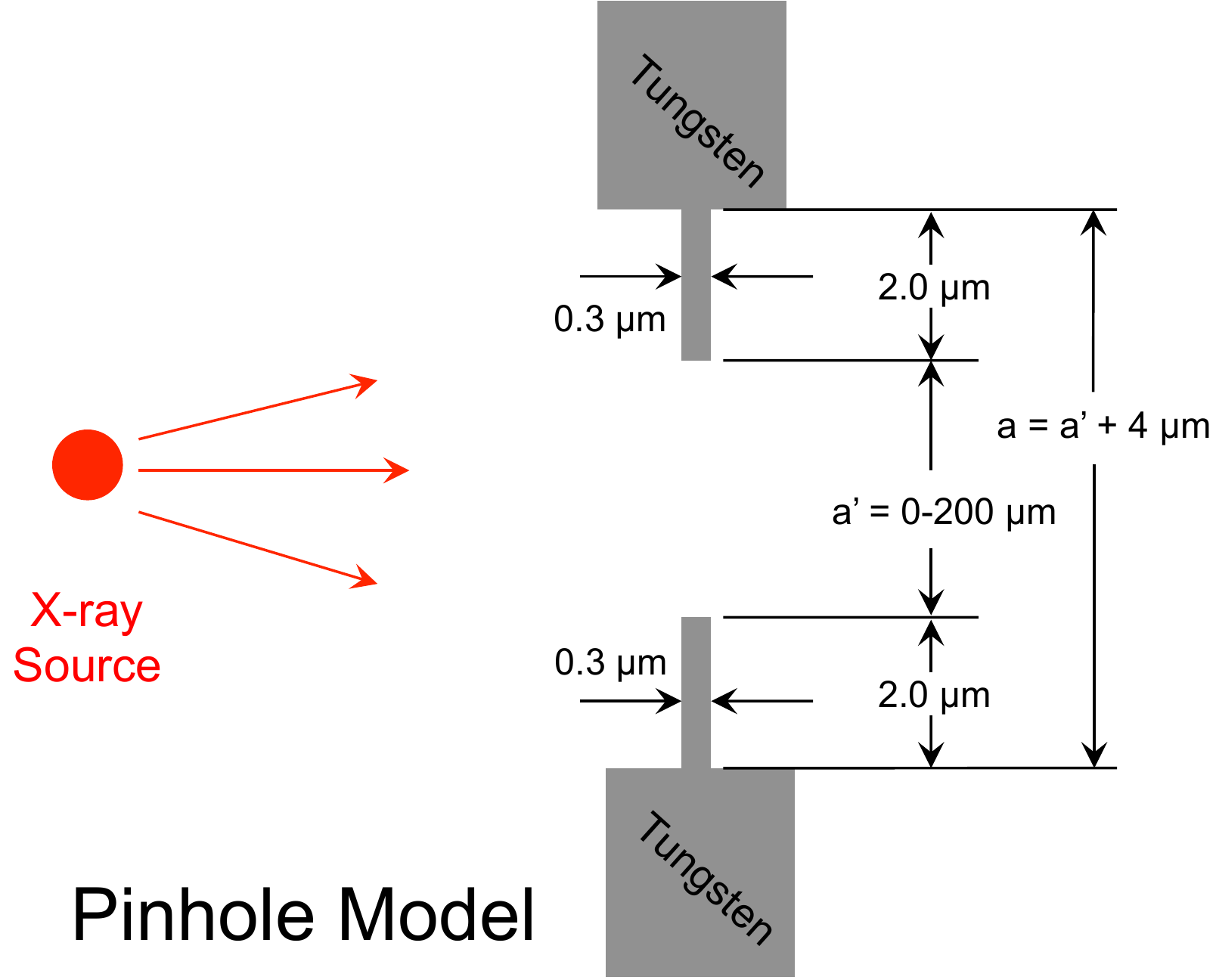}
   \caption{The adjustable pinhole model for \prftt\ generation (see text).}
\label{fig:lip}
\end{figure}

The adjustable pinhole, however, presented a challenge in two respects.
First, the separation of the blades, while reproducible through motor settings,
must be calibrated to an absolute scale. Secondly, 
we found that allowing for\mi-level imperfections in the machined blade
surfaces only narrows the single-peaked \prftt\ relative to a no-roughness
model (this is similar to the effects of masking on single-slit optical elements
discussed in \Sec{sec:design}). 
Hence we adopt a roughness model to generate \prftt s that are in the
``middle'' of the range of roughnesses that are possible. From the discussion
in \Sec{sec:design}, it becomes apparent that x-rays passing through any 
tungsten
on their way to the target will affect the \prftt\ in a similar manner as does
the gold masking on a coded aperture. So our model, shown in \Fig{fig:lip}, 
has two parameters: the
width of a uniformly thick tungsten lip on each edge of the pinhole opening,
and a thickness of this lip. If this lip is very
thick ($>1$\mi), it cannot substantially affect the \prftt\ because the lip
region transmits virtually no light. Conversely, if it is very thin 
($<0.05$\mi), it introduces a phase shift which is too small to matter, except 
insofar
as the lip width increases the effective gap between the blades (and which
will be included in the calibration described below). We chose an effective
tungsten lip thickness of 0.3\mi\ as a value which introduces a substantial
phase shift but transmits enough light for it to affect the \prftt. 
This then allows the lip width to be the single parameter in the model
which controls effects of partially transmitting material.
Our model must assume some nonzero lip width on each side, at a value where the
narrowing of the \prftt\ is roughly midway between the no-lip and the widest
credible lip in the context of this model. We found this midway point to be a
lip width of 2\mi\ on each side of the opening; 
this width narrows the \prftt\ by
about 6\% relative to the no-lip case. 

\begin{figure}[tb]
   \includegraphics*[width=\figwid]{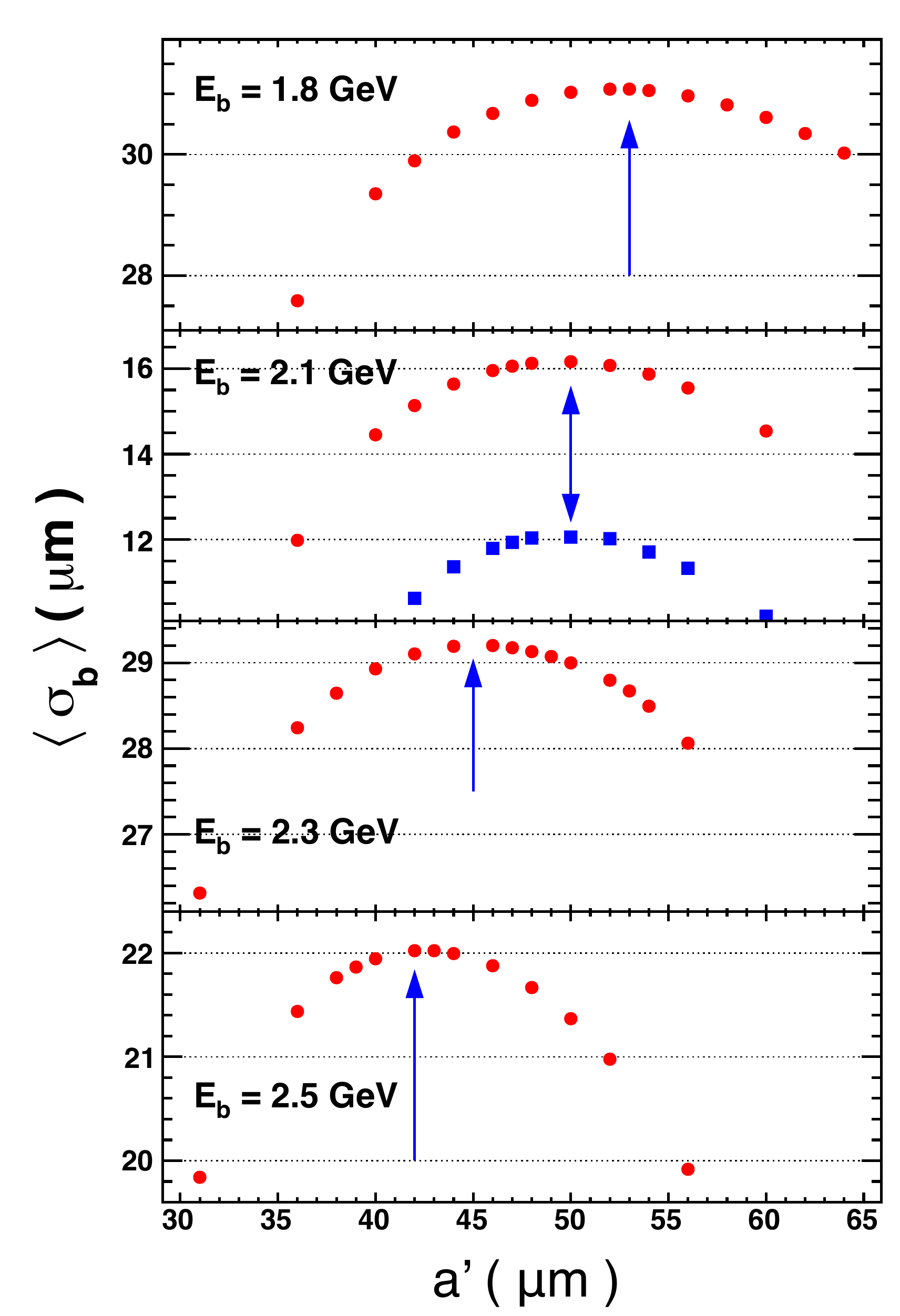}
   \caption{Turn-averaged beam size from $e^+$ beam data taken with the diamond 
   filter and the optimized pinhole optic vs.~pinhole gap size 
   used to generate 
   the template used to fit each point, for four beam energies (see text). 
   Each set of solid squares or circles 
   corresponds to data from the same run at that beam energy, but
   fitted with different \prftt-based templates.
   The arrows indicate the
   maxima of each distribution.}
\label{fig:phgapmodel}
\end{figure}

\begin{figure}[tb]
   \includegraphics*[width=\figwid]{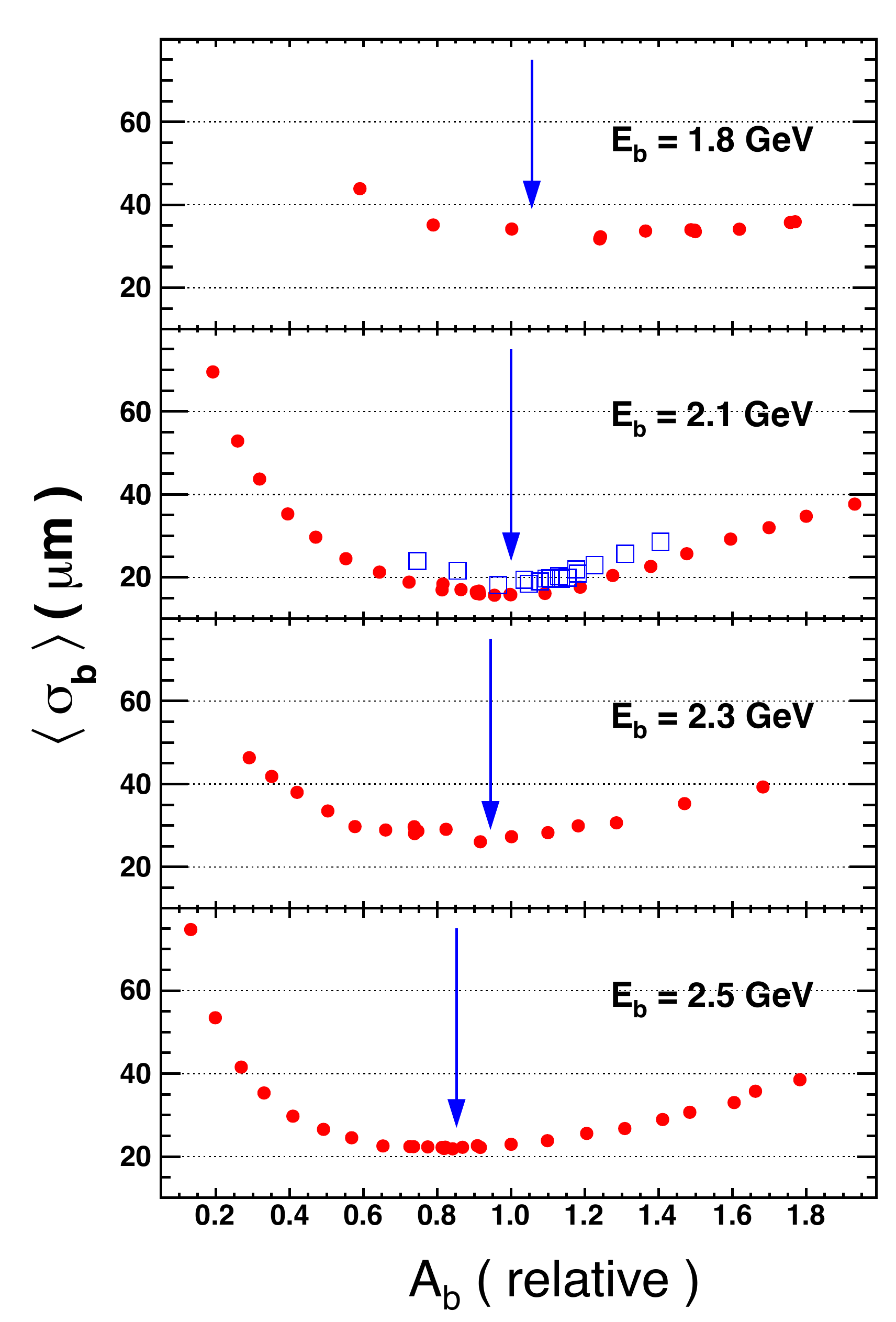}
   \caption{Turn-averaged beam size from $e^+$ beam data taken with the diamond 
   filter vs.~relative image area taken in the scan to optimize 
   the pinhole optic opening, for four beam energies (see text). 
   The arrow locations are placed so as to preserve the
   relative openings found in \Fig{fig:phgapmodel}, 
   and can be seen to point to 
   the minimim total width of the observed pinhole image at each beam energy.}
\label{fig:tasgap}
\end{figure}

What remains is to calibrate the pinhole motor setting to
an absolute gap size in our model. This calibration has
the following steps:
\begin{enumerate}
\item The model is verified to yield a predicted image area that varies linearly
with opening for gap sizes above 10\mi. The dependence is seen to be
that the size \ap, shown in \Fig{fig:lip}, depends on the relative
image amplitude (area) \amp\ as 
\beq
\ap(\mu{\rm m})=50.6\times \dfrac{\amp}{\amp(50\mi)} - 0.6.
\eeq
Thus, the two 2-$\mu$m-wide tungsten lips 
in our pinhole model correspond to an unobstructed opening of 0.6\mi\
in transmitted x-ray intensity.
\item The model is used to determine, for each beam energy and assuming
use of the diamond filter, the absolute gap size corresponding
to the narrowest \prftt. This has been done in two different ways
that obtain consistent results. In one method, each
predicted \prftt\ is fit to an {\it ad hoc} but conveniently analytic 
function with one floating parameter that controls its width.
The function used is a double Gaussian with both components
constrained to the same mean, one component with a 
narrow width $\sigma_N$ accounting for 80\% of the area
and a second with a broader width $\sigma_B$, where 
$\sigma_B\equiv{(100\mi)^2 + \sigma_N^2}$,
accounting for the remaining 20\%. The optimal gap size is that which 
yields a \prftt\ with the smallest $\sigma_N$.
The second method looks at a data run taken at the same beam energy.
A run with small beam size is chosen so as to obtain the best sensitivity. 
This same run is fit to 
templates based on \prftt s generated with different gap sizes; 
by definition, the narrowest \prftt\ occurs where
the largest beam size is obtained. Results from this latter
method appear in \Fig{fig:phgapmodel} for all four beam energies.
Note that two different runs with different beam sizes at $\Eb=2.1$\gev\
both result in maxima at the same gap of 50\mi,
and that all maxima are determined to within $\sim$1\mi.
The narrowest images when using the diamond filter
are obtained for model gap sizes of 53, 50, 45, and 42\mi\
for, respectively, $\Eb=1.8$, 2.1, 2.3, and 2.6\gev. 
\item The motor setting was measured to have linear
dependence upon the corresponding fitted amplitude \amp\ per unit beam current,
locally in the vicinity of the minimum \prftt\ width,
with reproducible slope and intercept on a given CESR fill.
Since the \prftt\ has unit normalized area, \amp\ corresponds
to area, which, from item (1) above, corresponds to a particular motor setting.
\item For each beam energy, runs are taken over a range
of pinhole motor settings on the same CESR fill just minutes apart, 
and the fitted \siz\ and \amp\ per unit beam current recorded for each
using templates from a single \prftt\ for all data at
that beam energy (it does not matter exactly what gap
size was used for the \prftt, as this step is not determining
absolute beam size). The values of fitted \amp\ per unit current 
are then scaled to that obtained for the motor setting
known to give the narrowest image at $\Eb=2.1$\gev. The resulting
data are shown in \Fig{fig:tasgap}. Two separate datasets
are shown for $\Eb=2.1$\gev. The arrow position shown are fixed in relative
amplitude to those obtained from \Fig{fig:phgapmodel},
establishing a consistency between the model and the data. Since
each value of relative amplitude per unit current is known
to correspond to a particular motor setting, this allows
a calibration of motor setting to model gap size.
\end{enumerate}

\subsection{Systematic uncertainty in beam size}
\label{sec:beamsize}

While sensitivity of some sources of uncertainty in beam size
can be investigated from first principles
(\eg statistical sensitivity in \Sec{sec:design}) or models
(\eg pinhole roughness in \Sec{sec:models}), others
must be addressed with measurements. We use data for this
purpose in two ways. First, we can quantitatively examine 
how a particular parameter, $P$, alters the
beam size in a given dataset as it varies around its likely
value and induces changes in the \prftt. With this technique 
one can determine whether the data are {\it self-calibrating}
in each such parameter, and what the effect of changing $P$ 
is on measured beam size. 
If the data are self-calibrating, the actual value, $P_0$
and its uncertainty $\Delta P$ can be extracted from the data and
no nominal or specified value for that parameter is needed;
if the data are not self-calibrating, reliable separate 
specifications of $P_0$ and $\Delta P$ are required.
In either case, $d\siz/dP$ in the vicinity of $P_0$ can be extracted
from the data, and the systematic uncertainty taken as
$d\siz/dP \times \Delta P$.
This use of data is quite effective, as the
same data are used for each parameter setting, meaning any variation
in fit quality or beam size is in fact due to the parameter value change.
With this method, we will describe tests of the sensitivity of beam size to
\begin{itemize}
\item the magnification, by varying it;
\item the pinhole roughness model, by varying the lip width;
\item the coded aperture model, by varying the slit widths;
\item the coded aperture model, by varying the gold masking thickness;
\item the \prftt\ approximation for a misaligned optical element, 
by varying the detector offset $\dtr$ as defined in \Eq{eqn:defined};
\item the predicted spectrum, by varying the assumed spectral response;
\end{itemize}

\begin{figure}[b]
   \includegraphics*[width=\figwid]{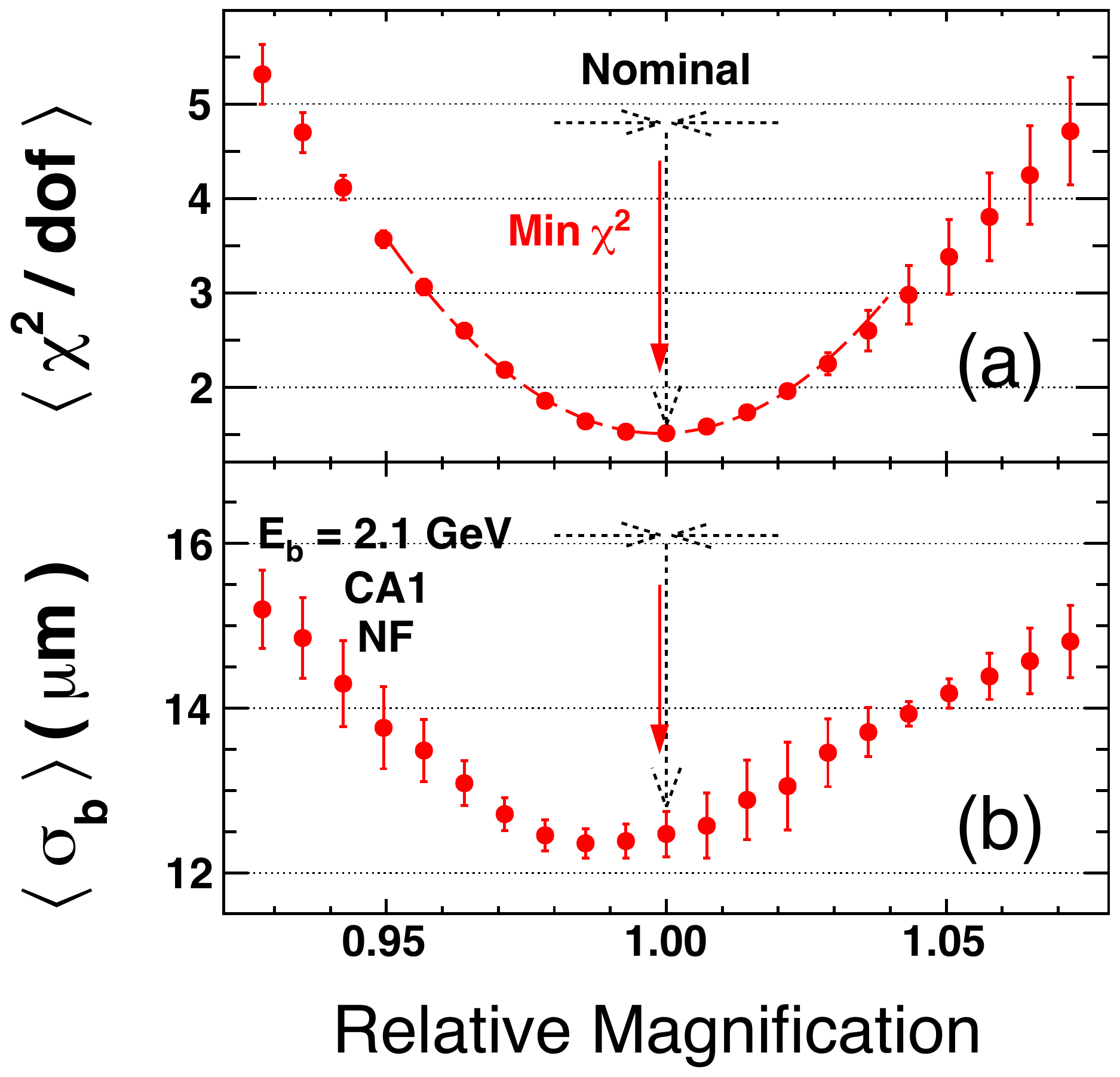}
   \caption{From coded aperture $e^+$ data for $\Eb=2.1$\gev\ with 
            no filter, turn-averaged (a) $\chi^2/$dof and (b) beam size 
            as a function of the magnification applied to the fits; 
            \ie the same data but different magnifications are used for 
            each point. The vertical arrows
            indicate the nominal and approximate best-fit magnifications as
            determined from a quadratic fit to $\left<\,\chi^2/{\rm dof}\,\right>$
            as shown by the dashed curve.
            Horizonal arrows indicate a span of $\pm0.1$\%,
            the nominal magnification uncertainty.}
\label{fig:mag}
\end{figure}

\begin{figure}[t]
   \includegraphics*[width=\figwid]{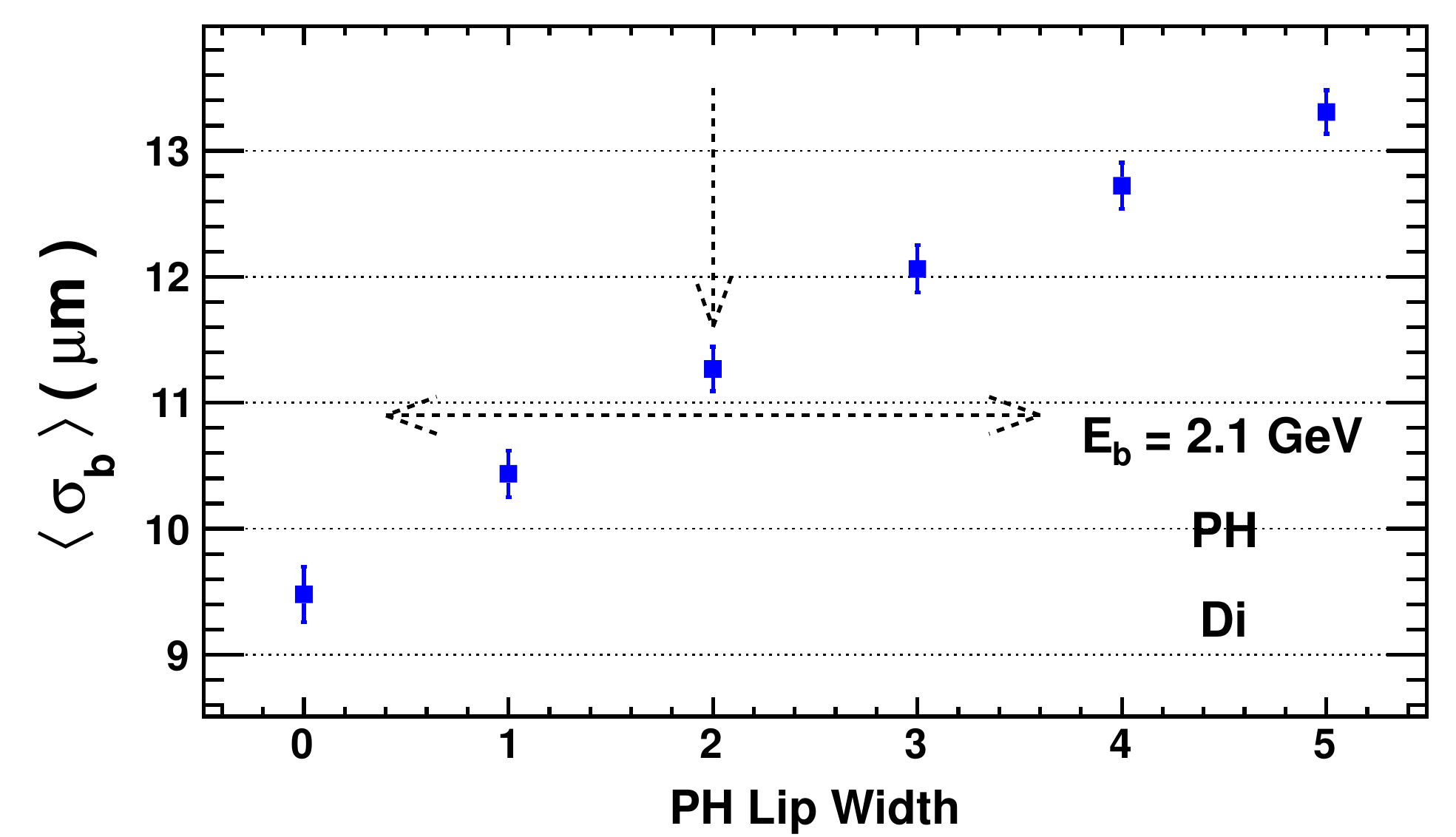}
   \caption{From pinhole $e^+$ data for $\Eb=2.1$\gev\ with 
            the diamond filter, turn-averaged beam size
            as a function of the lip width used in the \prftt;
            \ie the same data but different lip widths are used for 
            each point. The vertical arrow
            indicates the nominal value
            The horizonal arrow indicate a span of $\pm1.6$\mi,
            the estimated uncertainty.}
\label{fig:lipsiz}
\end{figure}
In the second method for determining sensitivity of measured beam size 
to various sources of
uncertainty, we compare data taken in different experimental conditions
during a given CESR fill with stable beam size .
Under the assumption that the beam size does not change in the
several seconds or minutes between measurements, stability
of beam size with respect to the particular change can be studied.
One obstacle limiting the precision of such tests is that
up until now, many, but not all, data runs contain only 1024 
turns each, and so are subject to
true beam size oscillations at low frequency. These oscillations are not
always present, but are not easily detectable without
runs containing 8192-32768 turns (as seen in \Sec{sec:size} and 
\Fig{fig:sizlongrun}). Such beam size variations
are frequently of magnitude $\sim$1\mi\ or larger. Sometimes this
effect can be mitigated by combining several 1024-turn runs
taken in essentially identical conditions. (For future
operations, we plan to
increase the standard number of turns from 1024 to at least 8192).
With this technique, we will describe beam size dependence upon
\begin{itemize}
\item the pinhole model, by varying the gap size;
\item the predicted spectrum and spectral response, by using different
filters and/or different beam energies; and
\item the overall systematic uncertainty of pinhole and coded aperture 
models, by changing optical element from one to another in various
conditions.
\end{itemize}
We will show the results of such tests, but due to the
aforementioned issue, the precision of these tests is generally
inferior to that of the first method.

\begin{figure}[b]
   \includegraphics*[width=\figwid]{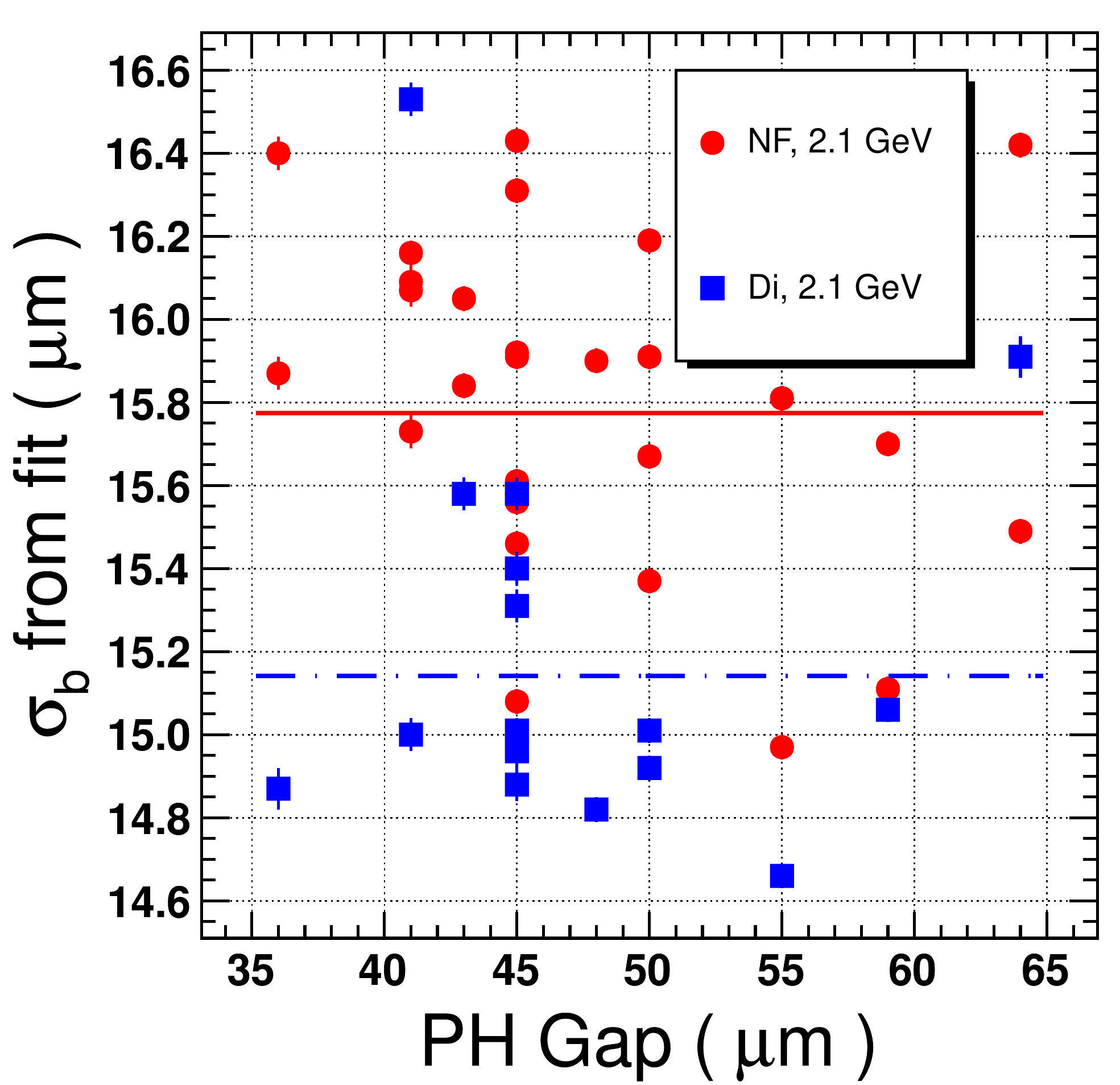}
   \caption{Fitted beam size from $e^+$ beam data taken with no filter (NF)
   and the diamond filter (Di) vs.~gap size at $\Eb=2.1$\gev.
   The solid (dot-dashed) 
   horizontal line
   represents the mean beam size of the NF (Di) data.
   }
\label{fig:gapscan}
\end{figure}

\subsubsection{Magnification}
\label{sec:mag}
Coded aperture image fit quality is sensitive to
knowledge of the geometry. Image features at the extreme ends
provide this sensitivity; as the magnification is raised
or lowered such features expand away from center or contract
toward it. \Fig{fig:mag} shows both the turn-averaged 
$\chi^2/$dof and beam size as a function of the magnification
assumed in the \prftt\ for $e^+$ CA1 data acquired at $\Eb=2.1$\gev\ 
with no filter.
The best-fit value is 0.12\% smaller than nominal, slightly outside
the uncertainty quoted in \Tab{tab:geom}.
Using the nominal 0.1\% as the uncertainty in magnification
 induces a 
beam size uncertainty for pinhole data of 0.1\% at any beam size because the
\prftt\ is single-peaked. For the CA1 optical element, the same
magnification uncertainty induces a beam size uncertainty of
$\sim$0.03\mi\ at $\siz=12$\mi, or 0.3\%; sensitivity to magnification 
decreases at larger beam sizes because the image features become
washed out. 
Hence, unlike the pinhole, relative beam size error for 
coded aperture measurements due to magnification 
uncertainty tends to shrink rather than grow with beam size.

\subsubsection{Pinhole lip width}
\Figure{fig:lipsiz} shows the variation of beam size with the
width of the tungsten lip in the pinhole roughness model.
We estimate a systematic uncertainty in this of $\pm1.6$\mi,
for which measured beam size is expected to change
by about $\pm1.33$\mi\ at $\siz\approx 12$\mi\ and $\Eb=2.1$\gev

\begin{figure}[t]
   \includegraphics*[width=\figwid]{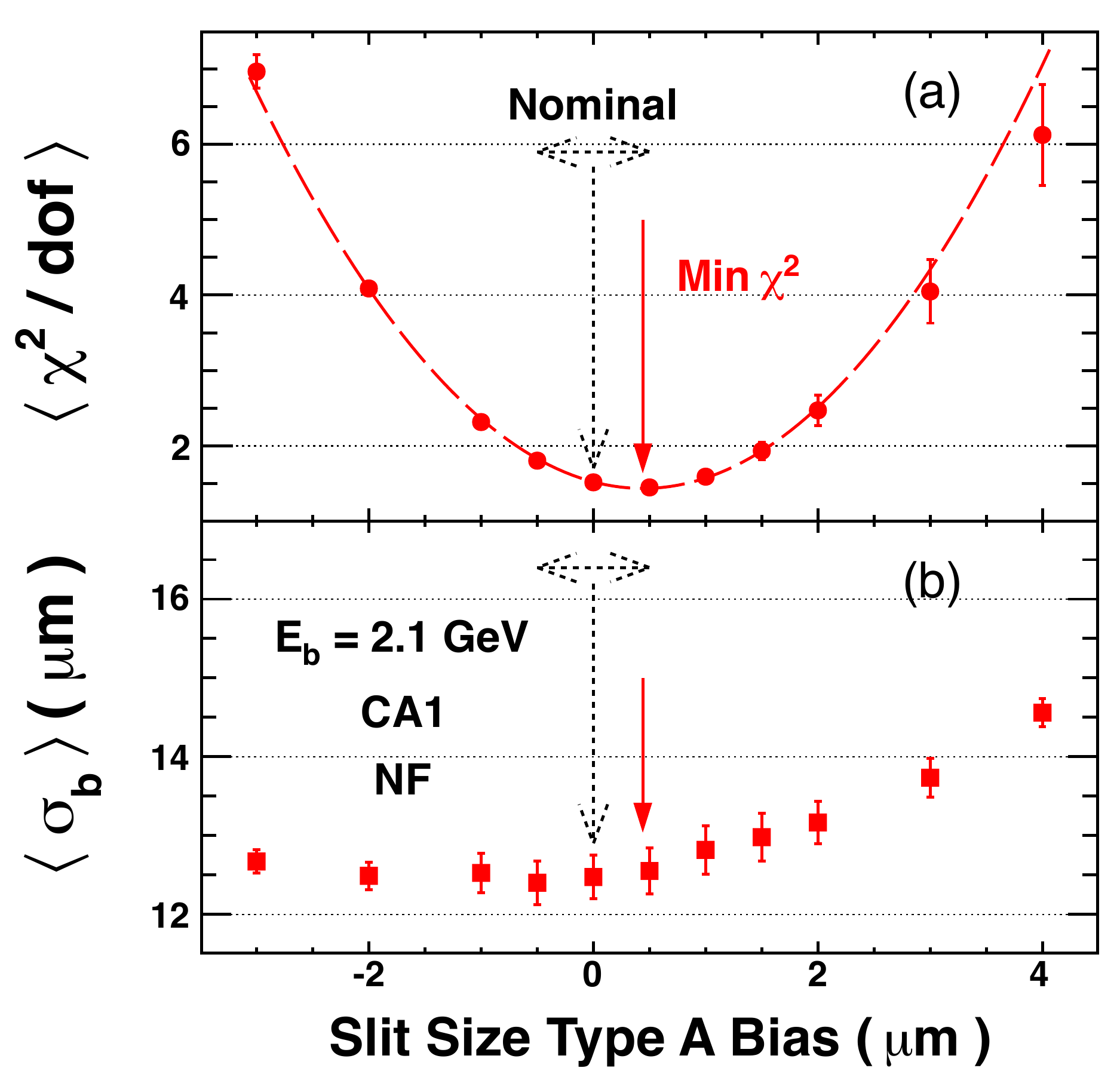}
   \caption{From coded aperture $e^+$ data for $\Eb=2.1$\gev\ with 
            no filter, turn-averaged (a) $\chi^2/$dof and (b) beam size 
            as a function of a slit size bias (Type A, see text) used in
            constructing the \prftt\ used in the fits; \ie the same 
            data but different \prftt s are used for each point. The vertical 
            arrows indicate the nominal (dashed) and best-fit (solid)
            variations from nominal as determined from a quadratic fit 
            shown by the dashed curve in (a). Horizontal (dashed) arrows 
            indicate the nominal uncertainty of $\pm0.5$\mi.}
\label{fig:slita}
\end{figure}

\begin{figure}[t]
   \includegraphics*[width=\figwid]{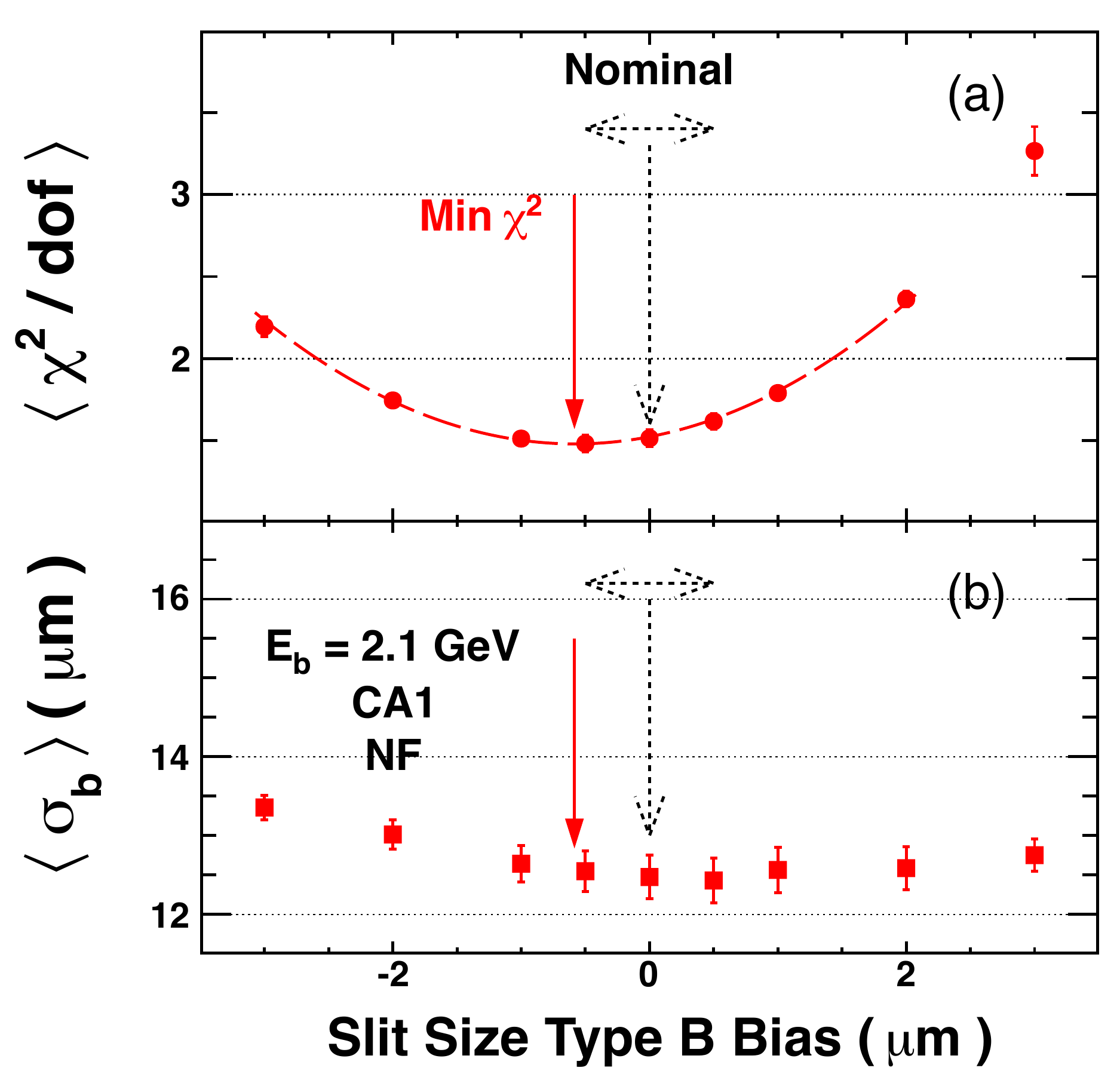}
   \caption{From coded aperture $e^+$ data for $\Eb=2.1$\gev\ with 
            no filter, turn-averaged (a) $\chi^2/$dof and (b) beam size 
            as a function of a slit size bias (Type B, see text) used in
            constructing the \prftt\ utilized in the fits; \ie the same 
            data but different \prftt s are used for each point. The vertical 
            arrows indicate the nominal (dashed) and best-fit (solid)
            variations from nominal as
            determined from a quadratic fit shown by the dashed curve in
            (a). Horizontal (dashed) arrows indicate the 
            nominal uncertainty of $\pm0.5$\mi.}
\label{fig:slitb}
\end{figure}

\subsubsection{Pinhole gap opening}

The pinhole gap calibration method described in \Sec{sec:models}
results in sensitivities to beam size shown in \Fig{fig:phgapmodel}.
It is apparent that sensitivity increases quickly as the PH gap
moves away from that which minimizes the width of the \prftt.
For purposes of estimating the relevant sensitivity from
\Fig{fig:phgapmodel}, we assume
that on average the setting is $\sim$10\mi\ from the optimal one,
but that the calibration of \Sec{sec:models} specifies the
setting within $\pm$2\mi. This leads to an assigned uncertainty
in beam size at $\Eb=2.1$\gev\ and $\siz=12$\mi\ of $\pm$0.60\mi.

Sometimes production data were mistakenly taken at settings
not near the optimal settings.
In order to verify that such data remain viable,
and to investigate sensitivity to our knowledge of
absolute pinhole gap size,
data from a scan over various openings were fit with the templates from 
the \prftt\ corresponding to the calibrated motor setting,
as described in \Sec{sec:models}.
Such a test
is shown in \Fig{fig:gapscan}. Variations at fixed gap size
show systematic run-to-run beam size variations of $\sim$0.5\mi,
consistent with real changes that can happen in these 1024-turn runs.
Where there are at least three runs at a given gap size, their
average \siz\ agrees with other such data with the same filter (but
different gap size) to
within less than 0.5\mi. This falls within the assigned systematic uncertainty.
The two filter datasets disagree on average by $\sim$0.6\mi;
since they were taken with the same gap opening, this disagreement cannot be due
to uncertainty in the gap setting, but rather must be due to other factors.

\begin{figure}[t]
   \includegraphics*[width=\figwid]{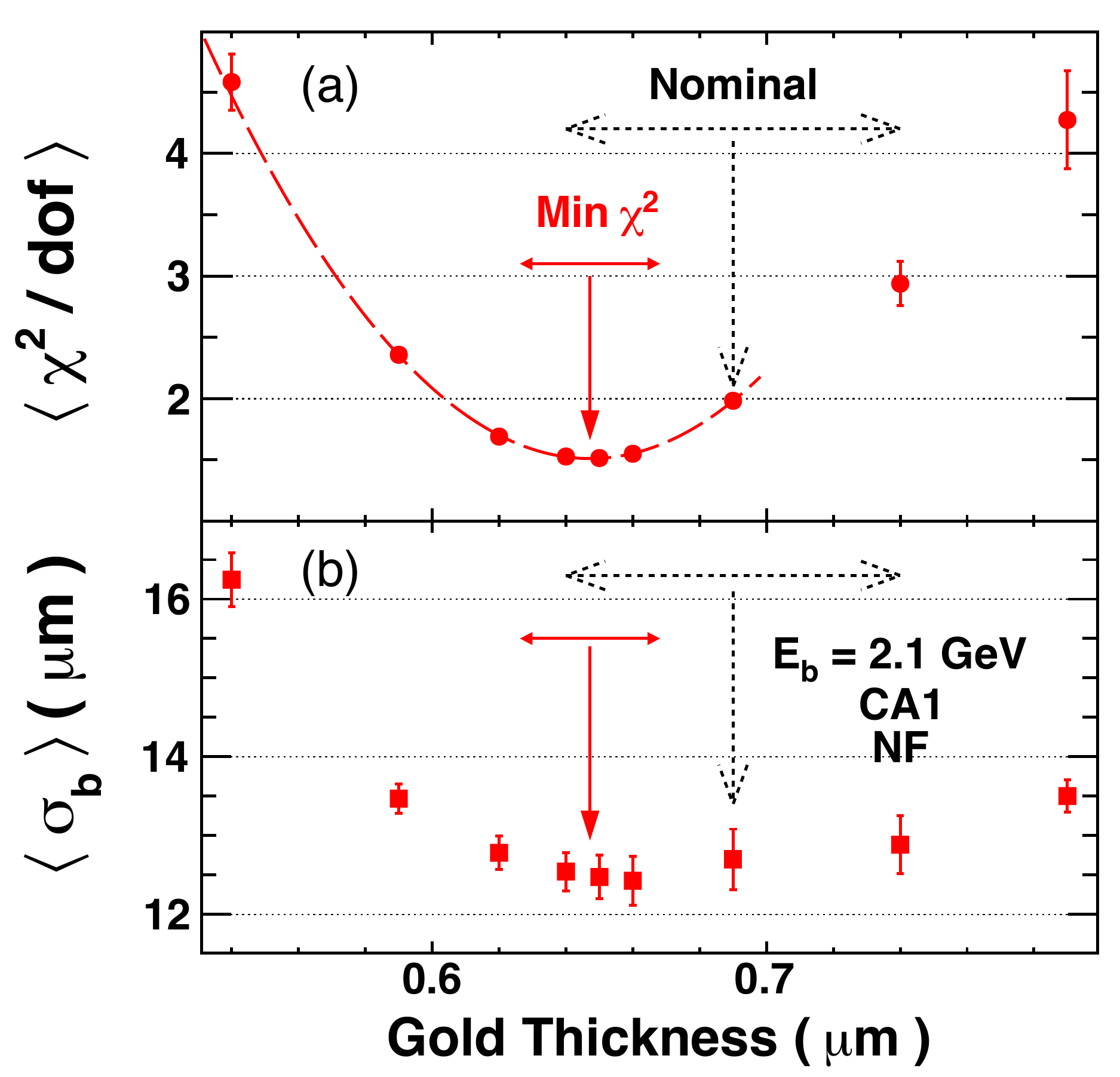}
   \caption{From coded aperture $e^+$ data for $\Eb=2.1$\gev\ with 
            no filter, turn-averaged (a) $\chi^2/$dof and (b) beam size 
            as a function of the gold masking thickness used in
            constructing the \prftt\ utilized in the fits; \ie the same 
            data but different \prftt s are used for each point. The vertical 
            arrows indicate the nominal (dashed) and best-fit (solid)
            thicknesses,  the latter 
            determined from the quadratic fit shown by the dashed curve.
            Horizontal arrows indicate nominal uncertainty (dashed) on thickness
            from the manufacturer's specification ($\pm0.05$\mi) and
            an estimated upper limit on the uncertainty (solid) 
            on self-calibrated thickness.}
\label{fig:thick}
\end{figure}

\subsubsection{Coded aperture slit size}
The manufacturer's specified tolerance on CA1 slit size and position is given
as $\pm0.5$\mi. We explore two scenarios, each likely to be more extreme
than any actual variations from nominal dimensions: 
one, Type A, in which all slits are
systematically larger than specification by the same amount (with
adjacent mask size compensating exactly to preserve overall height of
the pattern), and another, Type B, wherein adjacent slits alternate
being larger or smaller than nominal (with mask sizes unchanged).
\Figs{fig:slita} and \ref{fig:slitb} show both the turn-averaged 
$\chi^2/$dof and beam size as a function of the slit size bias
assumed in calculating the \prftt\ for $e^+$ data acquired at $\Eb=2.1$\gev\ 
with no filter. The best fits differ from the 
specification by $\sim0.5$\mi; if we use this as the uncertainty
on slit size and add in quadrature the change in beam size
from the two types of bias, the resulting beam size uncertainty is 0.28\mi.

\subsubsection{Coded aperture masking thickness}
The manufacturer's specification on CA1 masking thickness is given
as $0.69\pm0.05$\mi. Because the mask both attenuates intensity and
introduces a phase shift, this level of uncertainty has potential to 
affect beam size precision. Once again we find that the image fit 
quality allows the coded aperture to be self-calibrating, this time 
in gold thickness.  \Fig{fig:thick} shows both the turn-averaged 
$\chi^2/$dof and beam size as a function of the masking thickness 
assumed in the \prftt\ for $e^+$ data acquired at $\Eb=2.1$\gev\ 
with no filter. The best fit differs from the 
specification by 0.043\mi. If we use the best-fit value of thickness in
the generation of the \prftt\ and assume an uncertainty of 0.02\mi, 
the resulting uncertainty in beam size
is $\pm$0.40\mi.

\subsubsection{Misaligned optical element}
\label{sec:detoff}
When the x-ray source is significantly off-center with respect to the optic, 
the intensity profile on the optical element becomes asymmetric, and
the approximation of \Eq{eqn:prfvee} affects the shape of the \prftt.
For such data, it is important to know how, if at all, the approximation
affects the fitted beam size. Despite attempts to follow the alignment
procedures described in \Sec{sec:align}, several datasets with large
vertical offsets (requiring detector offsets $\dtr>200$\mi,
as defined by \Eq{eqn:defined}) are available.
Such runs were discovered in coded aperture data; the signature is 
significant asymmetry of the fit$-$data residual on extreme edges of 
the image. Variations of images acquired with different alignments
can be compared. When the pinhole is operated in minimum-\prftt-width 
conditions, its opening is small, and 
there is no significant change in incident intensity over its vertical extent.
(This is a positive feature of the pinhole optic: its insensitivity to 
misalignment.)
Conversely, the coded aperture is at least five times taller, and thus,
for large $\dtr$, the variation of intensity from one end to the other
can become apparent in the image. 

\begin{figure}[t]
   \includegraphics*[width=\figwid]{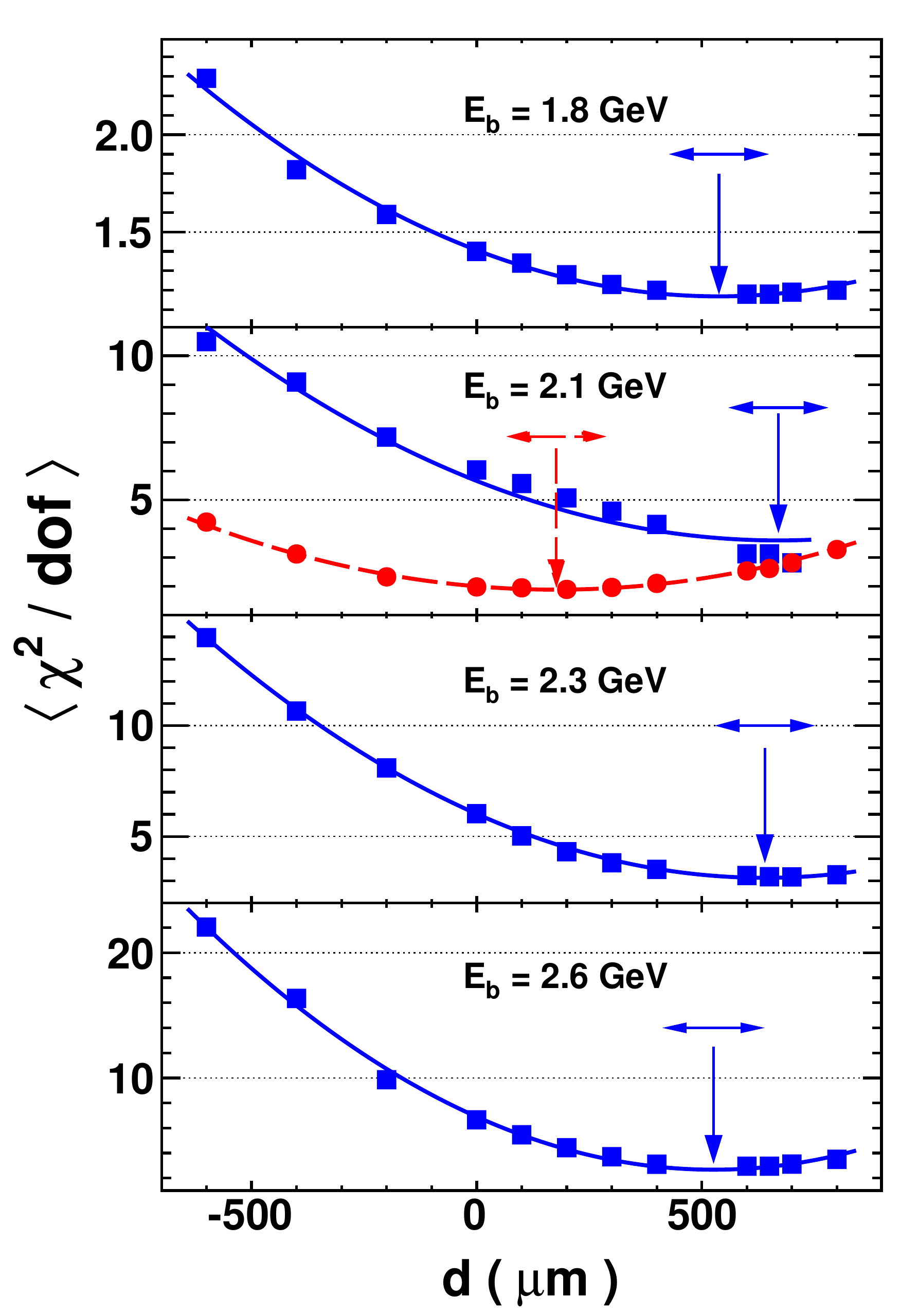}
   \caption{Turn-and-run-averaged $\chi^2/$dof of image fits
            as a function of the detector offset, $\dtr$, from
            the center of the coded aperture that is used to generate
            templates, for four beam energies and five sets of
            beam conditions. 
            Curves indicate
            quadratic fits and vertical arrows the minimum-$\chi^2/$dof
            location according to the fits. 
            Horizontal arrows indicate a $\pm100$\mi\ span,
            an estimated upper limit on the $\dtr$ uncertainty from this 
            measurement. The filled circles,
            dashed curve, and dashed arrows at $\Eb=2.1$\gev\ are 
            from a different
            set of beam conditions than the filled squares/solid curve/solid arrows.
           }
\label{fig:chisqd}
\end{figure}

\begin{figure}[t]
   \includegraphics*[width=\figwid]{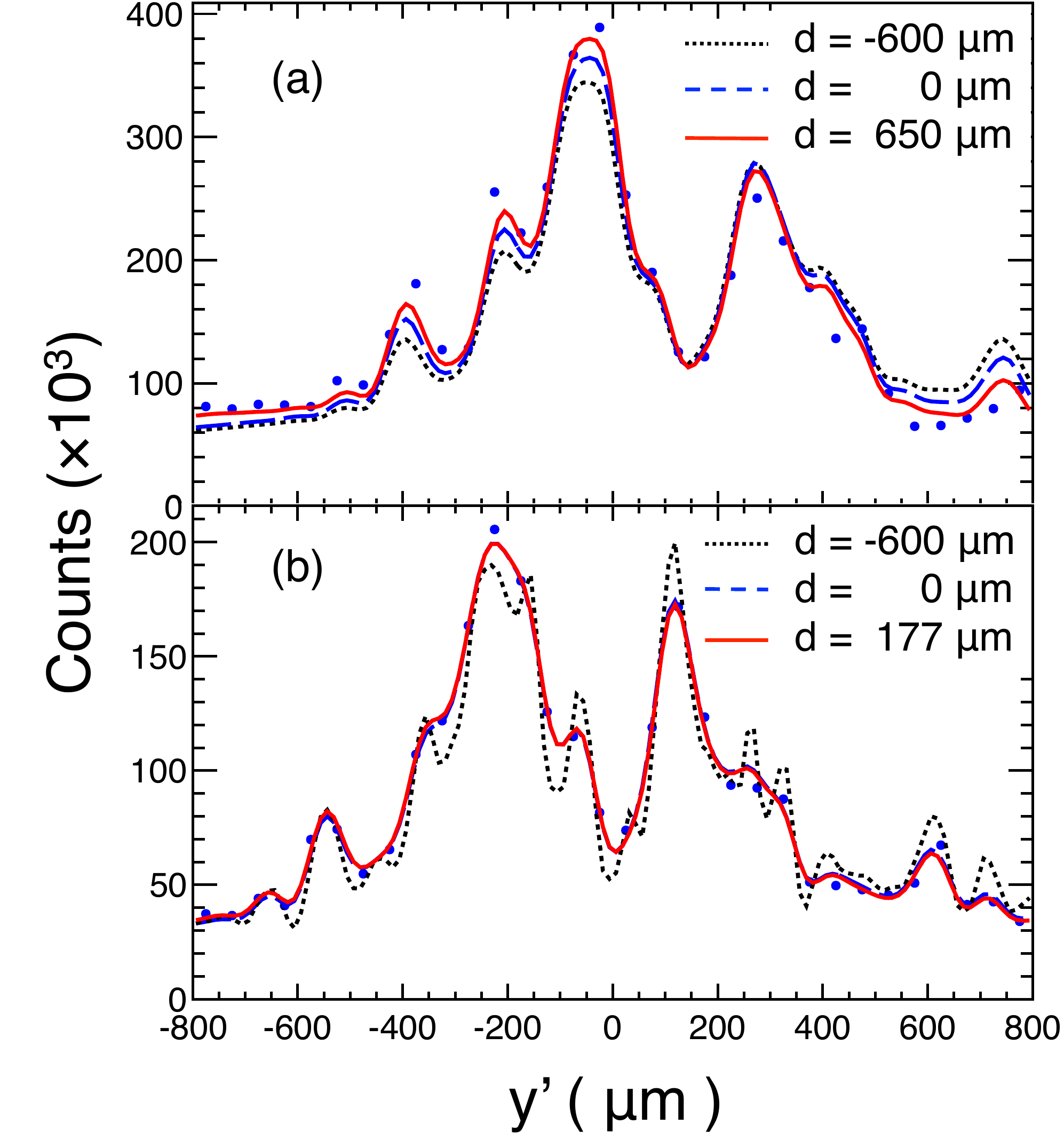}
   \caption{Detector image fits for $e^+$ two runs taken with different
            alignments, both acquired at $\Eb=2.1$\gev\
            with the diamond filter and CA1 optical element.
            The best $\chi^2/$dof occurs for detector
            offset (a) $\dtr=650$\mi\ and (b) $\dtr=177$\mi.
            Fits for different $\dtr$ values are indicated by curves that are
            solid (best fit), $\dtr=-600$\mi\ (dotted),
            or $\dtr=0$\mi\ (dashed).}
\label{fig:cadet}
\end{figure}

\begin{figure}[t]
   \includegraphics*[width=\figwid]{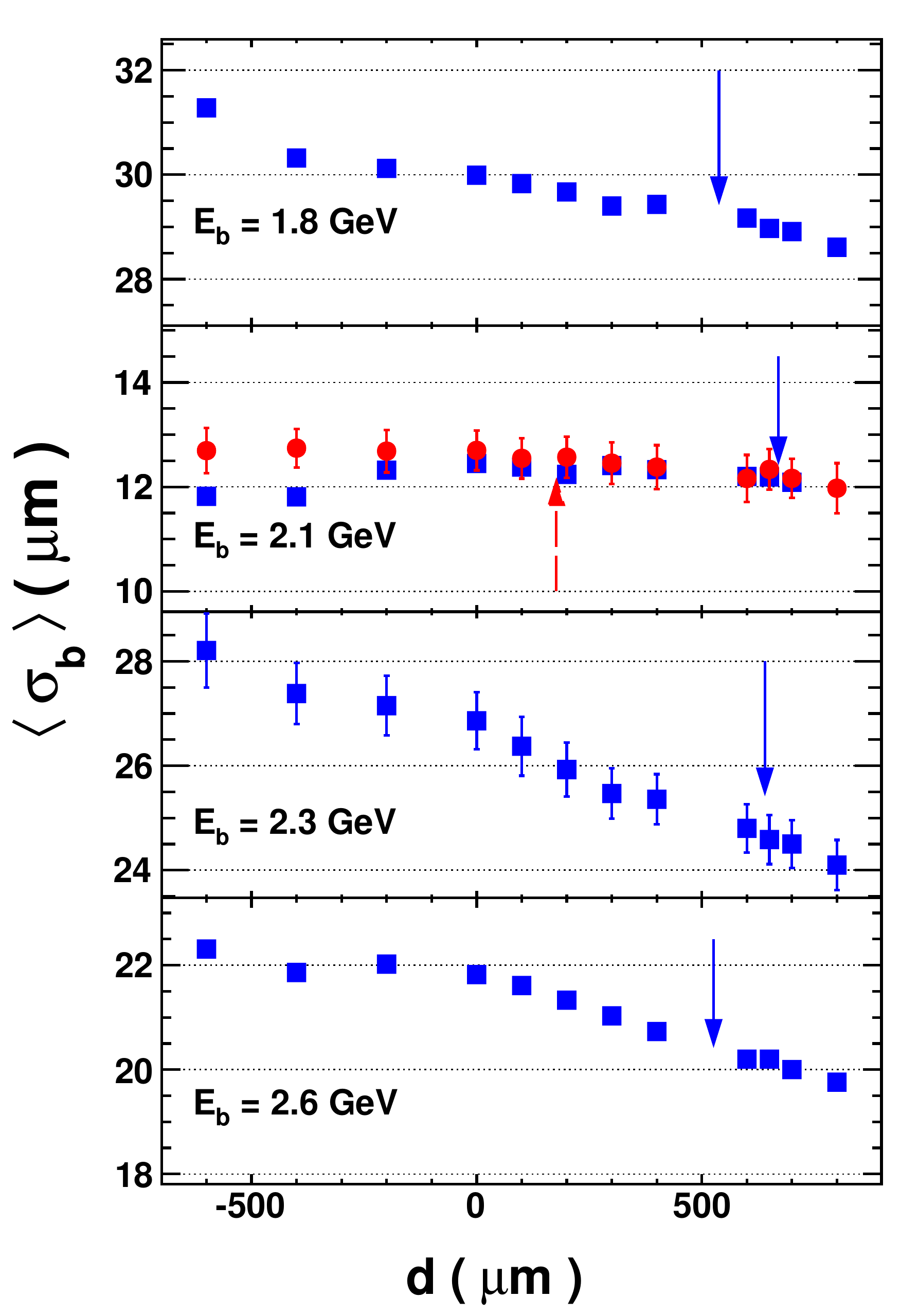}
   \caption{Turn-and-run-averaged \siz\ for $e^+$ data image fits
            as a function of the detector offset, $\dtr$, from
            the center of the pinhole that is used to generate
            templates, for four beam energies and five sets of
            beam conditions. The filled circles
            at $\Eb=2.1$\gev\ are from a different
            set of beam conditions than the filled squares.
            Error bars indicate either statistical error from
            a single run (typically very small), or the 
            standard error on the mean of several runs taken
            in the same beam conditions. Arrows indicate 
            the minimum-$\chi^2/$dof
            location according to the fits, as shown in
            \Fig{fig:chisqd};
            dashed upward arrow applies to filled circles,
            solid downward arrows to filled squares.}
\label{fig:sizdet}
\end{figure}

We have found that the value of $\dtr$ is measurable from coded aperture
data by examining the dependence of the image fit quality, $\chi^2/$dof,
on the assumed value of $\dtr$; this is shown
in \Fig{fig:chisqd} for $e^+$ data taken at four beam energies.
The optimal value of \dtr\ occurs where the fit quality is minimized,
and its uncertainty is determined by its parabolic shape.
Variations in image shape attributable to misalignment 
can be seen in \Fig{fig:cadet}, where
the asymmetry induced by the misalignment is apparent. 
Effects of the approximation in \Eq{eqn:prfvee} 
are visible where the optimal $\dtr$ is large,
as in \Fig{fig:cadet}(a), where even the best-fit misses several features
in the data. However, for a smaller $|\dtr|$ value,  as in \Fig{fig:cadet}(b),
the best-fit shape does a much better job of representing nearly
every nuance of the image shape.

Using an inaccurate value of $\dtr$ in the fit induces significant
variation in the resulting \sizta, as shown in \Fig{fig:sizdet}, 
where variation is seen to be approximately linear in $\dtr$. 
The value of \sizta\ for $\dtr=0$ differs from that for $\dtr=650$ 
by 1-2\mi; to avoid incurring such a bias one must either avoid optic 
misalignment altogether,
or perform this procedure on all coded aperture data (\ie find and then use
the optimal value of \dtr). In the examples of \Fig{fig:chisqd},
precisions in optimal \dtr\ are estimated as $\pm100$\mi. 
The uncertainty
in beam size due to this uncertainty in $\dtr$ (even for large $\dtr$) is 
only $\pm0.3$\mi.

\subsubsection{Spectral sensitivity}
\label{sec:specvar}
\begin{figure}[tbp]
   \includegraphics*[width=\figwid]{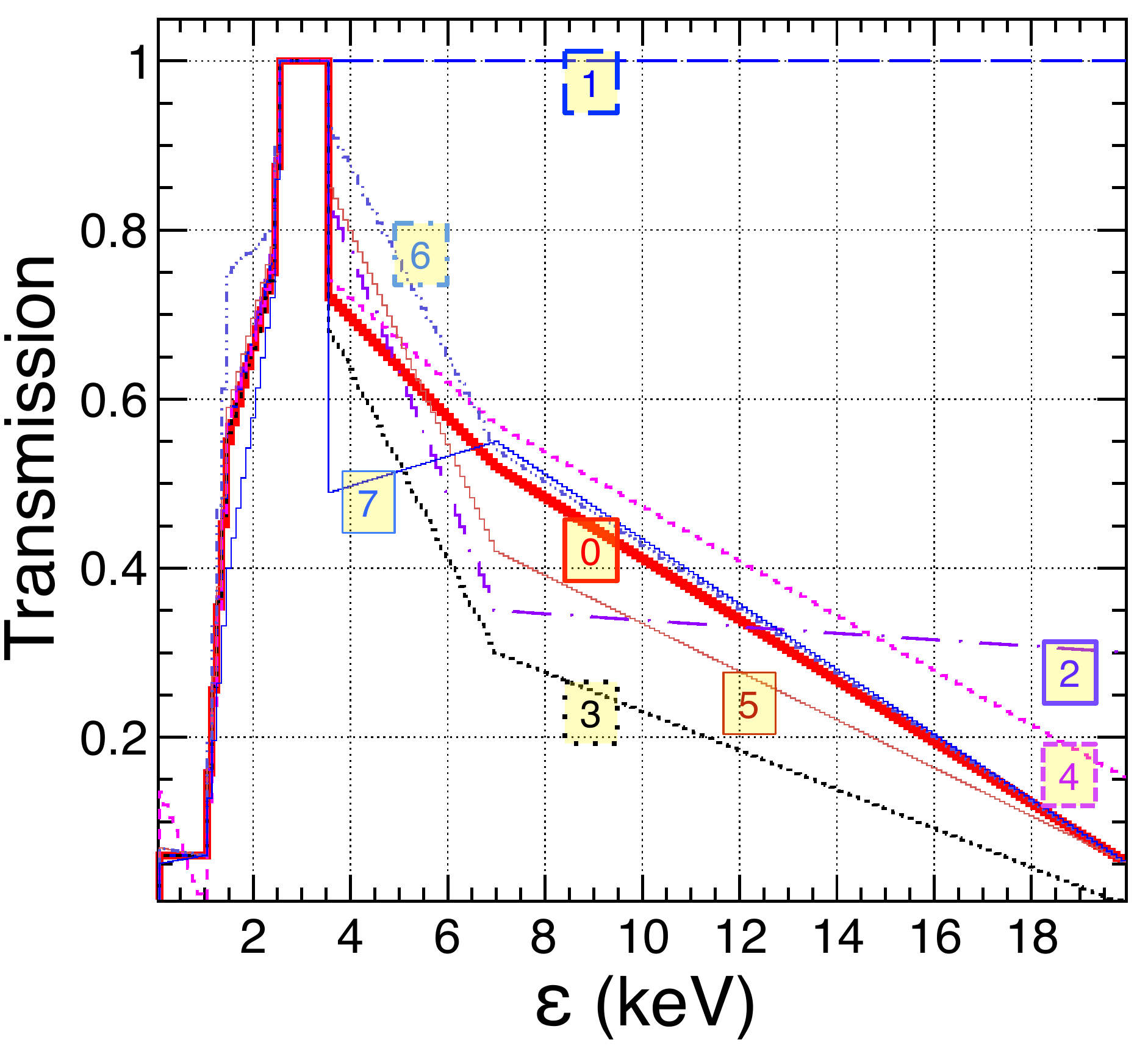}
   \caption{Spectral response curves used to explore sensitivity of measured
     beam size to the detected x-ray spectrum. The thickest solid curve
     represents the nominal spectrum and is labeled as ``0''; seven other
     alternative responses are also shown and labeled with numbers from 1-7.}
\label{fig:specvar}
\end{figure}

The most straightforward test of the spectral sensitivity of measured beam size
is to measure the same dataset with each optic with different assumed
spectra. Seven alternate spectral response functions were extracted from
the filter scan data as in \Sec{sec:specresponse} by fixing certain of the
points to extremes of their error bars and refitting. None of these spectral
response functions, shown in \Fig{fig:specvar}, can
be excluded by more than one standard deviation 
as being closer to reality than the nominal one. Both coded aperture
(with no filter) and pinhole (with the diamond filter) data were analyzed using
\prftt s from these alternate spectral responses. The extreme variations from
nominal were obtained for spectra applied to CA1 data from curves \#1 and \#6,
giving beam sizes that were $\sim$0.35\mi\ below and above the nominal result, 
respectively. For PH data, the extremes were obtained from curves \#6 and \#7,
giving beam sizes that are $\sim$0.5\mi\ smaller and larger, respectively, than
the nominal PH run-and-turn-averaged beam size. 
We use these extremes as
estimates of systematic uncertainty, which apply only to the no-filter (NF) and
diamond-filter (Di) data, for which a broad spectrum is sampled.

\subsubsection{Data-to-data comparisons}

\begin{figure}[tbp]
   \includegraphics*[width=\figwid]{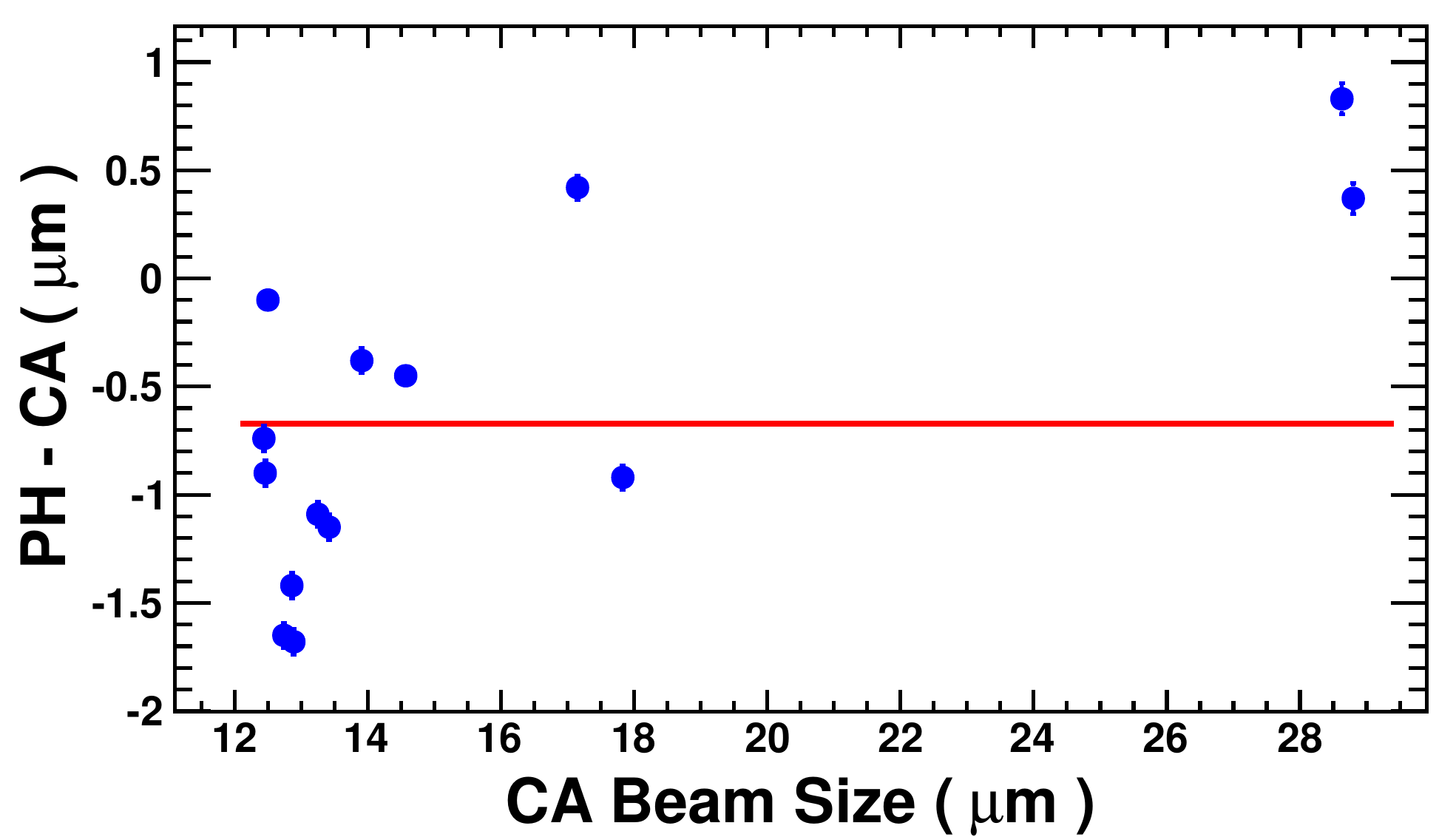}
   \caption{Difference between measured beam sizes as determined by the 
   pinhole (PH) optic using the diamond filter 
   and a coded aperture (CA1) using no filter vs.~the CA1-determined
   beam size for a selection of CESR settings which result in
   a range of beam sizes. The solid horizontal line represents the mean
   difference in beam size. }
\label{fig:coupeight}
\end{figure}

A comparison of data using CA1 with no filter to PH with the diamond filter
at $\Eb=2.1$\gev\ appears in \Fig{fig:coupeight}. For this CESR fill, 
steering parameters were varied to sample a range of vertical beam sizes,
 from 12-30\mi.
At each setting, data from CA1 and the PH were obtained, and so can be
directly compared. Variations with beam size are too small to consider
significant, given run-to-run variations. The overall average difference
is of similar magnitude as the NF-Di difference observed for the PH optic 
alone in \Fig{fig:gapscan}.

\begin{figure}[tbp]
   \includegraphics*[width=\figwid]{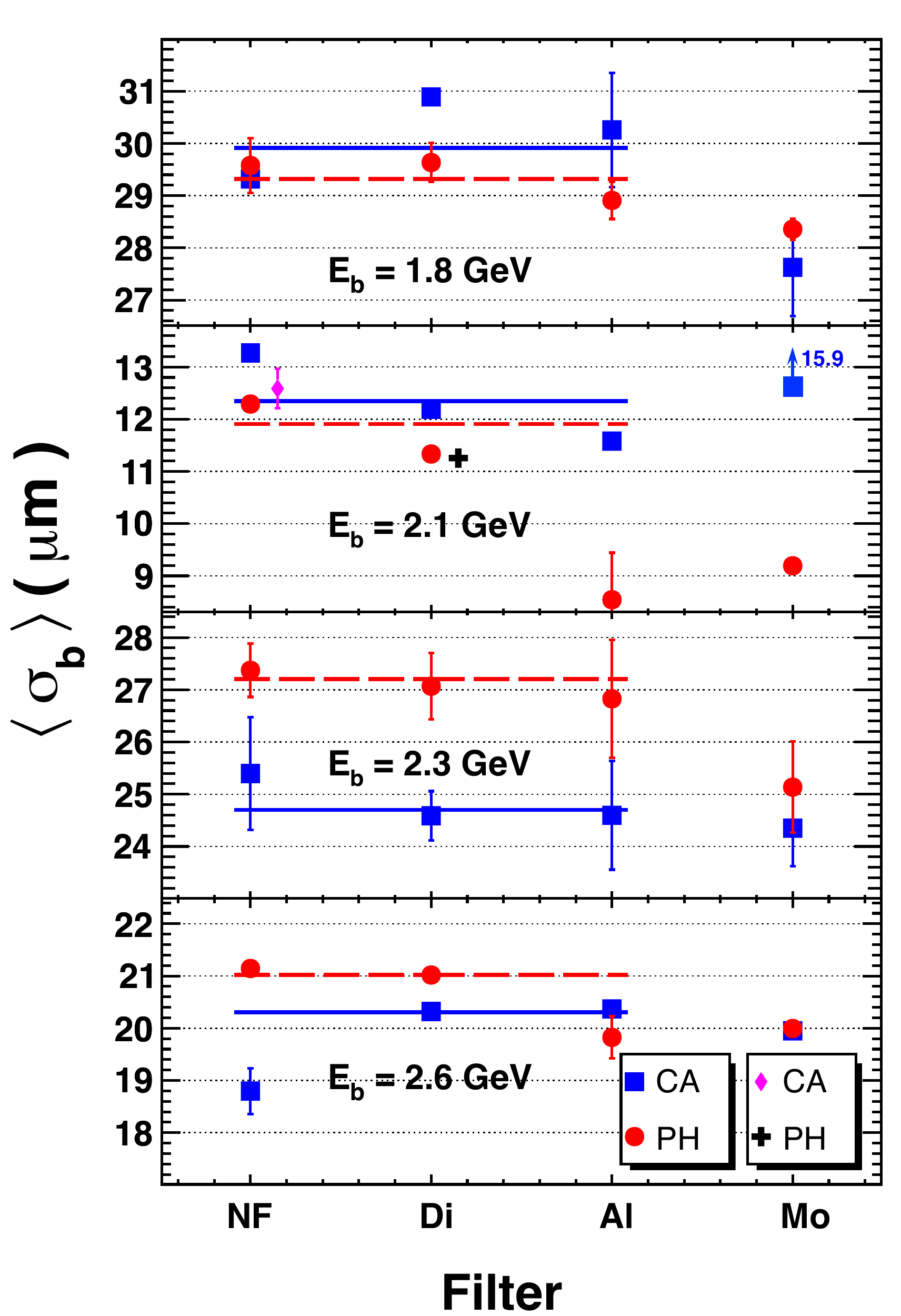}
   \caption{Turn-and-run-averaged \siz\ for $e^+$ data image fits
            as a function of optical element and filter, 
            for four beam energies and five sets of
            beam conditions.
            Circles represent results taken with the pinhole (PH)
            optic, and solid squares those taken with
            coded aperture CA1. A second dataset
            at $\Eb=2.1$\gev\ is shown: the diamond
            represents use of CA1 with NF, and 
            the cross applies to the PH optic with the Di filter.
            Horizontal lines represent fits to three filter (NF, Di, Al) points 
            for CA1 (solid for squares) and PH (dashed for circles) results.}
\label{fig:sizflt}
\end{figure}

Finally, a comparison of beam sizes obtained with different filters 
at different beam energies, and with the two optical elements are shown in
\Fig{fig:sizflt}. Agreement is generally good, given known $\sim$1\mi\ 
variations in beam size during a run (the points with small error bars include 
only one run with 1024 turns). 
Data from the molybdenum filter varies most in extracted beam size; this
is perhaps not surprising given that it selects only a narrow part of the
available spectrum, throwing away $\sim$90\% of the light relative to the
no-filter (NF) setting.

\subsubsection{Summary}
\label{sec:syssum}
A summary of expected systematic uncertainties appears in \Tab{tab:syserr},
where the uncertainties are listed at one energy and beam size, namely,
$\Eb=2.1$\gev\ and $\siz\approx 12$\mi, and apply only to the broad-spectrum
filter data (NF and Di). Note that these uncertainties assume
that the \prftt\ applied has been generated with the optimal parameters,
including geometry, partially absorptive material, pinhole opening 
or slit sizes, and that
for CA1 the optimal detector offset $\dtr$ is used, as described in
\Sec{sec:detoff}. Uncertainties can be much larger if any of these is not the
case. Note also that for PH data it assumes that the PH gap size 
is both known to within $\pm2$\mi\ and set at a size within 
5\mi\ of that which would give the
narrowest \prftt\ at that beam energy and filter setting; setting the opening
15\mi\ smaller than optimal increases the uncertainty five-fold.
This uncertainty on beam size is expected to scale approximately as 
$\propto 1/\siz$ and $\propto 1/\Eb$,
with the exception of the effect of magnification uncertainty, which scales 
as \siz. Taking the total uncertainties in quadrature can account for 
observed systematic differences between PH and CA1 results of $\sim1.6$\mi\ 
at this beam energy and beam size, as shown in \Fig{fig:coupeight}.

The foregoing analysis demonstrates that $\sim$1\mi\ precision on beam
size can be attained with the xBSM, but only after acquiring detailed
understanding of geometry, alignment, design and modeling of optical
elements, detector calibration and spectral response, and effective 
operational procedures.

\begin{table}[t]
\caption{Estimated systematic uncertainties in measured beam size
(in\mi), and their quadrature sums, for $\siz\approx 12.0$\mi\ and
$\Eb=2.1$\gev, as measured using no filter (NF) or the diamond 
filter (Di), of the pinhole (PH) and coded aperture (CA1) optical 
elements, listed by their characteristics $P$, for which 
variations $\Delta P$ away from nominal induce beam size uncertainty.}
\label{tab:syserr}
\setlength{\tabcolsep}{0.80pc}
\begin{center}
\begin{tabular}{lccc}
\hline\hline
\rule[10pt]{-1mm}{0mm}
Characteristic $P$ & $\Delta P$ & PH & CA1 \\
\hline
\rule[10pt]{-1mm}{0mm}
Pinhole gap size         & 2.0\mi   &   0.60  &  -   \\
Pinhole lip width        & 2\mi     &   1.33  &  -   \\
Etched slit size         & 0.5\mi   &   -     & 0.28 \\
Au mask thickness        & 0.02\mi  &   -     & 0.40 \\
Optic vertical alignment & 44\mi    & $\sim$0 & 0.30 \\
X-ray spectrum           &  -       &   0.50  & 0.35 \\
Magnification            &  0.1\%   &   0.01  & 0.03 \\
Quadrature sum           &  -       &   1.54  & 0.67 \\
\end{tabular}
\end{center}
\end{table}

\subsection{Bunch Oscillations}
\label{sec:osc}

Turn-by-turn measurements occupy the time domain, but can be displayed
in the frequency domain via a discrete Fourier transform. 
Because each turn represents
a time interval of 2.56~$\mu$s at CESR, the frequency domain is 
$[0,390.6]$~kHz, for which the result is symmetric about (390.6/2)~kHz and
higher frequencies are mapped into this interval as harmonics. The
intensity at each frequency is commonly referred to as the power and
expressed in dB, relative to an arbitrary reference value. Because beam
oscillations at CESR are generally above half-integer 
the convention is to plot only in $[390.6/2, 390.6]$~kHz, with
the property that any frequency $f$ below this range is mapped to an
observed frequency $390.6$~kHz$\;-f$. If there are $N_t$ turns of data in the
time domain, the bin size in the frequency domain is 390.6~kHz$/N_t$.

\begin{figure}[t]
   \includegraphics*[width=\figwid]{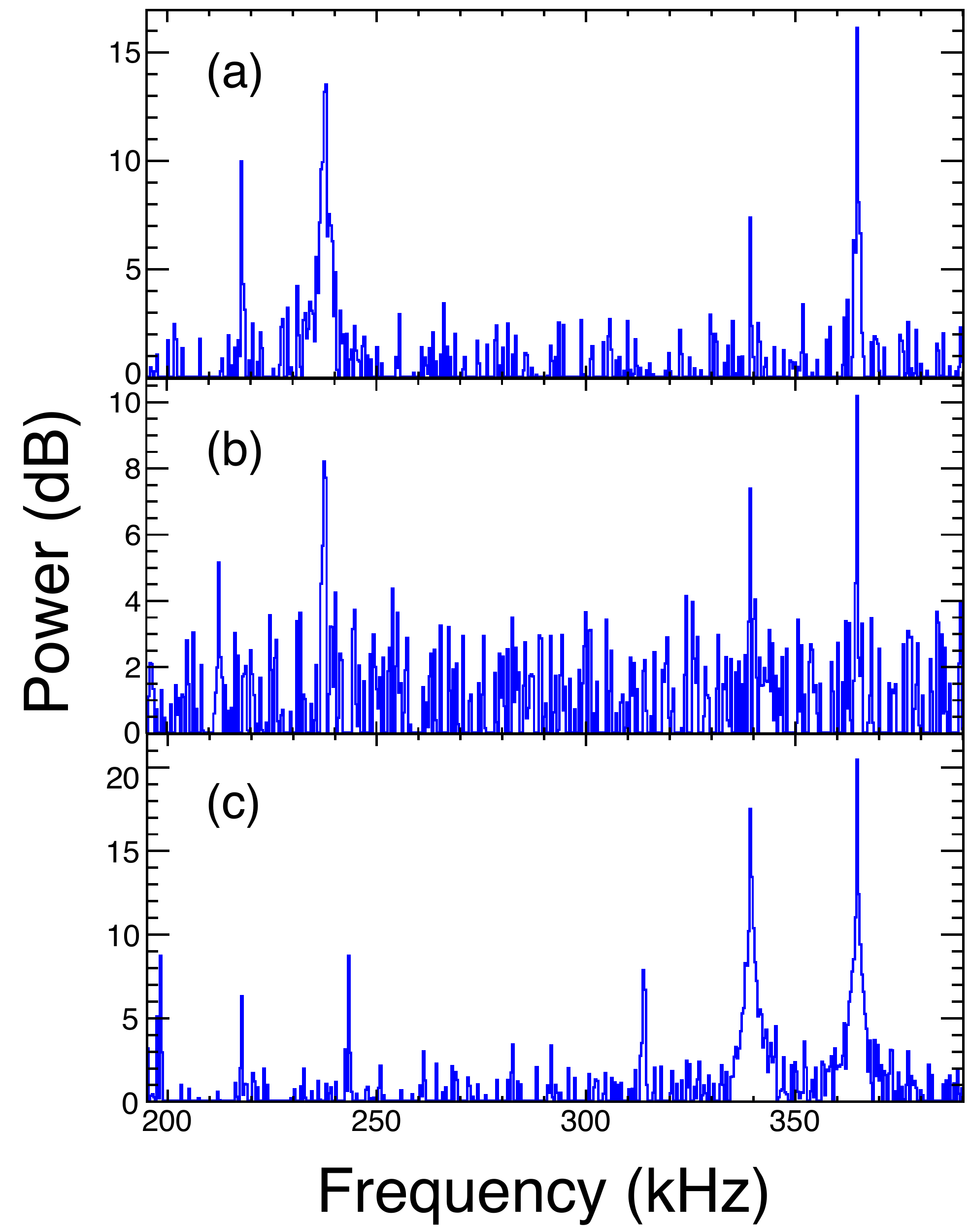}
   \caption{Fourier transforms of measured (a) vertical beam position, 
     (b) vertical beam size, and
     (c) intensity of the image for an $e^+$ bunch at $\Eb=2.1$\gev\
     roughly in the middle of
     train of 30 bunches spaced by 4\ns.
     For reference for this data set the horizontal, vertical and synchrotron
     tunes were 227, 243, and 25~kHz, respectively,
     with the synchrotron dipole and quadrupole tunes visible as peaks at 
     365 and 340~kHz, respectively. The revolution frequency of CESR is 390.6~kHz.}
\label{fig:fft}
\end{figure}

Three quantities measured by the xBSM lend themselves to frequency
analysis: vertical beam position, vertical beam size, and image
intensity. The interpretation of oscillations in image intensity is 
ambiguous; it could indicate vertical, horizontal, and/or longitudinal beam 
oscillations. Vertical intensity oscillations will occur due to the
Gaussian profile in vertical angle of the synchrotron radiation and
because for large vertical excursions the image can move 
partially or completely off the detector.
Horizontal excursions of the beam relative to the nominal orbit
could result in some of the synchrotron radiation being blocked by one or 
more apertures; the width in the horizontal plane of the aperture 
at the optical element is $\sim$1~mm at $\sim$4.4~m from the source point. 
Longitudinal oscillations could not only couple with lateral
oscillations, changing the direction of the emitted radiation, but also 
move the arrival time of the radiation away from the timing
synchronization point, thus reducing recorded pulse height.

\Fig{fig:fft} shows these three quantities for data acquired with the
pinhole optical element at $\Eb=2.1$\gev\ and 1024 turns. 
Such measurements are critical for electron cloud beam dynamics studies.  

\begin{figure}[t]
   \includegraphics*[width=\figwid]{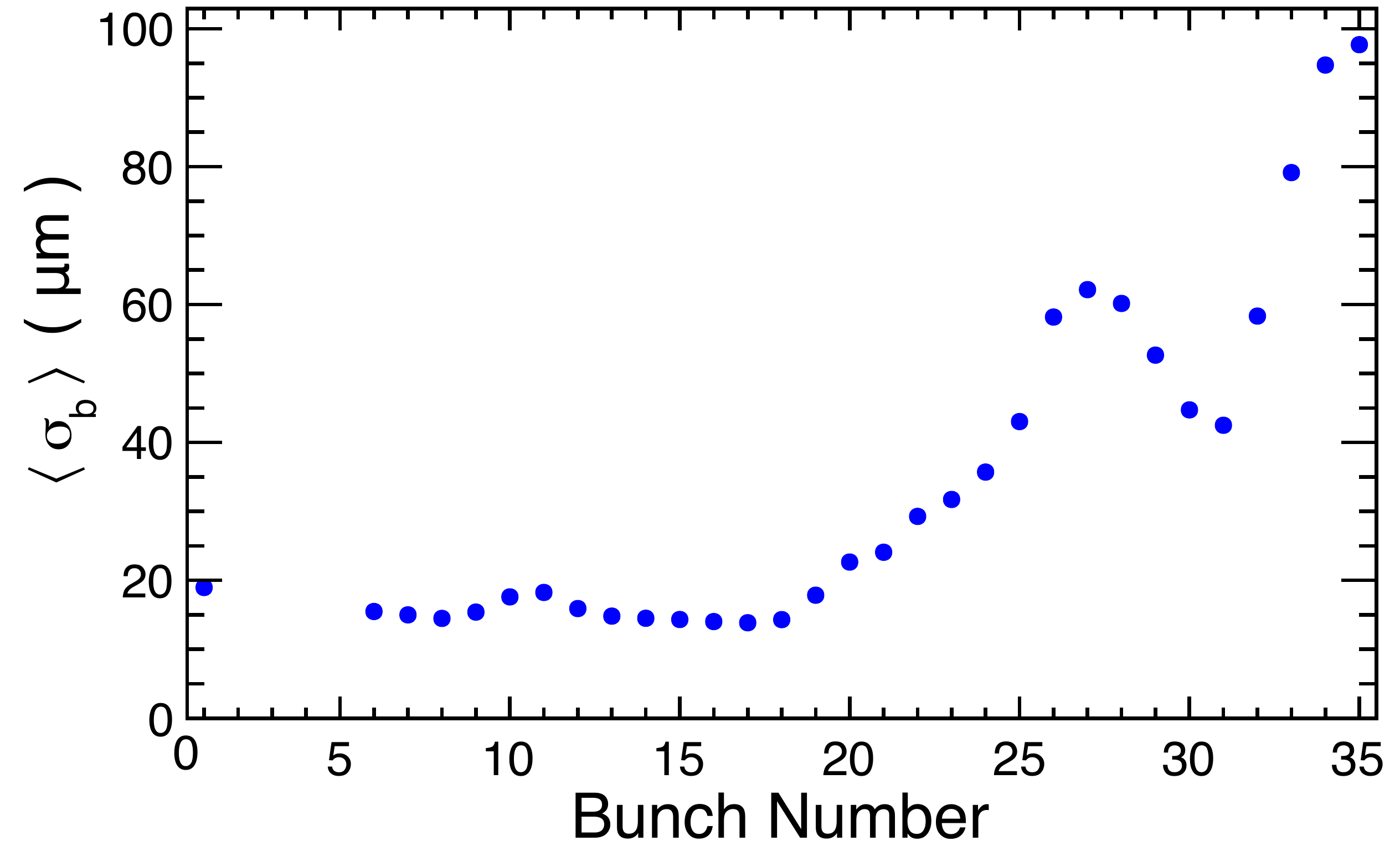}
   \caption{Turn-averaged beam size by bunch number for a run with 30
     bunches in a train with 4~ns bunch separation and an additional
     precursor bunch 20~ns ahead of the train.}
\label{fig:sizbybunch}
\end{figure}

\subsection{Multi-bunch performance}
\label{sec:mbp}

\begin{figure}[tbp]
   \includegraphics*[width=\figwid]{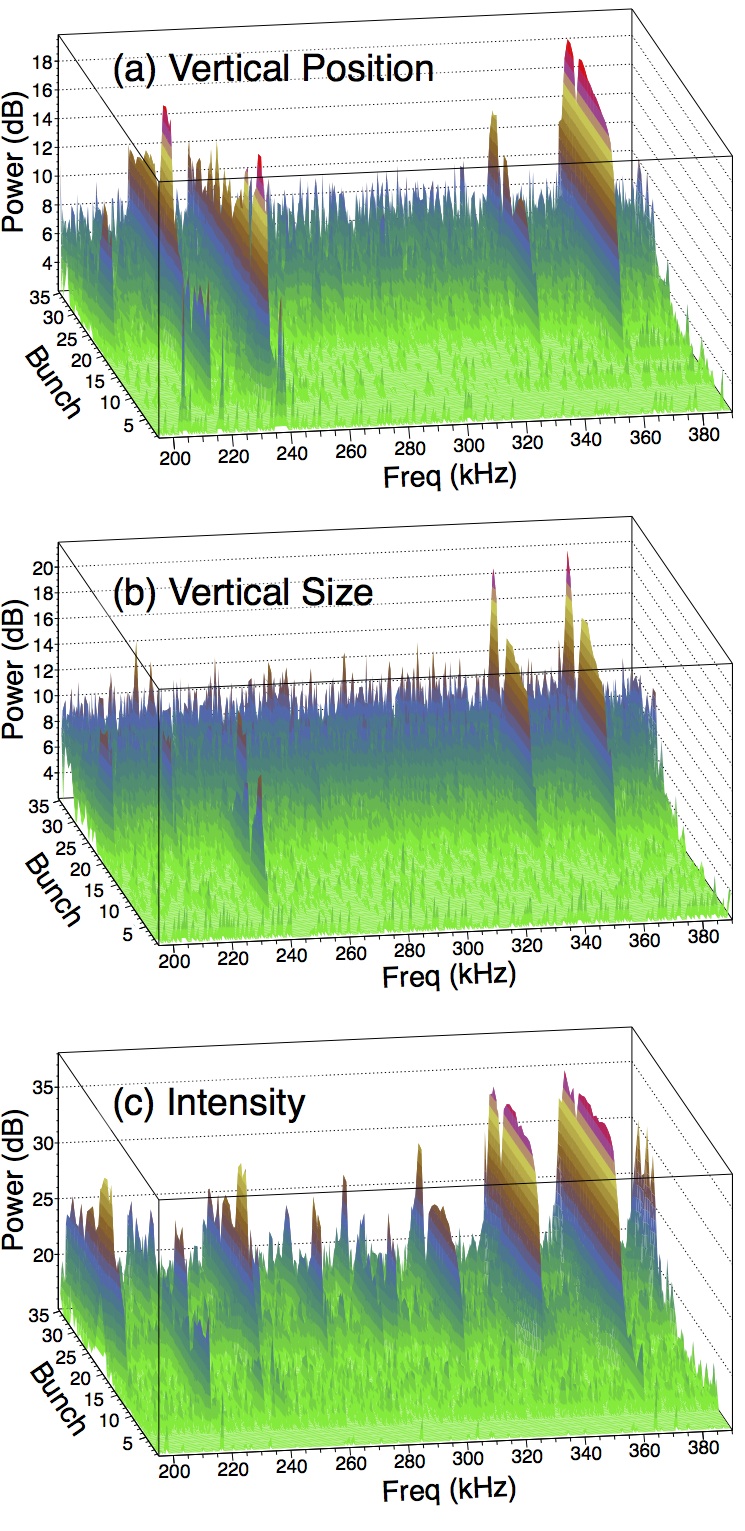}
   \caption{Fourier transforms, by bunch number, of measured 
     (a) vertical beam position, 
     (b) vertical beam size, and
     (c) intensity of the image.}
\label{fig:fftbybunch}
\end{figure}

Dynamics of multi-bunch running with short intervals between bunches is
of substantial interest, in order to learn how to maintain
bunch properties for all bunches in a train despite multi-bunch instability
effects due to the
electron cloud buildup for positrons. 
For example, \Fig{fig:sizbybunch} shows beam size as a
function of bunch number for 30 bunches with 4~ns spacing, preceded
by a {\it precursor} bunch 20~ns ahead of the train; without the precursor
bunch, which is meant to temporarily sweep away the electron cloud, 
bunch \#6 tends to have larger beam size. One of the goals of multi-bunch
studies is to understand the bunch size increase at and after the sixteenth
bunch and the
effects of various spacings of precursor bunches on the beam size of 
the first bunch in the train. 
The frequency-domain quantities
described in \Sec{sec:osc} are well-matched to study these
phenomena. \Fig{fig:fftbybunch} shows the position,
size, and intensity oscillations by bunch number. Beam size
blow-up is correlated with introduction of oscillations in all three
quantities at known beam oscillation frequencies. The xBSM is,
by design, uniquely suited to analyze these behaviors under a variety of
conditions. 

\section[Conclusions]{Conclusions}
\label{sec:conc}

We have described successful installation and operation of $e^\pm$ x-ray 
beam size monitors (xBSM) for CESR-TA. The xBSM uses synchrotron radiation
from a hard bend magnet as the light source.
For $E_b\approx 2$\gev, the xBSM images x-rays of $\enr\approx1$-10\kev\ 
($\lambda\approx 0.1-1$~\nm) onto a photodiode array.
The xBSM has measured single-shot vertical 
beam sizes from 10-100\mi\ with absolute systematic uncertainty of $\pm1$\mi.
Single-shot performance holds
down to 0.1~mA/bunch and 4\ns\ bunch separation.
Acquiring and analyzing xBSM data of high quality has benefited
from several lessons learned during development of the project.
Having the detector and optics share the CESR vacuum has allowed
windowless operation, extending the operational ranges of beam energy
and beam current downward. Several startup protocols
have proved to be essential for ensuring reliable operation, including
geometrical scans of pinhole opening; 
positions of detector, optical element, and filter elements;
timing scans of the electronics; and real-time adjustment of $e^\pm$ 
orbits. Calibrations of the detector electronics in pulse height
and bunch-to-bunch crosstalk were routinely acquired and utilized
in order to properly interpret the data.
Precision measurements of beam size and oscillations
were obtained only after development of extensive
analysis machinery to calibrate the apparatus,
model geometrical and spectral response, and fit
bunch-by-bunch, turn-by-turn beam crossing data
using different optical elements and spectral filters.
Study of beam-size resolving-power has improved
our ability to optimize pinhole openings and coded
aperture patterns for best performance.
The xBSM has already enabled important
investigations of beam dynamics,
including electron cloud studies~\cite{ref:mb},
intra-beam scattering~\cite{ref:ibs},
low-emittance tuning~\cite{ref:let},
and horizontal crabbing~\cite{ref:crabbing}.
Future development will include development of more
effective optical elements, measuring performance up
to $\sim$5\gev, and further work aimed toward making
the device a real-time or nearly real-time machine-tuning tool.

\clearpage
\section*{Acknowledgments}

{\small
This work would not have been possible without the meticulous efforts of
the CESR Operations Group and support of the Cornell Laboratory for
Accelerator-based Sciences and Education (CLASSE) and Cornell High Energy 
Synchrotron Source (CHESS). This research was supported under the National 
Science Foundation awards PHY-0734867, PHY-1002467, PHYS-1068662, Department 
of Energy contracts DE-FC02-08ER41538 and DE-SC0006505, and by the NSF and 
National Institutes of Health/National Institute of General Medical Sciences 
under NSF award DMR-0936384.

The authors thank M.~Billing, N.~Eggert, B.~Kreis, H.~Liu, M.~McDonald,
and M.~Palmer for their participation in and contributions to the xBSM project.
}


\end{document}